\documentstyle[a4,cite,12pt,aps,epsfig]{revtex}
\input axodraw.sty

\def\nnd{\end{document}}
\def\dgr{\dagger}
\def\nn{\nonumber}
\def\nnb{\nonumber}

\def\be{\begin{equation}}
\def\ee{\end{equation}}
\def\mn{\mu\nu}
\newcommand{\bea}{\begin{eqnarray}}
\newcommand{\eea}{\end{eqnarray}}

\def\dd#1{ 8 \pi^2 {d #1\over d t}}
\def\wbra#1{\left  ( #1  \right ) }
\def\sbra#1{\Big   ( #1  \Big   ) }
\def\mbra#1{\bigg  [ #1  \bigg  ] }
\def\bbra#1{\Bigg \{ #1  \Bigg \} }
\def\wpng#1#2{\frac{1}{#2} \wbra{#1}}

\def\mpng#1#2{\frac{1}{#2} \mbra{#1}}

\def\fwpng#1#2#3{\frac{#1}{#3} \wbra{#2}}
\def\fspng#1#2#3{\frac{#1}{#3} \sbra{#2}}
\def\fmpng#1#2#3{\frac{#1}{#3} \mbra{#2}}

\def\cma{\,,}
\def\hA{\widehat A}

\def\hWp{\widehat W^{+}}
\def\bWp{\overline W^{+}}
\def\hWm{\widehat W^{-}}
\def\bWm{\overline W^{-}}
\def\hZ{\widehat Z}
\def\bZ{\overline Z}

\def\wsep{ \nnb \\ &&}
\def\ssep{\right. \nnb\\ && \left.}
\def\eed{\end{document}}
\def\ch{\overline h}
\def\lvec#1{{\stackrel{\leftharpoonup}{#1}}}
\def\rvec#1{{\stackrel{\rightharpoonup}{#1}}}
\def\al{\alpha}
\def\ab{\alpha\beta}
\def\al{\alpha}
\def\be{\beta}

\makeatother

\begin{document}

\draft

\title{The renormalization of the effective gauge theory
	with spontaneous symmetry breaking:
	the $SU(2)\times U(1)$ case}

\author{
             Qi-Shu YAN\footnote{
        E-mail Address: yanqs@mail.ihep.ac.cn} and
		Dong-Sheng Du\footnote{
        E-mail Address: duds@mail.ihep.ac.cn}\\
	Theory division,
        Institute of high energy physics,
	Chinese academy of sciences, Beijing 100039,
	Peoples' Republic of China
}
\bigskip

\address{\hfill{}}

\maketitle


\begin{abstract}
We formulate the electroweak chiral Lagrangian in its
mass eigenstates, and study the its one-loop renormalization
and provide its renormalization group equations
to the same order, so as to complete
it as the low energy effective theory of the
standard model below a few TeV. 
In order to make our computation consistent,
we have provided a modified power counting rule
to estimate the contributions of
higher loop and higher operators.
As one of the application
of its renormalization group equations, we
analyze the solution to the effects
of the Higgs scalar. We find that similar to the
$SU(2)$ case, that the triple anomalous couplings
are sensitive to the quartic couplings (here $\al_5$). While the
quadratic anomalous couplings are not sensitive, due to
the large leading contributions and the accidental
cancellation. The differences in
the triple anomalous couplings between the direct method
and renormalization group equation method are well within
the detection power of the LHC and LC, if the Higgs scalar
is not too heavy (say, 300 or 400 GeV).
We also suggest a new mechanism to generate the
negative $S$ parameter through the radiative corrections
of the anomalous couplings.
Comparison of the renormalization group equation method and direct methods is provided
in the full theory, the standard model, to reveal the basic differences of them.
The problem of the unitarity violation is also addressed
for our assumption in the modified power counting rule.
\end{abstract}
\pacs{}

\section{Introduction}

The effective field theory (EFT) method \cite{wein, georgi, pich}
is a universal and powerful
theoretical framework for us to understand the laws and rules
of the nature. According to its spirit \cite{georgi, pich},
the practical ingredients of this method should at least include
the following two basic constituents: 1) the most important dynamic
degrees of freedom (DOFs) and 2) the most important interactions among
these DOFs. For some cases where quantum corrections are considerable,
we need the third ingredient, 3) the renormalization group equations (RGEs),
which is indispensable to efficiently sum up large logarithms, like in
B physics. 4) The only remnants of the higher dynamics
are reflected by the initial conditions of the anomalous couplings (ACs)
at the matching scale, which can affect low energy phenomenologies
when the RGEs develop from the ultraviolet cutoff ( the matching scale )
of the effective theory down to the infrared cutoff ( below which some of
the low energy DOFs of the effective theory will decouple and the
effective description will break down, and a new effective theory should
be introduced).

There are two important effective theories, the QCD chiral Lagrangian \cite{qcdcl}
and the electroweak chiral Lagrangian (EWCL)\cite{long,appel}. 
These two theories are suitable to describe hadronic
dynamics from $100$ MeV to $1$ GeV ($4\pi v_0$, with $v_0=92$ MeV) and
electroweak dynamics from $90$ GeV to 2.5 TeV ($4\pi v_0$, with
$v_0=246$ GeV), respectively. The RGEs of the Chiral perturbation theory (ChPT)
have been studied up to $O(p^6)$ order \cite{bijnens}. While
the RGEs of the EWCL is still lack. 

The RGEs of the EWCL are necessary
due to the fact that most of the ACs (the anomalous in ACs
means the deviation from the requirement of the renormalizability,
since after integrating out the heavy DOF, the divergences generated by
low energy DOF can not be canceled out so that the anomalous operators (AOs) must be introduced )
are induced by loop processes in the standard model (SM)
and are small, so that if the ACs from
new physics are not small, the radiative corrections
of low energy DOF might be relatively
important. Since we do not know the actual mechanism of the
electroweak symmetry breaking, and
the ACs can be considerable large, as in the not too heavy
Higgs case we consider below. The present experimental
constraint on the
ACs \cite{ypkss} can be summarized and estimated as
\bea
\label{expcon}
(\al_1\,\,\al_8\,\,\beta) \sim O(0.01),\,\,
(\al_2\,\,\al_3\,\,\al_9)\sim O(0.1),\,\,
(\al_4\,\,\al_5\,\,\al_6\,\,\al_7\,\,\al_a)\sim O(1)\,,
\eea
which indicates that the radiative corrections from the permitted ACs
could be large. Another pure theoretical reason for us to consider the
RGEs of the EWCL is that up to the $O(p^4)$, the 11 extra operators
belong to the marginal operators in the Wilsonian renormalization
method \cite{wilson, pol}, we just want to know the behavior of these ACs
under the drive of quantum fluctuations. More practically,
considering the fact that the new machines, the LHC and future linear
colliders (TESLA or JLC), can increase their measurement precision
up to two orders than the LEP, it is urgent to upgrade the theoretical
prediction precisions of the electroweak theory to the same order.
Furthermore, once these machines will start to run, it
is quite necessary for the experimentalists
to conduct their analysis in a universal and model-independent way,
then the RGEs can meet this end, and can provide a powerful tool to
comparatively study most of the new physics candidates.
So, deriving the RGEs of EWCL and taking into account the radiative
corrections of low energy DOF become important, at least should be
as important as to consider the contributions of operators at $O(p^6)$ order.

The study of the EWCL is started from quite early time, and
can be traced back to the references \cite{long}, where the author began to
study the independent operator set of the EWCL up to $O(p^4)$, 
and in 1993, the authors of \cite{appel}, A. Appliquist and
G. H. Wu, established the relations of the ACs
in the EWCL with the usual precision test quantities \cite{precision}.
The ACs in various models have been derived in \cite{effcoup}.
The authors of the reference \cite{ewhiggs} also extended the EWCL
to include one light Higgs, and recently, the reference \cite{hjhe}
explored the light Higgs case to the precision test constraints.
In 1996, the authors of \cite{hky}, H. J. He, Y. P. Kuang,
and C. P. Yuan, by using the equivalence theorem \cite{eqth} and the modified
power counting rule, qualitatively and quasi-quantitative estimated
the detection power of several possible new machines to these effective operators.
The authors of \cite{fermion} have extended the bosonic EWCL to the
fermionic part. Recently, the authors \cite{mass} have explored to formulate
EWCL in the partial mass eigenstates.
Several groups even have studied the complete set of operators
of $O(p^6)$, and their possible effects in the LHC and LC are also 
analyzed \cite{op6}.

People expect that these operators up to $O(p^6)$ might be important in LHC and
LC. Considering the possible situation for the measurement at the LHC and
LC, the ACs can be determined at 200 GeV, 500 GeV, and 1 TeV, for instance.
The values of these couplings might be definitely measured
different considering the precision that these
machines can reach, just like that $\alpha_{em}=1/137$ at $m_e$ 
and $\alpha_{em}=1/128$ at
$m_Z$. Then it is urgent for us to understand the underlying reason
for these differences.

There are two obstacles for us to study the renormalization of the EWCL
and derive its RGEs. 1) The non-linear effective gauge theory is
a non-renormalizable theory (Non-renormalizability here means that there will be
the tower of infinite divergence structure and quartic divergences even
at one-loop level), how to
consider its renormalization order by order? 2) There are so
many interaction vertices, how to efficiently evaluate the loop contributions
of those large amount of Feynman diagrams? We have shown the basic conceptions in
our paper \cite{our1} as to overcome the first obstacle in the framework of
effective theory, and the second one by using the background field
method \cite{bfm} (BFM).
In our previous work \cite{our2}, by using the related conceptions and
tools, we have found that when the quartic ACs
are large (In the Higgs model, corresponding to the case that the
Higgs is not too heavy), the contributions of low energy dynamic DOF
can be quite significant, and the predictions of the RGE method
and the direct method (DM) (where the mass squared terms are dropped due to
the fact that in the decoupling limit those terms can be safely neglected)
are quite different. So it is natural and logic for
us to examine these properties in the EWCL case.

In this paper, we will
study the one-loop renormalization of the EWCL and
derive its RGEs. In order to systematically and consistently
control the contributions of the
higher loop and higher operators, we provide a modified power counting
rule. Considering the fact that the complete RGEs are
very complicated (we will provide
in our next paper \cite{our3}), here we only provide the simplified but workable
ones. But we would like to emphasize that our numerical analysis
is based on the complete RGEs and is reliable.
We will use the conceptions developed
in our previous works. The basic computational skills include
the BFM \cite{bfm}, the Stueckelberg transformation \cite{stuck},
the Schwinger proper time and heat kernel method \cite{htkl}, and the covariant
short distance expansion technology \cite{wkb}.
With these conceptions and methods, we will extract the desired
one loop RGEs of EWCL, so, theoretically, as to complete it as a realistic
theoretical framework to describe the SM below a few TeV. We will
also examine the Higgs' contribution to the low energy
precision test parameters. We find that similar to the
$SU(2)$ case, the triple gauge vertices are quite sensitive to
the large quartic ACs. While the quadratic ones
are not sensitive for the EWCL of Higgs model, and the underlying
reason is due to the large leading contributions and
the accidental cancellation between $\al_2$ and $\al_3$.
For those cases where there is no such cancellation, the
quadratic vertices might also be sensitive to the large quartic ACs
couplings. We also compare the predictions given by the DM.
The results given by these two methods are quite different in the
case of EWCL of Higgs model, and the differences are well
within the detection power of the LHC and LC. The basic reason
for the differences are revealed in the full theory, the SM.
We also address the problem of the unitarity violation related with
our modified power counting rule.

For the conventions of the computation below,
we would like to emphasize that, in order to avoid inconsistency,
all formula are provided in the Euclidean space, including
the partition functional from the beginning.

This paper is organized as follows.
In Sec. II, the bosons sector of the standard model
is introduced. In Sec. III, the EWCL ${\cal L}_{ew}$
up to $O(p^4)$ are introduced, we introduce a new basis
of the mass eigenstates for the effective operators.
In Sec. IV, the renormalization of the SM in BFM
is conducted to provide a reference for the effective ones.
In Sec. V, we use the BFM to extract the
quadratic terms of quantum fields, to evaluate
the logarithm and trace, and to construct the RGEs.
In Sec. VI, the ACs of
the EWCL of the Higgs model is analyzed and
the Higgs' contribution to electroweak precision test
parameters are studied. We end this paper with several discussions
and conclusions. The appendix is devoted to provide the related matrices
of the quadratic terms of the standard form.

\section{The standard model}
The Lagrangian of the standard $SU(2)\times U(1)$ gauge theory
can be formulated as
\bea
\label{eq:sm}
{ \cal L} &=& - H_1 - H_2 - (D\phi)^{\dagger}\cdot(D \phi) \wsep
+\mu^2 \phi^{\dagger} \phi - {\lambda \over 4} (\phi^{\dagger} \phi)^2
+ {\cal L}_{\psi}+{\cal L}_{Yukawa}\cma\\
H_1&=&{1\over 4}  W_{\mn}^a W^{a \mn}\cma\\
H_2&=&{1\over 4}  B_{\mn}   B^{\mn}\cma
\eea
where the $W$ and $B$ are the vector bosons of $SU_L(2)$ and $U_Y(1)$ gauge
groups, respectively. The $\phi$ is the Higgs field, a weak doublet scalar.
The $\mu^2$ and $\lambda$ are two variables of the Higgs potential, which
determine the spontaneous breaking of symmetry. The ${\cal L}_{\psi}$ and
${\cal L}_{Yukawa}$ are the standard gauge interactions of Fermions
and the Yukawa coupling between the Higgs field and Fermions, respectively.
For the sake of simplicity, the interactions of Fermions are neglected below.
And the relevant definitions are listed below
\bea
W_{\mn}^a&=&\partial_{\mu} W_{\nu}^a - \partial_{\nu} W_{\mu}^a +g f^{abc} W_{\mu}^b W_{\nu}^c\,,\\
B_{\mn} &=&\partial_{\mu} B_{\nu} - \partial_{\nu} B_{\mu} \cma \\
D_{\mu} \phi&=&\partial_{\mu} \phi - i g W_{\mu}^a T^a - i y_{\phi} g' B_{\mu} T^3 \phi\,,\\
\phi^{\dgr}&=&(\phi_1^*,\phi_2^*)\,,
\eea
where $T^a$ are the generators of the Lie algebra of
$SU(2)$ gauge group, and $a=1,\,2,\,3$. The
Y charge of the field $\phi$ and $y_{\phi}=-1$.
The $g$ and $g'$ are the couplings of the corresponding
gauge interactions, respectively.

The spontaneous symmetry breaking is induced by
the positive mass square $\mu^2$ in the Higgs potential.
The vacuum expectation value of Higgs
field is solved from the Higgs potential
as $|\langle \phi \rangle| = v/{\sqrt 2}$.
And by eating the corresponding
Goldstone bosons, the vector bosons $W$ and $Z$ obtain their masses, while
the vector bosons $A$ of the unbroken $U(1)_{em}$ gauge symmetry are still massless.
The Lagrangian given in Eq. (\ref{eq:sm})
with the Higgs mechanism can be reformulated in its nonlinear
form by changing the variable $\phi$
\bea
\label{eq:chv}
\phi={1\over \sqrt{2}} (v + h) U\,,\,
U=\exp\left ({\bf 2} {i \xi^a T^a\over v }\right )\,,\,
v=2 \sqrt{\mu^2 \over \lambda}\,,
\eea
where the $h$ is the Higgs scalar, $v$ is the vacuum expectation value.
The $U$ is a phase factor, and the $\xi^a,\,a=1,\,2,\,3$ are the corresponding
Goldstone bosons as prescribed by the Goldstone theorem.

As we know, the change of variables in Eq. (\ref{eq:chv}) induces
a determinant factor in the functional integral ${\cal Z}$
\bea
{\cal Z} = \int {\cal D} W_{\mu}^a {\cal D} h {\cal D}\xi^b
\exp\sbra{- {\cal S}'[W,h,\xi]} \det\bbra{ \sbra{1 + {1\over v} h} \delta(x-y) }\,.
\eea
and correspondingly modifies the Lagrangian density to
\bea
\label{hml}
{\cal L} &=& - H_1 - H_2 - {(v+h)^2 \over 4} tr[D U^{\dagger} \cdot D U] \nnb\\
&& - {1\over 2} \partial h \cdot \partial h
+ {\mu^2 \over 2 } (v + h)^2 - {\lambda \over 16} (v+h)^4
- \delta^4(0) ln \left \{ 1 + {1\over v} h \right \}\,.
\label{uglag}
\eea
As pointed out by several references \cite{ugauge},
this determinant containing quartic divergences is
indispensable and crucial to cancel exactly the quartic
divergences brought into by the longitudinal part
of vector bosons, and is important in verifying the
renormalizability of the Higgs model in the U-gauge
and the equivalence of U-gauge to other gauges.

\section{The effective Lagrangian up to $O(p^4)$ (the relevant and marginal operators)}
The most general effective Lagrangian ${\cal L}_{EW}$,
which respects the Lorenz invariance, the $SU(2)\times U(1)$ gauge
symmetry, and the discrete symmetries (the charge, parity, and the combined
CP symmetries), can be formulated as
\bea
{\cal L}_{EW} &=&{\cal L}_{EW}^{p^2} + {\cal L}_{EW}^{p^4} +\cdots+{\cal L}_{qd}\label{ewla}\\
{\cal L}_{EW}^{p^2}&=&{\cal L}_B\cma\\
{\cal L}_{EW}^{p^4}&=&\beta {\cal L}_0
+\sum_{i=1}^a{\alpha_i } {\cal L}_i\,
\eea
where
\bea
{\cal L}_{B}&=&- H_1 - H_2 + {\cal L}_{WZ}\cma\\
{\cal L}_{WZ}&=&{v^2 \over 4} tr(V\cdot V)=- {v^2 \over 8} (G^2 Z\cdot Z + 2 g^2 W^+ \cdot W^-)\,,
\eea
where $G$ is defined as $G=\sqrt{g^2 + g'^2}$. After using the relations of Lie algebra and
the classic equation of motion to eliminate
the redundant operators, the complete
Lagrangian ${\cal L}_{EW}^{p^4}$ includes the
following independent operators \cite{long, appel}:
\bea
\label{ewclb}
{\cal L}_0&=&  {v^2 \over 4}[tr({\cal T} V_\mu)]^2\cma\nn\\
{\cal L}_1&=&i {g g'\over 2} B_{\mu\nu}tr({\cal T} W^{\mu\nu})\cma\nn\\
{\cal L}_2&=&i {g'  \over 2} B_{\mu\nu}tr({\cal T} [V^\mu,V^\nu])\cma\nn\\
{\cal L}_3&=&i    g        tr( W_{\mu\nu}[V^\mu,V^\nu])\cma\nn\\
{\cal L}_4&=&             [tr(V_\mu V_\nu)]^2\cma\nn\\
{\cal L}_5&=&             [tr(V_\mu V^\mu)]^2\cma\nn\\
{\cal L}_6&=&             tr(V_\mu V_\nu)tr({\cal T} V^\mu)tr({\cal T} V^\nu)\cma\nn\\
{\cal L}_7&=&             tr(V_\mu V^\mu)[tr({\cal T} V^\nu)]^2\cma\nn\\
{\cal L}_8&=&  {g^2\over 4} [tr({\cal T} W_{\mu\nu})]^2\cma\nn\\
{\cal L}_9&=&i {g  \over 2} tr({\cal T} W_{\mu\nu})tr({\cal T} [V^\mu,V^\nu])\cma\nn\\
{\cal L}_a&=&             [tr({\cal T} V_\mu)tr({\cal T} V_\nu)]^2\,.
\label{ewcle}
\eea
where the auxiliary variable $V_{\mu}$ and ${\cal T}$ is defined as
\bea
V_{\mu}&=&U^{\dagger} (\partial_{\mu} - i W_{\mu}^a T^a) U  + i B_{\mu} T^3\,.\\
{\cal T}&=& 2 U^{\dagger} T^3 U=U^{\dagger} {\tau}^3 U\,,
\eea
with the ${\tau}^3$ is the third Pauli matrices.
The operators $H_1$, $H_2$, and ${\cal L}_i,\, i=1,\,\cdots,\,a$
contribute the kinetic, trilinear, and quartic interactions.
While operators ${\cal L}_{WZ}$ and ${\cal L}_0$
contribute to the mass terms.

The effective Lagrangian ${\cal L}_{EW}$
is invariant under the following local chiral transformation
 \bea
\label{}
U &\rightarrow& g_L U g_R^\dagger \cma\nnb\\
W_\mu &\rightarrow& g_L W_\mu g_L^\dagger + i g_L \partial_\mu g_L^\dagger \cma \nnb\\
W_{\mu\nu}&\rightarrow& g_L W_{\mu\nu} g_L^\dagger \cma \nnb \\
B_\mu &\rightarrow& B_\mu + i g_R \partial_\mu g_R^\dagger\cma \nnb \\
B_{\mu\nu}&\rightarrow&  B_{\mu\nu}\cma
\eea
where the gauge transformation factor $g_L$ and $g_R $ are defined as
\bea
g_L = \exp{\bbra{ i {\alpha^{a}}_L T^a } } \,\cma\,\,
g_R = \exp{\bbra{ i        \beta_R T^3 } } \,.
\eea

The ACs $\alpha_i$ form the effective parameter space,
when the effective scale $\mu$ runs from its ultraviolet cutoff down to
its infrared cutoff, each theory will depict a characteristic curve
in this space. The initial conditions of the ACs at the
ultraviolet cutoff (the matching scale of the effective theory and the full theory)
reflect the remnant of high energy dynamics, while the effects of the heavy
DOF to the low energy dynamics can be solved out from the corresponding RGEs.
By measuring the ACs at different energy scales,
we can extract important information of the possible underlying theory
and induce the actual mechanism of the spontaneous symmetry breaking.

The Higgs model ( a full and renormalizable theory ) given in
Eq. (\ref{uglag}) can be effectively described
by the effective Lagrangian ${\cal L}_{EW}$ if the Higgs field
is heavy and integrated out. The equation of motion of
the Higgs field $h$ can express it into the low energy DOFs,
which reads
\bea
\label{eom}
h &=& - {v \over 2 m_0^2} tr(D U^{\dagger}\cdot D U) + \cdots\,,\nn\\
&=&{v \over 2 m_0^2} tr(V \cdot V) + \cdots \cma \\
m_0^2  &=& {1\over 2} \lambda v^2\,,
\eea
where $m_0$ is the mass of Higgs boson. The omitted
terms contain at least four covariant partials and belong
to operators $O(p^6)$.

At the matching scale, after matching the full theory and
the effective theory by integrating out the heavy Higgs scalar,
we get the following initial condition for the ACs
\bea
\beta(m_0) = 0\,,\,\,
\al_5(m_0)   = {v^2 \over 8 m_0^2}={1 \over 4 \lambda}\,,\,\,
\al_i(m_0)   = 0\,, i \neq 5,
\label{init}
\eea

Since the couplings of these AOs are dimensionless,
naively from the power law we expect that the complete
Lagrangian with both relevant and marginal operators is renormalizable. 
But, as well known, the longitudinal
part of the propagators of vector bosons will invalidate
this power counting law. Here we emphasize that
the terms in the ${\cal L}_{qd}$
make it possible in practical computation to simply and consistently
discard those quartic divergences, and make it possible to
renormalize the effective Lagrangian order by order.

The above set of operator is formulated in the interaction
eigenstates, and below we introduce
an equivalent basis (in U-gauge) represented in its mass
eigenstates $A$, $W^{\pm}$, and $Z$ (these particles are the ones
detected in experimental facilities), which read
\bea
L_{EW}&=&-\sum_{i=1}^{4} C_i O_i+\sum_{i=5}^{c} C_i O_i - O_{M_W} - \rho O_{M_Z} \cma\\
O_1 & = &{1\over 4} A_{\mu\nu} A^{\mu\nu}\cma\nn\\
O_2 & = &{1\over 2} A_{\mu\nu} Z^{\mu\nu}\cma\nn\\
O_3 & = &{1\over 4} Z_{\mu\nu} Z^{\mu\nu}\cma\nn\\
O_4 & = &{1 \over 2} W^{+}_{\mu\nu} W^{-\mu\nu}\cma\nn\\
O_5 & = &{i \over 2} A_{\mu\nu} W^{+\mu} W^{-\nu}\cma\nn\\
O_6 & = &{i \over 2} Z_{\mu\nu} W^{+\mu} W^{-\nu}\cma\nn\\
O_7 & = &{i \over 2} \sbra{W^{+}_{\mu\nu} Z^{\mu} W^{-\nu}
	- W^{-}_{\mu\nu} Z^{\mu} W^{+\nu}}\cma\nn\\
O_8 & = & Z \cdot Z W^{+} \cdot W^{-}\cma\nn\\
O_9 & = & Z \cdot W^{+} Z \cdot W^{-}\cma\nn\\
O_a & = & Z \cdot Z Z \cdot Z\cma\nn\\
O_b & = & W^{+} \cdot W^{-}  W^{+} \cdot W^{-}\cma\nn\\
O_c & = & W^{+} \cdot W^{+}  W^{-} \cdot W^{-}\cma\nn\\
O_{M_W}& = & {v^2 \over 4} W^{+} \cdot W^{-}\cma\nn\\
O_{M_Z} &= & {v^2 \over 8} Z \cdot Z\cma
\eea
where the first $12$ contribute to the kinetic, triple, and
quartic interactions, and the last two contribute to the masses
of vector bosons. This set of operators are all expressed
in the mass eigenstates, while the operators given in
\cite{mass} are only partially expressed in the mass eigenstates,
where the authors have only chosen $Z$ and $W^{\pm}$,
and used $B$ (which is not mass eigenstates). That's why we call
their basis as operators in partial mass eigenstates.
And the relevant definitions are given as
\bea
A_{\mn}&=&\partial_{\mu} A_{\nu} -\partial_{\nu} A_{\mu}\cma \nn\\
Z_{\mn}&=&\partial_{\mu} Z_{\nu} -\partial_{\nu} Z_{\mu}\cma\nn\\
W^{\pm}_{\mn}&=&d_{\mu} W^{\pm}_{\nu} - d_{\nu} W^{\pm}_{\mu} \nn\\
d_{\mu} W^{\pm}_{\nu} &=& \partial_{\mu} W^{\pm}_{\nu} \mp i e A_{\mu} W^{\pm}_{\nu}\,.
\eea

The fundamental relations between the mass eigenstates and the weak interaction
eigenstates are determined as
\bea
A & =&  \sin \theta_W W^{3} + \cos \theta_W B,\,\,
Z  = -\cos \theta_W W^{3} + \sin \theta_W B,\,\,\nnb\\
W^+&=&{1\over \sqrt{2}}(W^{1} - i W^{2}),\,\,
W^-={1\over \sqrt{2}}(W^{1} + i W^{2}),\,\,\nnb\\
e  &= &{g' g \over G},\,\,
\tan \theta_{W} ={g'\over g}
\,,
\eea
where the $theta_W$ is also called the Weinberg angle.
There are also some important relations, which we list
below
\bea
{\widetilde W^{+}_{\mn}} &=&W^{+}_{\mn}-i {g^2 \over G} F^{+}_{\mn}\,,\\
{\widetilde W^{-}_{\mn}} &=&W^{-}_{\mn}-i {g^2 \over G} F^{-}_{\mn}\,,\\
{\widetilde W^3_{\mn}}   &=&{g'\over G} A_{\mn} - {g \over G} Z_{\mn} - i g F^Z_{\mn}\,,\\
B_{\mn}                  &=&{g \over G} A_{\mn} + {g'\over G} Z_{\mn}\,,
\eea
where the definitions of $F^{+}_{\mn}$, $F^{-}_{\mn}$, and $F^Z_{\mn}$,
read
\bea
F^{+}_{\mn} &=& W^{+}_{\mu} Z_{\nu} - W^+_{\nu} Z_{\mu}\,,\\
F^{-}_{\mn} &=& Z_{\mu} W^{-}_{\nu} - Z_{\nu} W^{-}_{\mu}\,,\\
F^Z_{\mn}   &=& W^{+}_{\mu} W^{-}_{\nu} - W^{+}_{\nu} W^{-}_{\mu}\,.
\eea

These $14$ operators are linearly independent with
each other, and are equivalent with those $14$ bosonic terms
in the effective Lagrangian given in Eq. (\ref{ewla}).

Transformation relations of the masses operators of these two independent sets
are determined as
\bea
{\cal L}_{WZ}=O_{M_W} + O_{M_Z}\cma\,{\cal L}_0=- 2 O_{M_Z}\,.
\eea
And the relations of rest operators read
\bea
H_1&=& -\wpng{-2  g^{'2} O_1 + 2 g g^{'} O_2 - 2 g^2 O_3 - 2 G^2 O_4 + 4 g G g^{'} O_5 \ssep -
     4 g^2 G O_6  + 4 g^2 G O_7 - 2 g^4 O_8 + 2 g^4 O_9 - g^2 G^2 O_b +
     g^2 G^2 O_c}{2 G^2}
	\cma\nnb\\
H_2&=&-\wpng{- g^2 O_1 - g g^{'} O_2 -  g^{'2} O_3}{G^2}\cma\nnb\\
{\cal L}_1&=&-\wpng{-2 g^2  g^{'2} O_1 + g^3 g^{'} O_2 - g g^{'3} O_2 \ssep + 2 g^2  g^{'2} O_3 +
     2 g^3 G g^{'} O_5 + 2 g^2 G  g^{'2} O_6}{G^2}\cma\nnb\\
{\cal L}_2&=&\wpng{-2 g^3 g^{'} O_5 - 2  g^2  g^{'2} O_6}{G}\cma\nnb\\
{\cal L}_3&=&-2 {g^3 g^{'} \over G} O_5 + 2 {g^4 \over G} O_6 - 2 g^2 G O_7 +
   2 g^4 O_8 - 2 g^4 O_9 + g^4 O_b - g^4 O_c\cma\nnb\\
{\cal L}_4&=&\wpng{4 g^2 G^2 O_9 + G^4 O_a + 2 g^4 O_b + 2 g^4 O_c}{4}\cma\nnb\\
{\cal L}_5&=&\wpng{4 g^2 G^2 O_8 + G^4 O_a + 4 g^4 O_b}{4}\cma\nnb\\
{\cal L}_6&=&\wpng{2 g^2 G^2 O_9 + G^4 O_a}{2}\cma\nnb\\
{\cal L}_7&=&\wpng{2 g^2 G^2 O_8 + G^4 O_a}{2}\cma\nnb\\
{\cal L}_8&=&\fspng{g^2} {2  g^{'2} O_1 - 2 g g^{'} (O_2 + 2 G O_5) \wsep +
     g^2 \left (2 O_3 + G (4 O_6 + G O_b - G O_c) \right)} {2 G^2}\cma\nnb\\
{\cal L}_9&=&\fwpng{g^3} {-2 g^{'} O_5 + 2 g O_6 + g G O_b - g G O_c}{G}\cma\nnb\\
{\cal L}_a&=&{G^4 \over 2} O_a\,.
\eea
The reverse relations among these operators, which are quite helpful for us
to extract the standard structures of the EWCL given in Eq. (\ref{ewclb}), read
\bea
O_{M_Z}&=&-{1\over 2} {\cal L}_0\cma\nn\\
O_{M_W}&=&{\cal L}_{WZ} + {1 \over 2} {\cal L}_0\cma\nn\\
O_1&=&\wpng{g^4 H_2 + g^2 ({\cal L}_1 - {\cal L}_2) \ssep + g^{'2} (-{\cal L}_4 + {\cal L}_5
+ {\cal L}_6 - {\cal L}_7 + {\cal L}_8 - {\cal L}_9)}{g^2 G^2}\cma\nnb\\
O_2&=&\mpng{g^2 (2 g^{'2} H_2 - {\cal L}_1 + {\cal L}_2) \wsep +
    g^{'2} \left ( {\cal L}_1 - {\cal L}_2 + 2 ({\cal L}_4
- {\cal L}_5 - {\cal L}_6 + {\cal L}_7 - {\cal L}_8 + {\cal L}_9)\right )}{g G^2 g^{'}}\cma\nnb\\
O_3&=&\wpng{g^{'2} H_2 - {\cal L}_1 + {\cal L}_2 - {\cal L}_4 + {\cal L}_5
+ {\cal L}_6 - {\cal L}_7 + {\cal L}_8 - {\cal L}_9}{G^2}\cma\nnb\\
O_4&=&-\wpng{g^4 ({\cal L}_6 - {\cal L}_7) + G^4 {\cal L}_8
- g^2 G^2 (G^2 H_1 - {\cal L}_3 + {\cal L}_9)}{g^2 G^4}\cma\nnb\\
O_5&=&-\wpng{g^2 {\cal L}_2 + g^{'2} (2 {\cal L}_4 - 2 {\cal L}_5
- 2 {\cal L}_6 + 2 {\cal L}_7 + {\cal L}_9)}{2 g^3 G g^{'}}\cma\nnb\\
O_6&=&\wpng{-{\cal L}_2 + 2 {\cal L}_4 - 2 {\cal L}_5
- 2 {\cal L}_6 + 2 {\cal L}_7 + {\cal L}_9}{2 g^2 G}\cma\nnb\\
O_7&=&-\wpng{2 g^2 ({\cal L}_6 - {\cal L}_7)
+ G^2 ({\cal L}_3 - {\cal L}_9)}{2 g^2 G^3}\cma\nnb\\
O_8&=&\wpng{{\cal L}_7 - {\cal L}_a}{g^2 G^2}\cma\nnb\\
O_9&=&\wpng{{\cal L}_6 - {\cal L}_a}{g^2 G^2}\cma\nnb\\
O_a&=&\wpng{2 {\cal L}_a}{G^4}\cma\nnb\\
O_b&=&\wpng{2 {\cal L}_5 - 2 {\cal L}_7 + {\cal L}_a}{2 g^4}\cma\nnb\\
O_c&=&\wpng{4 {\cal L}_4 - 2 {\cal L}_5 - 4 {\cal L}_6
+ 2 {\cal L}_7 + {\cal L}_a}{2 g^4}\,.
\label{op4}
\eea

One of the advantages of the set of operators given in Eq. (\ref{op4}) is
that it is helpful to discuss different symmetry breaking patterns.
By setting all terms with $Z$ vanished we get the pattern
$SU(2)\rightarrow U(1)$ \cite{su221}; by setting all terms with $A$
vanished we get the pattern $SU(2)$ breaks to a global $U(1)$ if the mass
of $Z$ different from that of $W^{\pm}$.
In our paper \cite{our2}, we have studied the pattern that
a local $SU(2)$ breaks to a global $SU(2)$. For the different
symmetry breaking patterns, the corresponding RGEs can be
obtained by taking the limits to eliminate some
of the ACs from the effective Lagrangian.

The $\rho$ is related to $\beta$ as
\bea
\beta = {\rho - 1 \over 2}\,.
\eea
The relations of ECs between $\alpha_i$ and $C_i$
are determined as
\bea
\alpha_1&=&-\frac{C_1}{G^2} + \frac{g C_2}{G^2 g^{'}} - \frac{g^{'} C_2}{g G^2} + \frac{C_3}{G^2}\,,\\
\alpha_2&=&\frac{C_1}{G^2} - \frac{g C_2}{G^2 g^{'}} + \frac{g^{'} C_2}{g G^2} - \frac{C_3}{G^2} - \frac{C_5}{2 g G g^{'}} -
 \frac{C_6}{2 g^2 G}\,,\\
\alpha_3&=&\frac{C_4}{G^2} - \frac{C_7}{2 g^2 G}\,,\\
\alpha_4&=&\frac{g^{'2} C_1}{g^2 G^2} - \frac{2 g^{'} C_2}{g G^2} + \frac{C_3}{G^2} - \frac{g^{'} C_5}{g^3 G} +
 \frac{C_6}{g^2 G} + \frac{2 C_c}{g^4}\,,\\
\alpha_5&=&-\frac{g^{'2} C_1}{g^2 G^2} + \frac{2 g^{'} C_2}{g G^2} - \frac{C_3}{G^2} + \frac{g^{'} C_5}{g^3 G} -
 \frac{C_6}{g^2 G} + \frac{C_b}{g^4} - \frac{C_c}{g^4}\,,\\
\alpha_6&=&-\frac{g^{'2} C_1}{g^2 G^2} + \frac{2 g^{'} C_2}{g G^2} - \frac{C_3}{G^2} + \frac{g^2 C_4}{G^4} +
\wsep + \frac{g^{'} C_5}{g^3 G} - \frac{C_6}{g^2 G} - \frac{C_7}{G^3} + \frac{C_9}{g^2 G^2} - \frac{2 C_c}{g^4}\,,\\
\alpha_7&=&\frac{g^{'2} C_1}{g^2 G^2} - \frac{2 g^{'} C_2}{g G^2} + \frac{C_3}{G^2} - \frac{g^2 C_4}{G^4}
\wsep - \frac{g^{'} C_5}{g^3 G} + \frac{C_6}{g^2 G} + \frac{C_7}{G^3}
 + \frac{C_8}{g^2 G^2} - \frac{C_b}{g^4} + \frac{C_c}{g^4}\,,\\
\alpha_8&=&-\frac{g^{'2} C_1}{g^2 G^2} + \frac{2 g^{'} C_2}{g G^2} - \frac{C_3}{G^2} + \frac{C_4}{g^2}\,,\\
\alpha_9&=&\frac{g^{'2} C_1}{g^2 G^2} - \frac{2 g^{'} C_2}{g G^2} + \frac{C_3}{G^2} - \frac{C_4}{G^2}
- \frac{g^{'} C_5}{2 g^3 G} + \frac{C_6}{2 g^2 G} + \frac{C_7}{2 g^2 G}\,,\\
\alpha_a&=&-\frac{C_8}{g^2 G^2} - \frac{C_9}{g^2 G^2} + \frac{2 C_a}{G^4} + \frac{C_b}{2 g^4} + \frac{C_c}{2 g^4}\,.
\eea
The inverse relations between the ACs
$\alpha_i$ and the ECs $C_i$ read
\bea
 C_1 &=& 1 - \fwpng{g^2 g^{'2}} {2 \alpha_1 + \alpha_8} {G^2},\label{c2aa}\\
 C_2 &=& \fwpng{g g^{'}} {\alpha_1 g^2 - \alpha_1 g^{'2} + \alpha_8 g^2 }{G^2},\\
 C_3 &=& 1 - \fwpng{g^2}{\alpha_8 g^2 - 2 \alpha_1 g^{'2}} {G^2},\\
 C_4 &=& 1,\\
 C_5 &=& \fwpng{2 g g^{'}} {1 - (\alpha_1 + \alpha_2 + \alpha_3 +
    \alpha_8  + \alpha_9 ) g^2 } {G},\\
 C_6 &=& - \fwpng{2 g^2} { 1 -  (\alpha_3 + \alpha_8 + \alpha_9 )g^2 + (\alpha_1
    + \alpha_2) g^{'2}}{G},\\
 C_7 &=& \fwpng{2 g^2 }{1 - \alpha_3 G^2}{G},\\
 C_8 &=& -\frac{g^4}{G^2} + 2 \alpha_3 g^4  + (\alpha_5 + \alpha_7) g^2 G^2,\\
 C_9 &=&  \frac{g^4}{G^2} - 2 \alpha_3 g^4  + (\alpha_4 + \alpha_6) g^2 G^2,\\
 C_a &=& \fwpng{G^4} {\alpha_4 + \alpha_5 + 2 \alpha_6 + 2 \alpha_7 + 2 \alpha_a}{4},\\
 C_b &=& -\frac{g^2}{2} + \frac{g^4}{2} \left ( 2 \alpha_3 + \alpha_4  + 2 \alpha_5  + \alpha_8 + 2 \alpha_9 \right ),\\
 C_c &=&  \frac{g^2}{2} - \frac{g^4}{2} \left ( 2 \alpha_3 - \alpha_4  + \alpha_8  + 2 \alpha_9  \right )\,.
\label{c2ab}
\eea
There is an equation among $C_1$, $C_2$, and $C_3$, which reads
\bea
G^2 = C_1 g^2 + 2 g g' C_2 + C_3 g^{'2}\,.
\eea
Such a relation indicates that there are only two of these three
parameters are free. The coupling $C_4$ is a fixed constant.
It is worthy to mention that the factors before the gauge kinetic terms of
vector field $A$ and $Z$ are not unit. For a large $\al_1$ and $\al_8$, with
large couplings $g$ and $g'$, $C_1$ and $C_3$ might be very small (
it is possible for some strong couplings theories), which
in effect is equivalent to the strong coupling of $A$ and $Z$ after normalizing
the kinetic terms. The ECs $C_i$ contain not only the
contributions of the SM, but also those of the ACs.
When $\alpha_i$ vanish, these $C_i$ reduce to
their values of the SM at tree level without including the contribution
of Higgs.

There is a remarkable feature, that the ACs $\alpha_i$ always appear
in the combination with $g^2,\,\,g^{'2},\,\, G^2$. Such a feature will
be quite helpful for us to establish the modified power counting rule.

Thus, there are $12$ free parameters
in $C_i$, including $g$ and $g'$
(please remember $G^2 = g^2 + g^{'2}$).
While there are also $12$ free parameters in
$\alpha_i$, including $g$ and $g'$.
Such a fact also indicates that these two
bases are equivalent.

The $C_i$ are just the ECs of the effective vertices
among vector bosons when we
calculate the $S-$matrix by using the traditional
Feynman diagram method in the mass eigenstates,
and the effects of the ACs
of the EWCL and the background of the SM
have been taken into account in these ECs,
as shown in Eq. (\ref{c2aa}---\ref{c2ab}).

\section{The one loop renormalization of the SM}

In the spirit of the background field gauge quantization \cite{bfm},
we can decompose the Goldstone field into the classic part ${\overline U}$
and quantum part $\xi$
as
\bea
U\rightarrow {\overline U} {\widehat U}\,,\,\, {\widehat U} = \exp\{{i 2 \xi \over ( v+ \ch) }\}\,.
\label{upara}
\eea
To parameterize the quantum Goldstone field in the above form is to simplify
the presentation of the standard form of quadratic terms.
The vector fields in the mass eigenstates are split as
\bea
V_\mu \rightarrow {\overline V_\mu} + {\widehat V_\mu}\cma
\eea
where ${\overline V_\mu}$ represents the classic background
vector fields and ${\widehat V_\mu}$ represents the quantum
vector fields.

By using the Stueckelberg transformation \cite{stuck} for the
background vector fields,
\bea
{\overline W}^{s\, a} \rightarrow U^{\dagger} {\overline W} U + i {\overline U}^{\dgr} \partial{\overline U} \cma
{\overline B} \rightarrow {\overline B}\cma
{\widehat W}^{s\, a} \rightarrow U^{\dagger} {\widehat W} U \cma
{\widehat B} \rightarrow {\widehat B}\cma
\eea
So the background Goldstone fields can completely be
absorbed by redefining the background vector fields,
and will not appear in the one-loop effective Lagrangian.
The Stueckelberg fields is invariant under the
gauge transformation of the background gauge fields,
such a property guarantees that the following computation
is gauge invariant from the beginning if we can express
all effective vertices into the Stueckelberg fields.
After the loop calculation, by using the inverse
Stueckelberg transformation, the Lagrangian can be
restored to the form represented by its low energy DOFs.

Similarly, the Higgs scalar is split as
\bea
h = {\overline h} + {\widehat h}\,.
\eea

We would like to comment on the relations
between the interaction and mass eigenstates.
The mass eigenstates should be understood as
the combination of the Stueckelberg fields, and
read
\bea
Z= - \cos\theta_W \, W^{s\,3} + \sin\theta_W\, B\,,\,\, A=\sin\theta_W\, W^{s\,3} + \cos\theta_W\, B\,,\nnb\\
W^{+} = {1\over \sqrt{2}} (W^{s\,1} - i W^{s\,2})\,,\,\,W^- ={1\over \sqrt{2}} (W^{s\,1} + i W^{s\,2})\,.
\eea
Then by using these relations and the definition of Stueckelberg transformation,
we can formulate the set of independent operators in the mass eigenstates in
U-gauge back to that of the EWCL in the interaction eigenstates.

The equation of motion of the background vector fields is determined as
\bea
D_{\mu} {\cal \widetilde W}^{\nu\mu} = - \sigma_{0,VV} V^{\nu}\,,
\eea
with ${\cal \widetilde W}^{\mu\nu,T}=\{ A^{\mu\nu} + i e F_Z^{\mu\nu},
Z^{\mn} - i {g^2 \over G} F_Z^{\mn},
{\widetilde W^{+,\mn}},{\widetilde W^{-,\mn}} \}$. The EOM of vector bosons
derives the following relations
\bea
\partial \ln (v + \ch) \cdot Z &=& - { 1\over 2} \partial \cdot Z\,,\\
\partial \ln (v + \ch) \cdot W^{+} &=& - {1 \over 2} d \cdot W^{+} + i {1 \over 2} {g'^2\over G} Z \cdot W^{+}\,,\\
\partial \ln (v + \ch) \cdot W^{-} &=& - {1 \over 2} d \cdot W^{-} - i {1 \over 2} {g'^2\over G} Z \cdot W^{-}\,,
\eea
The equation of motion of the background Higgs field is given as
\bea
\partial^2 \ch = ( v+\ch) [{G^2 \over 4} Z \cdot Z + {g^2 \over 2} W^{+} \cdot W^{-}
- \mu^2 + {\lambda \over 4} ( v+ \ch)^2 ]\,.
\label{eomhiggs}
\eea

The gauge fixing term for the quantum fields are chosen as below
in order to make the quadratic terms have the standard form
\bea
{\cal L}_{GF,A}&=&-{1 \over 2} (\partial\cdot\hA  - i e (\hWm \cdot \bWp - \hWp \cdot \bWm))^2\cma\\
{\cal L}_{GF,Z}&=&-{1 \over 2} (\partial\cdot\hZ - {1\over 2} G (v + \ch)  \xi_Z
+ i {g^2 \over G} (\hWm \cdot \bWp - \hWp \cdot \bWm))^2\cma\\
{\cal L}_{GF,W}&=&- (d\cdot\hWp + {1\over 2} g ( v + \ch) \xi_W^+
	+ i {g^2 \over G} \bZ \cdot \hWp - i {g^2 \over G} \bWp \cdot \hZ
	+ i e \bWp \cdot \hA)\nnb\\&&(d\cdot\hWm + {1\over 2} g ( v + \ch) \xi_W^-
	- i {g^2 \over G} \bZ \cdot \hWm + i {g^2 \over G} \bWm \cdot \hZ
	- i e \bWm \cdot \hA)\,.
\eea

Compared with the Dyson-Feynman method, the number of diagrams in
BFM \cite{bfm} for the loop corrections
can be greatly reduced. Another remarkable advantage is that, in the BFM,
each step of calculation is manifestly gauge
covariant with reference to the gauge transformation of the background gauge
field, and the Ward identities have been naturally incorporated in
the calculation procedure. The method is quite powerful to deal with the
theories with many vertices, gravity and the nonlinear effective
gauge theories (the EWCL given below), for instance. At the same time, the
freedom for choosing the different gauges for the classic and quantum
gauge field makes the procedure of calculation simple.
The Schwinger proper time and heat kernel method \cite{htkl} in per se is
the Feynman integral. Combining with the covariant short distance
expansion \cite{wkb} in the coordinate space,
these methods can considerably simplify the loop calculation.
In the next subsections, we will use these concepts and methods to
help us to extract the RGEs of the EWCL.

Several groups of authors have conducted the renormalization of the SM in the
BFM \cite{smbfm}. Different from their procedures which are performed in the momentum
space, here we conduct our calculation in the coordinate space and we
only consider up to the one loop renormalization ( For using the
BFM to consider the two loop renormalization, please refer to
the literature of C. Lee in \cite{wkb} and \cite{bfm1} ).
Our purpose here is to check our method and to provide a comparison
to the renormalization of the effective one given in the
next section.

\subsection{The quadratic forms of the one-loop Lagrangian}
We can cast the quadratic terms of the one-loop Lagrangian
into its standard form, as prescribed in \cite{our2},
which read
\bea
{\cal L}_{quad}&=&{1\over 2 } {\widehat V_{\mu}^{\dagger a}} \Box^{\mu\nu,ab}_{V\,V} {\widehat V_{\nu}^b}
+ {1\over 2} \xi^{\dagger i}  \Box_{\xi\,\xi}^{ij} \xi^j
+ {\bar c}^a  \Box_{{\bar c}c}^{ab} c^b
+ {1\over 2} {\hat h} \Box_{hh} {\hat h}\wsep
+ {1\over 2} {\widehat V_{\mu}^{\dagger,a}} {\stackrel{\leftharpoonup}{X}}_{\xi}^{\mu,aj} \xi^j
+ {1\over 2} \xi^{\dagger,i} {\stackrel{\rightharpoonup}{X}}_{\xi}^{\nu,ib} {\widehat V_{\nu}^b}
\wsep+ {1\over 2} {\widehat V_{\mu}^{\dagger,a}} {\stackrel{\leftharpoonup}{X}}_{h}^{\mu,a} {\widehat h}
+ {1\over 2} {\widehat h} {\stackrel{\rightharpoonup}{X}}_{h}^{\mu,a} {\widehat V_{\mu}^{a}}
+ {1\over 2} \xi^{\dagger,i} X_{\xi h}^i {\widehat h}
+ {1\over 2} {\widehat h} X_{h \xi}^i \xi^{i}\,,\\
\Box^{\mu\nu,ab}_{V\, V} &=& D^{2,ab} g^{\mu\nu} + \sigma_{0,VV}^{ab} g^{\mu\nu} + \sigma_{2,VV}^{\mu\nu,ab}\,\,,\\
\Box_{\xi\,\xi}^{ij}&=& d^{2,ij} +  \sigma_{0,\xi\xi}^{ij}  + \sigma_{2,\xi\xi}^{ij}\,,\\
\Box_{hh} &=& \partial^2 + \sigma_{h h}\,,\\
\Box_{{\bar c}c}^{ab}&=& D^{2,ab} + \sigma_{0,VV}^{,ab}\,,\\
X_{h \xi}^i &=& X_{h \xi}^{\alpha, i} d_{\alpha}        + X_{h \xi,0}^{i}\,,\\
X_{\xi h}^i &=& X_{\xi h}^{\alpha, i} \partial_{\alpha} + X_{\xi h,0}^{i}
\label{stdfesm}
\eea
where $V^{\dagger}=(A,Z,W^{-},W^{+})$ and
$\xi^{\dagger}=(\xi_Z, \xi^{-}, \xi^{+})$,
the covariant differential operators $D=\partial+\Gamma_V$
and $d=\partial+\Gamma_{\xi}$, and the gauge connection of
vector bosons $\Gamma_V$ is defined as
\begin{displaymath}
\Gamma_{V,\mu} =\left (\begin{array}{cccc}
0&0& i e W^{-}_{\mu}&-i e W^{+}_{\mu}\\
0&0&-i {g^2 \over G} W^{-}_{\mu} & i {g^2 \over G} W^{+}_{\mu}\\
 i e W^{+}_{\mu} & - i {g^2 \over G} W^{+}_{\mu}& - i e A_{\mu} + i {g^2 \over G} Z_{\mu} &0\\
-i e W^{-}_{\mu} &   i {g^2 \over G} W^{-}_{\mu}&0&i e A_{\mu} - i {g^2 \over G} Z_{\mu}
\end{array}\right)\,,
\end{displaymath}
The gauge connection of Goldstone bosons $\Gamma_{\xi}$ is defined as
\begin{displaymath}
\Gamma_{\xi,\mu} =\left (\begin{array}{ccc}
0&i {g\over 2} W^{-}_{\mu}&-i {g\over 2} W^{+}_{\mu}\\
i {g\over 2} W^{+}_{\mu}& - i e A_{\mu}  &0\\
- i {g\over 2} W^{-}_{\mu}&0&i e A_{\mu}
\end{array}\right)\,.
\end{displaymath}
The mass matrices have the form $\sigma_{0, VV}^{ab}=dia\{0,G^2 {(v+\ch)^2}/{4},g^2 {(v+\ch)^2}/{4},g^2 {(v+\ch)^2}/{4}\}$
and $\sigma_{0,\xi\xi}^{ij}=dia\{G^2 {(v+\ch)^2}/{4},g^2 {(v+\ch)^2}/{4},g^2 {(v+\ch)^2}/{4}\}$.

The matrix $\sigma_{2,VV}$ is given below as
\begin{displaymath}
\sigma_{2,VV}^{\mu\nu,ab} =\left (\begin{array}{cccc}
\sigma_{2,AA}^{\mu\nu}&\sigma_{2,AZ}^{\mu\nu}&\sigma_{2,AW^+}^{\mu\nu} &\sigma_{2,AW^-}^{\mu\nu} \\
\sigma_{2,ZA}^{\mu\nu}&\sigma_{2,ZZ}^{\mu\nu}&\sigma_{2,ZW^+}^{\mu\nu} &\sigma_{2,ZW^-}^{\mu\nu} \\
\sigma_{2,W^-A}^{\mu\nu} &\sigma_{2,W^-Z}^{\mu\nu}  &\sigma_{2,W^-W^+}^{\mu\nu}  &\sigma_{2,W^-W^-}^{\mu\nu} \\
\sigma_{2,W^+A}^{\mu\nu} &\sigma_{2,W^+Z}^{\mu\nu}  &\sigma_{2,W^+W^+}^{\mu\nu}  &\sigma_{2,W^+W^-}^{\mu\nu}
\end{array}\right)\,,
\end{displaymath}
and the components read
\bea
\sigma_{2,AA  }^{\mu\nu}&=&\sigma_{2,AZ}^{\mu\nu}=\sigma_{2,ZA}^{\mu\nu}=\sigma_{2,ZZ}^{\mu\nu}=0\,,\nnb\\
\sigma_{2,AW^+}^{\mu\nu}&=&-\sigma_{2,W^+A}^{\mu\nu} =   2 i e {\widetilde W^{-,\mu\nu}}\,,\nnb\\
\sigma_{2,AW^-}^{\mu\nu}&=&-\sigma_{2,W^-A}^{\mu\nu} = - 2 i e {\widetilde W^{+,\mu\nu}}\,,\nnb\\
\sigma_{2,ZW^+}^{\mu\nu}&=&-\sigma_{2,W^+Z}^{\mu\nu} = - 2 i {g^2 \over G} {\widetilde W^{-,\mu\nu}}\,,\nnb\\
\sigma_{2,ZW^-}^{\mu\nu}&=&-\sigma_{2,W^-Z}^{\mu\nu} =   2 i {g^2 \over G} {\widetilde W^{+,\mu\nu}}\,,\nnb\\
\sigma_{2,W^- W^+} &=& - \sigma_{2,W^+ W^-} = 2 i g W^{3,\mu\nu}\,,\nnb\\
\sigma_{2,W^+ W^+}&=&\sigma_{2,W^- W^-} = 0\,,
\eea
The matrix $\sigma_{2,\xi\xi}$ is given as
\begin{displaymath}
\sigma_{2,\xi\xi}^{ij} =\left (\begin{array}{ccc}
\sigma_{2,\xi_Z\xi_Z}& \sigma_{2,\xi_Z\xi^+} & \sigma_{2,\xi_Z\xi^-}\\
\sigma_{2,\xi^-\xi_Z}& \sigma_{2,\xi^-\xi^+} &\sigma_{2,\xi^-\xi^-}\\
\sigma_{2,\xi^+\xi_Z}& \sigma_{2,\xi^+\xi^+} &\sigma_{2,\xi^+\xi^-}
\end{array}\right)\,,
\end{displaymath}
and its components read
\bea
\sigma_{2,\xi_Z\xi_Z}&=&{\lambda \over 4} [v^2 - (v+{\overline h})^2] -{G^2 \over 4}  Z \cdot Z\,,\nnb\\
\sigma_{2,\xi^+\xi^+}&=&- {g^2 \over 4} W^{-} \cdot W^{-}\,,\nnb\\
\sigma_{2,\xi^-\xi^-}&=&- {g^2 \over 4} W^{+} \cdot W^{+}\,,\nnb\\
\sigma_{2,\xi_Z\xi^+}&=&\sigma_{2,\xi^+\xi_Z}={g G \over 4} W^{-} \cdot Z\,,\nnb\\
\sigma_{2,\xi_Z\xi^-}&=&\sigma_{2,\xi^-\xi_Z}={g G \over 4} W^{+} \cdot Z\,,\nnb\\
\sigma_{2,\xi^+\xi^-}&=&\sigma_{2,\xi^-\xi^+}={\lambda \over 4} [v^2 - (v+\ch)^2] -{g^2 \over 4}  W^{+} \cdot W^{-}\,.
\eea
We have used the equation of motion of the background Higgs given in Eq. (\ref{eomhiggs}) in this step,
which is reflected by the terms proportional to $\lambda$ in $\sigma_{2,\xi_Z\xi_Z}$,
$\sigma_{2,\xi^+\xi^-}$, and $\sigma_{2,\xi^-\xi^+}$.

The $\sigma_{hh}$ is determined as
\bea
\sigma_{hh} &=&-{1\over 4} (G^2 Z\cdot Z + 2 g^2 W^+ \cdot W^-) + {\lambda \over 4} v^2 - {3\over 4} \lambda ( v+ \ch)^2\,.
\eea

The mixing terms between the vector and Goldstone bosons
are determined as
\begin{displaymath}
{\stackrel{\leftharpoonup}{X}}_{\xi}^{\mu,aj}  =\left (\begin{array}{ccc}
0&-i {g^2 g' \over G} (v + {\overline h}) W^{-,\mu} & i {g^2 g' \over G} (v + {\overline h}) W^{+,\mu}\\
G \partial^{\mu} {\overline h}& i {g \over 2 G} (g^2 - g^{'2}) (v + {\overline h}) W^{-,\mu}& - i {g \over 2 G} (g^2 - g^{'2}) (v + {\overline h}) W^{+,\mu}\\
- i {g^2 \over 2} ( v + {\overline h}) W^{+,\mu}&-g \partial^{\mu} {\overline h} - {1\over 2} i g G (v+ {\overline h}) Z^{\mu}&0\\
  i {g^2 \over 2} ( v + {\overline h}) W^{-,\mu}&0&-g \partial^{\mu} {\overline h} + {1\over 2} i g G (v+ {\overline h}) Z^{\mu}
\end{array}\right)\,,
\end{displaymath}
while the matrix ${\stackrel{\rightharpoonup}{X}}_{\xi}^{\mu,aj}$ is just
the rearrangement of the ${\stackrel{\leftharpoonup}{X}}_{\xi}^{\mu,aj}$, and
here we do not rewrite it.
The mixing terms between vector and Higgs bosons
are determined as
\bea
{\stackrel{\leftharpoonup}{X}}_{h}^{\mu,a} &=&\{0,-{1\over 2} G^2 (v + {\overline h}) Z^{\mu},-{1\over 2} g^2 (v + {\overline h}) W^{+,\mu},-{1\over 2} g^2 (v + {\overline h}) W^{-,\mu}\}\,,
\eea
The mixing terms ${\stackrel{\rightharpoonup}{X}}^{\mu,a}_h$
is the rearrangement of the ${\stackrel{\leftharpoonup}{X}}^{\mu,a}_h$.
The mixing terms between Higgs and Goldstone bosons
are determined as
\bea
X_{h \xi}^{\alpha, i}&=& \{-G Z^{\alpha}, g W^{-,\alpha}, g W^{+,\alpha}\}\,,\\
X_{h \xi,0}^{i}      &=& \{-{G \over 2} \partial \cdot Z,
{g \over 2} d \cdot W^- - i {1\over 2} ({g^2 \over G} - G) W^- \cdot Z,\nnb\\&&
{g \over 2} d \cdot W^+ + i {1 \over 2} ({g^2 \over G} - G) W^+ \cdot Z \}
\eea
The terms $X_{\xi h}^{\alpha, i}$ and $X_{\xi h,0}^{i}$
are omitted here.

\subsection{Evaluating the traces and logarithms}
By diagonalizing the quantum fields, we can integrate the quadratic terms
of the Lagrangian by using the Gaussian integral. And the ${\cal L}_{1-loop}$
can be expressed as the traces and logarithms
\bea
S_{1-loop}&=& Tr\Box_{\bar{c}c} - {1\over 2} \mbra{ Tr\ln\Box_{VV}
+Tr\ln\Box_{\xi\xi}' + Tr\ln\Box_{hh}''}\,\,,
\label{logtrsm}
\eea
where
\bea
\Box_{\xi\xi}^{'ij}& =& \Box_{\xi\xi}^{ij} - \rvec{X}_{\xi} \Box_{VV}^{-1} \lvec{X}_{\xi}\,,\\
\Box_{hh}'&=&\Box_{hh} - \rvec{X}_h \Box_{VV}^{-1} \lvec{X}_h \,,\\
\Box_{hh}''&=&\Box_{hh}' - X'_{h\xi} \Box_{\xi\xi}^{'-1} X_{\xi h}\,,\\
X'_{h\xi} &=& X_{h\xi} - \rvec{X}_h \Box_{VV}^{-1} \lvec{X}_{\xi}\,,\\
X'_{\xi h} &=& X_{\xi h} - \rvec{X}_{\xi} \Box_{VV}^{-1} \lvec{X}_h\,,
\eea
Expanding the $Tr\ln\Box_{\xi\xi}'$ and $Tr\ln\Box_{hh}''$ with the following
relations
\bea
Tr\ln\Box_{\xi\xi}' &=& Tr\ln\Box_{\xi\xi}
+ Tr\ln( 1 - \rvec{X}_{\xi} \Box_{VV}^{-1} \lvec{X}_{\xi} \Box_{\xi\xi}^{-1})\,,\\
Tr\ln\Box_{hh}''    &=& Tr\ln\Box_{hh}'
+ Tr\ln( 1 - X'_{h\xi} \Box_{\xi\xi}^{'-1} X_{\xi h}  \Box_{hh}^{'-1})\,.
\eea
Since we consider the renormalization, so we are only interested in those
divergent terms, which can be expressed as
\bea
\int_x {\cal L}_{1-loop}&=& Tr\Box_{\bar{c}c}  - {1\over 2} \mbra{ Tr\ln\Box_{VV}
+Tr\ln\Box_{\xi\xi}  + Tr\ln\Box_{hh}
\wsep - Tr(\rvec{X}_{\xi} \Box_{VV}^{-1} \lvec{X}_{\xi} \Box_{\xi\xi}^{-1})
 - Tr(\rvec{X}_h \Box_{VV}^{-1} \lvec{X}_h \Box_{hh}^{-1})
\wsep  - Tr(X_{h\xi} \Box_{\xi\xi}^{-1} X_{\xi h} \Box_{hh}^{-1})
\wsep - {1\over 2} Tr(X_{h\xi} \Box_{\xi\xi}^{-1} X_{\xi h} \Box_{hh}^{-1}X_{h\xi} \Box_{\xi\xi}^{-1} X_{\xi h} \Box_{hh}^{-1})
 + \cdots}
\,.
\label{divt}
\eea
Due to the property of the $Tr$, the above
equation is independent of the sequence of integrating-out quantum fields.
The omitted terms are finite and will not contribute to the one-loop divergence
structures.

\subsection{Counter terms}

To evaluate the traces in the Eq. (\ref{divt}), we use the Schwinger proper time
and heat kernel method \cite{htkl} with the covariant short distance expansion technique \cite{wkb}.
The detailed calculation steps are omitted here.
By using the heat kernel method directly,
we have the following divergence structures
from the contributions of $Tr\ln\Box$ in the Eq. (\ref{divt})
\bea
{1 \over 2} {\bar \epsilon} Tr\ln\Box_{VV}&=&- {20 \over 3} H_1 -{(2 g^2 + G^2) \over 16} (v + \ch)^4 \,,\\
{1 \over 2} {\bar \epsilon} Tr\ln\Box_{\xi\xi}&=&
+ {g^2 \over 12} H_1 + {g^{'2} \over 12} H_2 + {1\over 12} {\cal L}_1 - {1 \over 24} {\cal L}_2
 - {1 \over 24} {\cal L}_3
\wsep-{1 \over 12} {\cal L}_4 + {1 \over 48} {\cal L}_5
 - {G^2 g^{'2} \over 32} (v + \ch)^2 Z \cdot Z
\wsep + {1 \over 32} [\lambda v^2 - (g^2 + \lambda) (v+\ch)^2] (G^2 Z\cdot Z + 2 g^2 W^+ \cdot W^-)
\wsep + {1 \over 32} [3 \lambda + (2 g^2 + G^2)] \lambda v^2 ( v+\ch)^2
\wsep- {1\over 64} [(2 g^4 + G^4) + 2 ( 2 g^2 + G^2) \lambda + 12 \lambda^2] (v + \ch)^4
\wsep -{ 3 \over 64 } \lambda^2 v^4\,,\\
{1\over 2} {\bar \epsilon} Tr\ln\Box_{hh}&=&
-{1 \over 16} L_5
+{1 \over 32} \lambda [v^2 - 3 (v +\ch)^2 ] (G^2 Z\cdot Z + 2 g^2 W^+ \cdot W^-)
\wsep +{3 \over 32} \lambda^2 v^2 (v+\ch)^2 - { 9 \over 64} \lambda^2 (v+ \ch)^4
-{ 1 \over 16 } \lambda^2 v^4\,,\\
- {\bar \epsilon} Tr\ln\Box_{{\bar c}c}= &=& -{2 \over 3} H_1 + {G^2 + 2g^2 \over 32 } (v +\ch)^4\,,
\eea
where $1/{\bar \epsilon}=i/{16 \pi^2} (2/\epsilon - \gamma_E + \ln(4 \pi^2)$, $\gamma_E$ is
the Euler constant, and $\epsilon=4-d$. The terms $(v+\ch)^4$ in the $Tr\ln\Box_{\xi\xi}$ comes
from the EOM of the background Higgs field Eq. (\ref{eomhiggs}).
The divergence terms from the mixing terms with two
propagators are given as
\bea
-{\bar \epsilon \over 2} Tr(\rvec{X}_{\xi} \Box_{VV}^{-1} \lvec{X}_{\xi} \Box_{\xi\xi}^{-1})&=&
{g^2 g^{'2} \over 8} Z\cdot Z (v + \ch)^2
\wsep - {g^2 + G^2 \over 8} (v+\ch)^2 (G^2 Z\cdot Z + 2 g^2 W^+ \cdot W^-)
\wsep - {1\over 2} (2 g^2 +G^2 ) \partial \ch \cdot \partial \ch\,,\\
- {\bar \epsilon \over 2} Tr(\rvec{X}_h \Box_{VV}^{-1} \lvec{X}_h \Box_{hh}^{-1})&=&
- {g^2 g^{'2} \over 8} Z\cdot Z (v + \ch)^2
\wsep-{g^2 \over 8} (v + \ch)^2 (G^2 Z\cdot Z + 2 g^2 W^+ \cdot W^-)\,,\\
- {\bar \epsilon \over 2} Tr(X_{h\xi} \Box_{\xi\xi}^{-1} X_{\xi h} \Box_{hh}^{-1})&=&- {1\over 2} (t_{BB1} + t_{BB2} + t_{BC} + t_{CC})\,,\\
t_{BB1}&=&- {g^2 \over 6} H_1 - {g^{'2} \over 6} H_2 + {1\over 6} {\cal L}_1 + {1 \over 6} {\cal L}_2 + {1 \over 6} {\cal L}_3
+ {1 \over 6} {\cal L}_4 - {1 \over 6} {\cal L}_5
\wsep -{1\over 2} ({g^2 \over G^2 } -1)^2 {\cal L}_6 + {1\over 2} ({g^2 \over G^2 } -1)^2 {\cal L}_a
\wsep -{ G^2 \over 4 } (\partial \cdot Z)^2 -{g^2 \over 2} (d \cdot W^{+}) (d \cdot W^{-})
\wsep - i {g^2 g^{'2} \over G} [(d \cdot W^{+}) (W^{-} \cdot Z) - (d \cdot W^{-}) (W^{+} \cdot Z)]\,,\\
t_{BB2}&=&- { 1\over 4} {\cal L}_2 - {1\over 4} {\cal L}_3 - {1\over 2} {\cal L}_4
 - {G^2 g^{'2} \over 16} Z\cdot Z (v +\ch)^2
\wsep + {1 \over 16} [2 \lambda v^2 - (g^2 + 4 \lambda) (v+\ch)^2 ] (G^2 Z\cdot Z + 2 g^2 W^+ \cdot W^-)\,,\\
t_{BC}&=&
({g^2 \over G^2 } -1)^2 {\cal L}_6 - ({g^2 \over G^2 } -1)^2 {\cal L}_a
\wsep +{G^2 \over 2} (\partial \cdot Z)^2 + g^2 (d \cdot W^{+}) (d \cdot W^{-})
\wsep - i {g^2 g^{'2} \over G } [(d \cdot W^{+}) (W^{-} \cdot Z) - (d \cdot W^{-}) (W^{+} \cdot Z)]
 \,,\\
t_{CC}&=&
-{1\over 2} ({g^2 \over G^2 } -1)^2 {\cal L}_6 + {1\over 2} ({g^2 \over G^2 } -1)^2 {\cal L}_a
\wsep - {G^2 \over 4} (\partial \cdot Z)^2 -{g^2 \over 2} (d \cdot W^{+}) (d \cdot W^{-})
\wsep + i {g^2 g^{'2} \over 2 G}[(d \cdot W^{+}) (W^{-} \cdot Z) - (d \cdot W^{-}) (W^{+} \cdot Z)]
\,.
\eea
The divergences of the four propagators term is given as
\bea
- {\bar \epsilon \over 4} Tr(X_{h\xi} \Box_{\xi\xi}^{-1} X_{\xi h} \Box_{hh}^{-1}X_{h\xi} \Box_{\xi\xi}^{-1} X_{\xi h} \Box_{hh}^{-1})
&=&- {1\over 12} {\cal L}_4 - {1 \over 24} {\cal L}_5\,,
\eea
The sum over all contributions yields the following total divergence structures as
\bea
{\bar \epsilon} D_{tot} &=& -{ 43 \over 6} g^2 H_1 + {1 \over 6} g^{'2} H_2
\wsep -{2 g^2 + G^2 \over 8} ( v+ \ch)^2 (G^2 Z \cdot Z + 2 g^2 W^+ \cdot W^-)
\wsep -{2 g^2 + G^2 \over 2} \partial \ch \cdot \partial \ch
\wsep +{1 \over 32} (6 \lambda + 2 g^2 +G^2) \lambda v^2 (v+ \ch)^2
\wsep-{1 \over 64} [(6 g^4 + 3 G^4) + (2 g^2 + G^2) \lambda + 12 \lambda^2] (v+\ch)^4
\wsep -{1 \over 16} \lambda^2 v^4\,.
\eea

The $D_{tot}$ just indicates that the extra divergences just cancel out exactly
with each other, and even the terms like $(\partial Z)^2$ will not appear in the
total divergence structures. No gauge fixing term of the background
fields should be added to the Lagrangian, and
the equations of motion are just sufficient.

The coefficients of $H_1$ and $H_2$
have the correct value which contribute
to the $\beta$ functions of the gauge couplings $g$ and $g'$, respectively.
The coefficients of the terms
$( v+ \ch)^2 (G^2 Z \cdot Z + 2 g^2 W^+ \cdot W^-)/8$
and $\partial \ch \cdot \partial \ch/2$ are equal, and such a fact is not
accidental and should be the requirement of renormalizability. If
we reformulate the Lagrangian in its linear form,
the combination of these two terms just yields the term
$(D\phi)^{\dagger}\cdot(D \phi)$. From the requirement
of renormalizability of the theory, we know that
coefficients of divergences of these two terms should
be the same. The last constant divergences just contribute to
the unobservable vacuum, and can be dropped out. This constant
will also appear in other computational methods, and is not the
special feature of BFM.

Due to the parameterization
in the Eq. (\ref{upara}), we have found that there is no quartic divergences
in the one-loop Lagrangian, contrary to the expectation of \cite{quartic}.
As matter of fact, had the authors of the references \cite{quartic} used the
EOM of the background Higgs to sum the quartic divergences and higher divergence
structures, they would have gotten the same result as given by us.
In other words, such a fact is independent of the parameterization
of quantum Goldstone bosons.

Another remarkable feature is that the EOM of the background
Higgs makes the counter term of the quartic coupling of Higgs
potential in BFM different than that calculated in the usual
Feynman diagrams in linear representation. This difference
is caused directly by the term $\sigma_{\xi\xi}$, and might
be one of the special characters of the BFM.

To extract the divergences, we have used the following relation
\bea
H^{\mn}_{-} SF_{+,\mn} = 8 i O_6 + 4 i O_7
+ 4 Z \cdot W^{+} d \cdot W^{-} - 4 Z \cdot W^{-} d\cdot W^{+}
+ H_{+}^{\mn} SF_{-,\mn}\,,
\eea

\section{The one loop renormalization of the EWCL}
In this section we will conduct the one loop renormalization of the EWCL,
by taking the limit that all ACs vanish, we can check the following
calculation with the renormalization of the SM, as provided in the
above section. Before the actual computation, we would like to
establish a modified power counting rule, in order to control higher
corrections from higher loops and higher dimension operators.

\subsection{The modified power counting rule in EWCL}
Before establishing our power counting rule, we would like to
make a brief review on the framwork of the hadronic chiral perturbation
theory.

In the usual ChPT approach to low-energy hadronic,
the chiral Lagrangian is organized as an expansion in
powers of momenta $p^2$
\bea
L^{eff} &=& L_2 + L_4 + L_6 + ...
\eea
Each term $L_n$, in turn, is given by a certain number of operators
$O_i^{(n)}$ with low-energy constants $l_i^{(n)}$ that, a priori, are
determined by the underlying theory:
\bea
L_n &=& \sum_i l_i^{(n)} O_i^{(n)}\,.
\eea

The general expectation of the importance of an operator
is that the lower order it belongs, the more importance of it.
Therefore, in the ChPT, the $L_2$ is the most important operator, and
determines the propagators of massless
Goldstone bosons and the scattering interaction at tree level, which
can be expressed as $c_2 \,\, {p^2 \over v^2}$, ($c_2$ is a O(1) constant).
At one-loop level, the scattering amplitude will get the radiative
corrections from the loop with two of this vertex and with internal
lines of Goldstones. While after dropping the divergences of the loop integral,
we get the finite one-loop contribution of this interaction can be expressed
as $ \alpha \,\, {1 \over (4 \pi)^2 } \,\, {p^4 \over v^4}$, ($\alpha$
is a constant factor determined by the loop and $c_2$, which is of order 1). Such a
contribution has the same momentum power with those of operators in the
$L_4$, which can be expressed as $\alpha_0 \,\, {p^4 \over v^4}$.

In the ChPT, coincidentally (not necessary for general chiral perturbation
theories), $\alpha_0$, as determined from low energy phenomenologies,
like hadronic scattering and decay processes, etc, is of the order
${1 \over (4 \pi)^2}$ \cite{pcr}.

So, if we go to further higher order, say two-loop order, then
we should include three parts of contributions,
1) the two loop contributions of pure $O(p^2)$
vertices, 2) the one-loop contribution with one $O(p^2)$ vertex and one $O(p^4)$
vertex, and 3) the tree level contribution of $O(p^6)$.
The first part can be expressed as 
$\beta_2 \,\, {1 \over (4 \pi)^4 } \,\, {p^6 \over v^6}$,
the second part can be expressed as
$\beta_1 \,\, {1 \over (4 \pi)^2 } \,\, {p^6 \over v^6}$,
and the third part can be expressed as
$\beta_0 \,\, {p^6 \over v^6}$.
The first part contains two loop suppression factor ${1 \over (4 \pi)^4 }$, while
the second part contains only one loop suppression factor ${1 \over (4 \pi)^2 }$.
But due to the fact that the $\beta_1$ is determined by both $c_2$ and $\alpha_0$,
so not only on the momentum power, but also on the magnitude order controlled by the
loop factors, the second parts will share the same importance as the first parts.
We also expect that, coincidently (not necessary for general chiral perturbation
theories), the $\beta_0$ will have a magnitude 
like ${1 \over (4 \pi)^4 }$.
So that we expect that such a standard power counting rule will hold at any a specified higher order.

But for the EWCL,
it seems not easy to take into account the radiative corrections of low energy
quantum DOFs (which should include both the massive vector boson
and its corresponding Goldstone).

The first difficulty is more manifest when we represent the EGT
in their unitary gauge. The propagator of massive vector bosons
can be expressed as
\bea
i \Delta^{\mu\nu}&=&i \Delta_T^{\mu\nu}+i \Delta_L^{\mu\nu}\,,\\
\Delta_T^{\mu\nu}&=&\frac{1}{k^2-m_V^2} \left ( -g^{\mu\nu}+\frac{k^{\mu} k^{\nu}}{k^2} \right )\,,\\
\Delta_L^{\mu\nu}&=&\frac{1}{m_V^2} \frac{k^{\mu} k^{\nu}}{k^2}\,,
\eea
where $\Delta_T$ and $\Delta_L$ represent the transverse and
longitudinal parts, respectively.
The longitudinal part of the propagator can bring into
quartic divergences and lead to the well-known bad
ultraviolet behavior. Two direct consequences of this
fact are 1) that the quartic divergences will appear in
radiative corrections and 2) that low dimensions operators
can induce the infinite number of divergences of higher
dimension operators, even at one-loop level. In a renormalizable theory, the
Higgs model for instance,
these two problems do not exist. The quartic divergences produced by
the low energy DOF just cancel exactly with
those produced by the Higgs scalar, and no extra
divergence structure will appear.

The second difficulty, which is related with the first difficulty,
is about the counting rule. In the gauge
theories with spontaneous symmetry breaking, the marginal interaction
vertices are proportional to ECs (expressed in both
gauge couplings and ACs, as given in (\ref{c2ab})),
not to the momentum
power ${p^2 \over v^2}$ as in the hadronic ChPT.
Then by direct evaluating the Feynman diagrams,
radiative corrections the ACs (which are determined at matching scale
by the ultraviolet dynamics and there is no reason to assume that they must
be as small as ${1 \over (4 \pi)^2}$.) are of the ${1 \over (4 \pi)^2}$, 
not as ${1 \over (4 \pi)^4}$ as expected
from the SPCR in the hadronic ChPT. 
So the native power counting rule is not proper to be used in this case. 
While we know, in order
to collect and reliably estimate the contributions of higher orders 
(say, those of higher loops and higher dimension operators) in terms
of magnitude, a power counting rule is needed. 
So to find a consistent power counting rule for this case is necessary.

As we know, for the EGT with spontaneous symmetry breaking mechanism,
we have at least two ways to collect and classify operators.

The first way is to collect operators in terms of their dimensions, 
(not by the momentum power ${p^2 \over v^2}$, as in the above case). 
We can formulate the EWCL in the unitary gauge, then
restore the low energy DOFs with the inverse 
Stueckelberg transformation,
while the ECs are regarded as free parameters, of which the
magnitude at the ultraviolet cutoff
is determined by the underlying dynamics and the matching conditions.
In the most general assumption, we regard these ACs are of order $O(1)$.
Then, according to the Wilsonian renormalization scheme, EOs
can be classified into three groups: the relevant operators, marginal operators,
and irrelevant operators. The relevant operators have mass dimensions less than the
dimension of space-time, and have ECs with positive mass power.
The marginal operators have the same dimensions of that of space-time, 
and have massless ECs.
The irrelevant operators have dimensions larger than
the dimensions of space-time, and have couplings with negative mass power.
By study the running of the ECs, we can
determine the importance of operators, which is controlled by
the strength of their corresponding ECs.
The couplings of the relevant operators will be dependent on the ultraviolet
cutoff $\Lambda={UV}$ in positive powers;
those of the marginal operators will be logarithmically dependent on the $\Lambda={UV}$;
while those of the irrelevant operators will be dependent on the $\Lambda={UV}$
in negative powers. If the $\Lambda={UV}$ is large enough, the irrelevant operators
will become unimportant, and the relevant and marginal operators will
mainly determine the low energy dynamics. Such a conclusion is based on the most
general analysis of the behavior of RGEs without assuming the smallness of the
ECs of irrelevant operators, as shown in \cite{wilson, pol}. So we can truncate
the infinite operator towers permitted in the EGT to a specified order.
While for the quartic divergences, we can use the dimension regularization method, and
simply discard them.

As we know, the groups of relevant and marginal operators include both the renormalizable
operators and AOs up to $O(p^4)$.
Meanwhile, in the general cases, the relative importance of an operator might
be quite different and is determined by the relative magnitude of its
EC. For instance, if the coupling is zero or
much much smaller, in principle, we can drop its contributions and regard it
as higher order corrections; while if the coupling is much
much lager than others, this operator should be definitely important to the low energy
dynamics. Then we should classify it as a lower order operator, {\it i.e.} to
increase its importance for our consideration. So a practical and realistic power counting
rule must based on the actual information of the
relative magnitude of the ECs.

The second way is to mimic the ChPT by classifying according to the momentum power.
In this way, up to $O(p^4)$, without regarding any information on the
magnitude of the ECs, these operators are divided into two groups,
the renormalizable ones are classified to $p^2$ order, while the anomalous ones
are classified as $p^4$ order. In this way, to classify the gauge kinetic
terms into $O(p^2)$ is somewhat ambiguous to the momentum power counting rule. 
But it is unlikely not to include the kinetic terms in this $p^2$. So
the dimensionless gauge couplings have to be set to have momentum power. 
To classify the rest of marginal operators in the group of $O(p^4)$,
such a counting rule, borrowed from the hadronic ChPT, implicitly assumes
that the strength of their couplings
should be of order ${1 \over (4 \pi)^2}$.

We would like to point out that such assumptions
are too strong for the a general EGT. In the framework of EFT,
the magnitudes of the couplings of an operator is determined at the matching
scale. There is no reason to expect that the ACs must be
such small. Since at the matching scale, those ACs
are determined by the ultraviolet dynamics, and can get contributions from
either the tree level or loop level, or both. The magnitude of these ACs
is related with both the actual value of the matching scale and the
underlying dynamics. As we know, the ACs can receive the tree level contributions,
like in the Higgs model we show in the numerical analysis, in the left-right
hand model, etc.  Furthermore, even determined at loop level, if the ultraviolet
dynamics are strong coupling case, like in the Technicolor models,
these ACs can be estimated as 
${1 \over (4 \pi)^2} {g_s^2 \over g_w^2}$. If $g_s$ is much larger than
$g_w$, the ACs might still be one or two order larger than the
expectation of SPCR in the hadronic ChPT. 

So we regard that, in order to be more realistic and be consistent 
with the EFT method as
a general and universal method, we should abandon the second way of the
classification of the operators, and before knowing the actual information
on the magnitude of the ACs (equivalently, the underlying theories), 
we will treat all relevant and marginal AOs as operators in $O(p^2)$ 
order by implicitly assuming all these ACs are of $O(1)$ (This assumption
is a more general one, and the assumption of the second way of
classification is only one of its specific cases). So
we modify the momentum power counting
rule to include the ECs of all AOs in the Eqs. (\ref{ewclb}---\ref{ewcle})
$\alpha_i$ as momentum $p^{-2}$, 
like coefficient of the gauge kinetic terms $1/g^2$.
And in this way, when extracting the Feynman rules directly from the Lagrangian given in
, the combination of $g^2 \alpha_i$ in the trilinear and quartic couplings
is regarded as of $O(p^0)\sim O(1)$. 
Thus, this modified power counting
rule will possess the powerful potent of the SPCR,
and can be applied to estimate and control the contributions of higher loop
and higher dimension operators, just like in the hadronic ChPT.

With this modified power counting rule in mind, below we will study 
the renormalization of the EWCL up to relevant and marginal operators
and derive the one-loop RGE of its ECs.

\subsection{The gauge fixing terms in the BFM}

The equations of motion of the mass eigenstates are determined as
\bea
D'_{\mu} {\cal \widetilde W}^{\nu\mu} = \sigma_{0,VV} V^{\nu}\,,
\eea
These EOM, in effect, act as the gauge fixing of the background fields, and
can derive the following relations
\bea
\partial \cdot Z = 0,\,\,
d \cdot W^{\pm} = \pm i {ccw \over 2} Z\cdot W^{\pm},\nnb\\
ccw=-e {C_2 \over 4 C_1} - {C_5 C_2 \over 8 C_1} + {C_6 \over 8} + {C_7 \over 8}\,.
\eea

In the background field gauge, the covariant
gauge fixing term for the quantum fields can be chosen as
\bea
{\cal L}_{GF,A}&=&-{g_A \over 2} (\partial\cdot\hA + f_{AZ} \partial\cdot\hZ - i f_{AW} (\hWm \cdot \bWp - \hWp \cdot \bWm))^2\cma\\
{\cal L}_{GF,Z}&=&-{g_Z \over 2} (\partial\cdot\hZ + f_{Z\xi} \xi_Z - i f_{ZW} (\hWm \cdot \bWp - \hWp \cdot \bWm))^2\cma\\
{\cal L}_{GF,W}&=&- g_W (d\cdot\hWp + f_{W\xi} \xi_W^+
	+ i f_{WZ} \bZ \cdot \hWp + i p_{WZ} \bWp \cdot \hZ
	+ i p_{WA} \bWp \cdot \hA)\nnb\\&&(d\cdot\hWm + f_{W\xi} \xi_W^-
	- i f_{WZ} \bZ \cdot \hWm - i p_{WZ} \bWm \cdot \hZ
	- i p_{WA} \bWm \cdot \hA)\cma
\label{gftew}
\eea
where the parameters in these gauge fixing terms are determined by requiring the
quadratic terms of Lagrangian has the standard form specified in Eq. (\ref{stdfeew}),
and they read
\bea
 f_{Z\xi} &=& \frac{\rho}{g_Z} \frac{G v}{2}\,,\\
 f_{W\xi} &=& - \frac{1}{g_W}\frac{g v}{2}\,,\\
 g_A &=&  C_1\,,\\
 f_{AZ} &=&{C_2 \over C_1}\,,\\
 g_Z &=& C_3 - {C_2^2 \over C_1}\,,\\
 g_W &=& 1\,,\\
 p_{WA} &=& {C_5 \over 2}\,,\\
 f_{AW} &=& \frac{g g^{'}}{G C_1}\,,\\
 p_{WZ} &=& {C_6 \over 2}\,,\\
 f_{ZW} &=& -{C_7 \over 2 g_Z} - {1 \over g_Z} {g g' \over G} {C_2 \over C_1}\,,\\
 f_{WZ} &=& {C_7\over 2}\,,
\eea
When all the ACs are set to vanish, these parameters will
reduce to those of the SM.

\subsection{The quadratic terms of the effective Lagrangian}

Thus, after diagonalizing and normalizing the variables,
we can collect the quadratic terms of quantum boson fields
in the following standard form
\bea
{\cal L}_{quad}&=&{1\over 2 } {\widehat V_{\mu}^{\dagger a}} \Box^{\mu\nu,ab}_{V\,V} {\widehat V_{\nu}^b}
+ {1\over 2} \xi^{\dagger i}  \Box_{\xi\,\xi}^{ij} \xi^j
+ {\bar c}^a  \Box_{{\bar c}c}^{ab} c^b
\wsep + {1\over 2} {\widehat V_{\mu}^{\dagger,a}} {\stackrel{\leftharpoonup}{X}}^{\mu,aj} \xi^j
+ {1\over 2} \xi^{\dagger,i} {\stackrel{\rightharpoonup}{X}}^{\nu,ib} {\widehat V_{\nu}^b}\,,\\
\Box^{\mu\nu,ab}_{V\, V} &=& D^{2,ab} g^{\mu\nu} + \sigma_{0,VV}^{ab} g^{\mu\nu} + \sigma_{2,VV}^{\mu\nu,ab}\,\,,\\
\Box_{\xi\,\xi}^{ij} &=& \Box_{\xi\,\xi}'^{ij} + X^{\alpha,ii'} d_{\alpha}^{i'j} + X^{\alpha\beta,ii'} d_{\alpha}^{i'j'} d_{\beta}^{j'j}\,\,,\\
\Box_{\xi\,\xi}'^{ij}&=& d^{2,ij} +  \sigma_{0,\xi\xi}^{ij}  + \sigma_{2,\xi\xi}^{ij} + \sigma_{4,\xi\xi}^{ij}\,,\\
\Box_{{\bar c}c}^{ab}&=& D^{2,ab} + \sigma_{0,VV}^{,ab}\,,\\
{\stackrel{\leftharpoonup}{X}}^{\mu,aj}&=&
 {\stackrel{\leftharpoonup}{X}}^{\mu,ai}_{\alpha\beta} d^{\alpha,ii'} d^{\beta,i'j}
+{\stackrel{\leftharpoonup}{X}}^{\mu\alpha,aj'} d_{\alpha}^{j'j}
+{\stackrel{\leftharpoonup}{X}}^{\mu,aj}_{01}
+{\stackrel{\leftharpoonup}{X}}^{\mu,aj}_{03Z}
+\partial_{\alpha} {\stackrel{\leftharpoonup}{X}}^{\mu\alpha,aj}_{03Y}\,,\\
{\stackrel{\rightharpoonup}{X}}^{\nu,ib}&=&
 {\stackrel{\rightharpoonup}{X}}^{\nu,ia}_{\alpha\beta} D^{\alpha,ab'} D^{\beta,b'b}
+{\stackrel{\rightharpoonup}{X}}^{\nu\alpha,ib'} D^{b'b}_{\alpha}
+{\stackrel{\rightharpoonup}{X}}^{\nu,ib}_{01}
+{\stackrel{\rightharpoonup}{X}}^{\nu,ib}_{03Z}
+\partial_{\alpha} {\stackrel{\rightharpoonup}{X}}^{\nu\alpha,ib}_{03Y}\,,
\label{stdfeew}
\eea
where $V^{\dagger}=(A,Z,W^{-},W^{+})$ and
$\xi^{\dagger}=(\xi_Z, \xi^{-}, \xi^{+})$.
And the covariant differential operators
$D=\partial+\Gamma_V$ and $d=\partial+\Gamma_{\xi}$.
The gauge connections $\Gamma_V$ and $\Gamma_{\xi}$
are defined as
\bea
\Gamma_V = X_V^T {\widetilde \Gamma_V} X_V,\,\,
\Gamma_\xi = X_\xi^T {\widetilde \Gamma_\xi} X_\xi\,,
\eea
where the matrices $X_V$ and $X_\xi$ are determined
by the matrices $W_V$ and $W_\xi$ by
\bea
X_V^T W_V X_V =1_{4 \times 4},\,\,X_\xi^T W_\xi X_\xi =1_{3\times 3}\,.
\eea
The matrices $W_V$ and $W_\xi$ will be presented below.

About the matrices given in Eq. (\ref{stdfeew}), due to the fact that $A$ is still massless and
has no the corresponding Goldstone bosons, the number of vector bosons and of
Goldstone is different. So we deliberately present the
indices of the Goldstone and vector bosons with different letters.

The related quantities given in (\ref{stdfeew}) are defined as
\bea
\label{tbeg}
\sigma^{\mu\nu\,\,ab}_{2,VV} &=&
-\partial \cdot \Gamma_{V}^{ab} g^{\mn} - \Gamma_{V}^{ac} \cdot \Gamma_{V}^{cb} g^{\mn} + {\widetilde \sigma_{2,VV}^{\mn,ab}}\,,\\
\sigma_{2,\xi\xi}^{ij} & = &
-\partial \cdot \Gamma_{\xi}^{ij} - \Gamma_{\xi}^{ik} \cdot \Gamma_{\xi}^{kj} + {\widetilde \sigma_{2,\xi\xi}^{ij}}\,,\\
X^{\alpha\beta,ij} &=&- {\widetilde S}^{\alpha \beta,ij}\,,\\
X^{\alpha,ij} &=& {\widetilde X^{\alpha,ij}}
-  \partial_{\beta} {\widetilde X}^{\alpha \beta,ij}
+ 2 {\widetilde S}^{\alpha\beta,ik} \Gamma_{\xi,\beta}^{kj}\,,\\
\sigma_{4,\xi\xi}^{ij} & = & {\widetilde X}_4^{ij}
+ {\widetilde S}^{\alpha \beta,ik} ( \partial_{\beta} \Gamma_{\xi,\alpha}^{kj} - \Gamma_{\xi,\alpha}^{kl} \Gamma_{\xi, \beta}^{lj})\nnb\\&&
- {\widetilde X^{\alpha,ik}} \Gamma_{\xi,\alpha}^{kj}
+ \partial_{\beta} {\widetilde X}^{\alpha \beta,ik} \Gamma_{\xi,\alpha}^{kj} \,,\\
{\stackrel{\leftharpoonup}{X}}^{\mu,ai}_{\alpha\beta}&=&
-{\widetilde S}^{\mu,ai}_{\alpha\beta}\,,\\
{\stackrel{\leftharpoonup}{X}}^{\mu\alpha,ai}&=&
{\widetilde X}_1^{\mu\alpha,ai} - {\widetilde X}_2^{\mu\alpha,ai}
- \partial^{\beta} {\widetilde X}^{\mu,ai}_{\beta\alpha'} g^{\alpha\alpha'}
+2 {\widetilde S}^{\mu,ak}_{\alpha'\beta} \Gamma^{\beta,kj}_{\xi} g^{\alpha \alpha'}\,,\\
{\stackrel{\leftharpoonup}{X}}^{\mu,ai}_{01}&=&{\widetilde X}^{\mu,ai}_{01}\,,\\
{\stackrel{\leftharpoonup}{X}}^{\mu,ai}_{03Z}&=&
{\widetilde X}^{\mu,ai}_{03}
+ {\widetilde S}^{\mu,ak}_{\alpha\beta} ( \partial^{\alpha} \Gamma^{\beta,kj}_{\xi} - \Gamma^{\alpha,kl}_{\xi} \Gamma^{\beta,lj}_{\xi})
\nnb\\&&- ({\widetilde X}_1^{\mu\alpha,ak} - {\widetilde X}_2^{\mu\alpha,ak}) \Gamma^{ki}_{\xi,\alpha}
+ \partial^{\beta} {\widetilde X}^{\mu,ak}_{\beta\alpha} \Gamma^{\alpha,ki}_{\xi}\,,\\
{\stackrel{\leftharpoonup}{X}}^{\mu\alpha,ai}_{03Y}&=&- {\widetilde X}_2^{\mu\alpha,ai}\,,\\
{\stackrel{\rightharpoonup}{X}}^{\nu,ia}_{\alpha\beta}&=&
-{\widetilde S}^{\nu,ai}_{\alpha\beta}\,,\\
{\stackrel{\rightharpoonup}{X}}^{\nu\alpha,ia}&=&
{\widetilde X}_2^{\nu\alpha,ai} - {\widetilde X}_1^{\nu\alpha,ai}
-\partial^{\beta} {\widetilde X}^{\nu,ai}_{\alpha'\beta} g^{\alpha\alpha'}
+2 {\widetilde S}^{\nu,ci}_{\alpha'\beta} \Gamma^{\beta,ca}_{V} g^{\alpha \alpha'}\,,\\
{\stackrel{\rightharpoonup}{X}}^{\nu,ia}_{01}&=&{\widetilde X}^{\nu,ai}_{01}\,,\\
{\stackrel{\rightharpoonup}{X}}^{\nu,ia}_{03Z}&=&
{\widetilde X}^{\nu,ai}_{03}
+ {\widetilde S}^{\nu,ci}_{\alpha\beta} (\partial^{\alpha} \Gamma^{\beta,cb}_{V} -\Gamma^{\alpha,cd}_{V} \Gamma^{\beta,da}_{V})
\nnb\\&&- ({\widetilde X}_2^{\nu\alpha,ci} - {\widetilde X}_1^{\nu\alpha,ci}) \Gamma^{ca}_{V,\alpha}
 + \partial^{\beta} {\widetilde X}^{\nu,ci}_{\alpha\beta} \Gamma^{\alpha,ca}_{V}\,,\\
{\stackrel{\rightharpoonup}{X}}^{\nu\alpha,ia}_{03Y}&=&
- {\widetilde X}_1^{\nu\alpha,ai}\,,
\eea
The tilded quantities given in the above equations are different
the ones given below,
and these two have the relation as ${\widetilde X_{\xi\xi}^{above}}
=X_{\xi}^{\dagger} {\widetilde X_{\xi\xi}^{below}} X_{\xi}$,
and ${\widetilde X_{V\xi}^{above}} =X_V^{\dagger} {\widetilde X_{V\xi}^{below}} X_{\xi}$.

The tilded quantities are determined by the
following pre-standard forms prescribed in \cite{bfm1}
\bea
{1\over 2 } {\widehat V_{\mu}^{\dagger a}} \Box^{\mu\nu,ab}_{V\,V} {\widehat V_{\nu}^b}&=&
{1\over 2 } {\widehat V_{\mu}^{\dagger a}} \left [ W_V^{ab} \partial^2 g^{\mu\nu}
+ 2 g^{\mu\nu} {\widetilde \Gamma_V^{\alpha,ab}}  \partial_{\alpha}
+ {\widetilde \sigma_{0,VV}^{ab} } g^{\mu\nu}
+ {\widetilde \sigma_{2,VV}^{\mn,ab} }  \right] {\widehat V_{\nu}^b}\,,\\
{1\over 2} \xi^{\dagger i}  \Box_{\xi\,\xi}^{ij} \xi^j&=&
{1\over 2 } \xi^{\dagger i} \left[ W_{\xi}^{ij} \partial^2
+ 2 {\widetilde \Gamma_{\xi}^{\alpha,ij}} \partial_{\alpha}
+ {\widetilde \sigma_{0,\xi\xi}^{ij} }
+ {\widetilde \sigma_{2,\xi\xi}^{ij} }
+ {\widetilde \sigma_{4,\xi\xi}^{ij} }
\ssep + {\widetilde X^{\alpha,ij}_{3}} \partial_{\alpha}
+ \lvec{\partial_{\alpha}} {\widetilde X^{\alpha \beta,ij}_2} \partial_{\beta} \right ] \,,\\
{\widehat V_{\mu}^{\dagger,a}} {\stackrel{\leftharpoonup}{X}}^{\mu,aj} \xi^j
&=&\xi^{\dagger,i} {\stackrel{\rightharpoonup}{X}}^{\nu,ib} {\widehat V_{\nu}^b}\nnb\\
&=&\partial^{\alpha} {\widehat W^a_{\mu}} {\widetilde X}^{\mu,ai}_{\alpha\beta}\partial^{\beta} \xi^i
+ {\widehat W^a_{\mu}} {\widetilde X^{\mu\alpha,ai}_1} \partial_{\alpha} \xi^i
+ \partial_{\alpha} {\widehat W^a_{\mu}} {\widetilde X^{\mu\alpha,ai}_2} \xi^i
\nnb\\&&+ {\widehat W^a_{\mu}} {\widetilde X^{\mu,ai}_{01}} \xi^i
+{\widehat W^a_{\mu}} {\widetilde X^{\mu,ai}_{03}}\xi^i\,.
\eea
The matrix $W_V^{ab}$ determines
the mixing and normalization of the quantum vector
boson fields,
and reads
\begin{displaymath}
W_V^{ab} =\left (\begin{array}{cccc}
C_1&C_2& 0 & 0\\
C_2&C_3& 0 & 0\\
0 & 0 & C_4 &0\\
0 & 0 &0& C_4
\end{array}\right)\,,
\end{displaymath}
The matrix $W_{\xi}^{ij}$ determines the mixing
and normalization of Goldstone particles, and
reads
\begin{displaymath}
W_{\xi}^{ij} =\left (\begin{array}{ccc}
\rho& 0 & 0\\
0& 1 & 0\\
0 &0& 1
\end{array}\right)\,.
\end{displaymath}

The ${\widetilde \Gamma_V^{\alpha,ab}}$
will determine the gauge connection of the
covariant differential operators $D$,
which have the following form
\begin{displaymath}
{\widetilde \Gamma_{V}^{\alpha,ab}} =\nnb\\\left (\begin{array}{cccc}
0&0&i c_{AW} W^{-,\alpha}&-i c_{AW} W^{+,\alpha}\\
0&0& i c_{ZW} W^{-,\alpha} & -i c_{ZW} W^{+,\alpha}\\
 i c_{AW} W^{+,\alpha} & i c_{ZW} W^{+,\alpha}& - i e A^{\alpha} + i c_{W^+ W^-} Z^{\alpha} &0\\
-i c_{AW} W^{-,\alpha} &-i c_{ZW} W^{-,\alpha}&0&i e A^{\alpha} - i c_{W^+ W^-} Z^{\alpha}
\end{array}\right)\,,
\end{displaymath}
where $c_{aw}=e/2 + C_5/4$, $c_{zw}=(C_6 -C_7)/4$, and $c_{W^+ W^-}=C_7/2$.
The ${\widetilde \Gamma_{\xi}}$ is defined as
\begin{displaymath}
{\widetilde \Gamma_{\xi}^{\alpha,ij}} =\left (\begin{array}{ccc}
0&i {g\over 2} W^{-,\alpha}&-i {g\over 2} W^{+,\alpha}\\
i {g\over 2} W^{+,\alpha}& - i e A^{\alpha} + i c_{\xi^+ \xi^-} Z^{\alpha} &0\\
- i {g\over 2} W^{-,\alpha}&0&i e A^{\alpha}- i c_{\xi^+ \xi^-} Z^{\alpha}
\end{array}\right)\,.
\end{displaymath}
where the coefficient $c_{\xi^+ \xi^-}=\rho G/2 - g^{'2}/G$.

The mass matrices have the form ${\widetilde \sigma_{0, VV}^{ab}}=dia\{0,- \rho G^2 {v^2}/{4}, - g^2 {v^2}/{4}, - g^2 {v^2}/{4}\}$
and ${\widetilde \sigma_{0,\xi\xi}^{ij}}=dia\{- \rho^2 G^2 {v^2}/{4 cZ'}, - g^2 {v^2}/{4}, - g^2 {v^2}/{4}\}$.

From the pre-standard form, we can define covariant differentials, as
in the \cite{bfm1}, and collect all quadratic terms into the standard form.
We provide all matrices in the appendix.

\subsection{The Schwinger proper time and the heat kernel method}
Since the one-loop effective Lagrangian is Gaussian, we can use the
functional integral to integrate out all quantum fields, i.e. to take into
account the contributions of quantum corrections or quantum fluctuations of
low energy DOFs. And their contributions to the ${\cal L}^{eff}$ can be
elegantly and concisely expressed
\bea
\int_x {\cal L}_{1-loop}&=& Tr\Box_{\bar{c}c} - {1\over 2} \mbra{ Tr\ln\Box_{V\,V}
+Tr\ln\Box_{\xi\,\xi}
\wsep+Tr\ln\left (1-{\stackrel{\rightharpoonup}{X}^{\mu}} \Box^{-1}_{V\,V;\mu\nu} {\stackrel{\leftharpoonup}{X}^{\mu}} \Box^{-1}_{\xi\,\xi} \right  )}
\,\,.
\label{logtrew}
\eea

To extract the desired divergence structures, we need to expand this
compact expression. The expansion of logarithm is simply expressed by
the following formula
\bea
\langle x|\ln (1 - X)|y\rangle &=& - \langle x|X |y\rangle- {1\over 2} \langle x|X.X |y\rangle
\wsep- {1\over 3} \langle x|X.X.X |y\rangle- {1\over 4} \langle x|X.X.X.X |y\rangle+ ...\,,
\label{tayler}
\eea
and here the $X$ should be understood as an operator (a matrix) which acts on
the quantum states of the right side.

To evaluate the trace, we will use the Schwinger proper time method
and the heat kernel method \cite{htkl}. In this method, the standard propagators
can be expressed
as
\bea
\langle x|\Box^{-1,ab}_{V\,V;\mu\nu}|y\rangle=
\int_0^{\infty} \frac{d \tau}{(4 \pi \tau)^{d\over 2}}  \exp\{-\epsilon_F \tau \} \exp\left( - {z^2\over 4 \tau}\right) H^{\mu\nu,ab}_{V\,V}(x,y;\tau)\,\,,\label{vpro}\\
\langle x|\Box'^{-1,ab}_{\xi\,\xi}|y \rangle  =
\int_0^{\infty} \frac{d \tau}{(4 \pi \tau)^{d \over 2}}  \exp\{-\epsilon_F \tau \} \exp\left( - {z^2\over 4 \tau}\right) H_{\xi\xi,ab}(x,y;\tau)\,\,,
\label{spro}
\eea
where the $\epsilon_F$ is the Feynman prescription which will be taken to vanish,
and $z=y-x$.
The integral over the proper time $\tau$ and the factor
${1/(4 \pi \tau)^{d \over 2}} \exp\left( - {z^2/(4 \tau)}\right)$
conspire to separate the divergent part of the propagator.
And the $H(x,y;\tau)$ is analytic with reference to the arguments $z$ and
$\tau$, which means that $H(x,y;\tau)$ can be analytically
expanded with reference to both $z$ and $\tau$. Then we have
\bea
H(x,y;\tau)&=&H_0(x,y) +H_1(x,y) \tau + H_2(x,y) \tau^2 + \cdots\,,\\
H_i(x,y) &=& H_i(x,y)|_{x=y} + z^{\alpha} \partial_{\alpha} H_i(x,y)|_{x=y}
+ {1\over 2} z^{\alpha} z^{\beta} \partial_{\alpha}\partial_{\beta} H_i(x,y)|_{x=y}
+ \cdots
\eea
where $H_0(x,y)$, $H_1(x,y)$, and, $H_2(x,y)$ are
the Seeley-De Witt coefficients. The coefficient
$H_0(x,y)$ is the Wilson phase factor, which indicates the
phase change of a quantum state from the point $x$ to
the point $y$ and reads
\bea
H_0(x,y)= \exp\int_x^y \Gamma(z)\cdot dz,
\eea
where $\Gamma(z)$ is the affine connection defined on the
coordinate point $z$.
Higher order coefficients are determined by the
lower ones by the following recurrence relation
\bea
( 1+ n + z^{\mu} D_{\mu,x} ) H_{n+1}(x,y) + (D_x^2 + \sigma) H_{n}(x,y) =0\,.
\eea
All these Seeley-De Witt coefficients are gauge covariant with respect to the
gauge transformation.

The divergence counting rule of the integral
over the coordinate space $x$ and the proper
time $\tau$ can be established as
\bea
[z^{\mu}]_d=1\,\,,\,\,[\tau]_d=-2\,,
\label{dcnt}
\eea

It is easy to evaluate the $Tr\ln\Box_{V\,V}$ and $Tr\Box_{\bar{c}c}$
by directly using the result of the heat kernel method, which reads
\bea
{\bar \epsilon} Tr\ln\Box_{VV}&=&\int_x \left [tr[\sigma_{0,VV}  \sigma_{2,VV}]
+ {8\over 3} ({1\over 4} {\Gamma_{V,\mn}}  {\Gamma^{\mn,a}_V})
\right.\nnb\\&&\left.+ {1\over 2} tr[\sigma_{2,VV} \sigma_{2,VV}] \right ]\,,\label{vctt}\\
{\bar \epsilon} Tr\ln\Box_{{\bar c}c}&=&\int_x \left [
 {2\over 3} ({1\over 4} {\Gamma^{a}_{V,\mn}} {\Gamma^{\mn,a}_{V}})
 \right ]\,,\label{gstt}
\eea
from these results, to extract the divergences of quadratic
and logarithm is straightforward.

For the contributions of terms $Tr\ln\Box_{\xi\,\xi}$ and
$Tr\ln\left (1-{\stackrel{\rightharpoonup}{X}^{\mu}} \Box^{-1}_{V\,V;\mu\nu}
 {\stackrel{\leftharpoonup}{X}^{\mu}} \Box^{-1}_{\xi\,\xi} \right )$,
it needs somewhat labor.
Below we list some crucial steps of calculations. The first
two relations are about the action of the covariant differential
on the propagators, which read
\bea
\langle x| D_{\alpha}^x |\Box^{-1,ab}_{V\,V;\mu\nu}|y\rangle&=&
\int d\lambda {1\over (4 \pi \lambda)^{d/2}} \exp\{-\epsilon_F \tau \} \exp{(- {z^2 \over 4 \lambda})}\nnb\\
&&\times ({z_{\alpha} \over 2 \lambda} + D_{\alpha}) H(x,y;\lambda)\,,\nnb\\
\langle x| D_{\alpha}^x D_{\beta}^x \Box^{-1,ab}_{V\,V;\mu\nu}|y\rangle&=&
\int d\lambda {1\over (4 \pi \lambda)^{d/2}} \exp\{-\epsilon_F \tau \} \exp{(-{z^2 \over 4 \lambda})}\nnb\\
&&\times \left [({z_{\alpha}z_{\beta} \over 4 \lambda^2} - {g_{\alpha\beta}\over 2})
 + {1\over 2 \lambda} (z_{\alpha} D_{\beta} +z_{\beta} D_{\alpha})
+ D_{\alpha} D_{\beta}\right ] H(x,y;\lambda)
\label{eqs1}
\eea

The second relations are about the Short distance expansion, which make
it possible to covariantly expand the external fields over the coordinate
space, and are defined as
\bea
{\stackrel{\rightharpoonup}{X}} D {\stackrel{\leftharpoonup}{X}} &=& {\stackrel{\rightharpoonup}{X}} \partial {\stackrel{\leftharpoonup}{X}}
+ {\stackrel{\rightharpoonup}{X}} \Gamma_W {\stackrel{\leftharpoonup}{X}}
- {\stackrel{\rightharpoonup}{X}} {\stackrel{\leftharpoonup}{X}} \Gamma_{\xi}\,,\nnb\\
{\stackrel{\rightharpoonup}{X}} D D {\stackrel{\leftharpoonup}{X}} &=& {\stackrel{\rightharpoonup}{X}} \partial \partial {\stackrel{\leftharpoonup}{X}}
+ {\stackrel{\rightharpoonup}{X}} \Gamma_W \Gamma_W {\stackrel{\leftharpoonup}{X}}
+ {\stackrel{\rightharpoonup}{X}} {\stackrel{\leftharpoonup}{X}} \Gamma_{\xi} \Gamma_{\xi}
-2 {\stackrel{\rightharpoonup}{X}} \Gamma_W {\stackrel{\leftharpoonup}{X}} \Gamma_{\xi}
\nnb\\&&+2 {\stackrel{\rightharpoonup}{X}} \Gamma_W \partial {\stackrel{\leftharpoonup}{X}}
-2 {\stackrel{\rightharpoonup}{X}} \partial {\stackrel{\leftharpoonup}{X}} \Gamma_{\xi}
+  {\stackrel{\rightharpoonup}{X}} \partial \Gamma_W {\stackrel{\leftharpoonup}{X}}
-  {\stackrel{\rightharpoonup}{X}} {\stackrel{\leftharpoonup}{X}} \partial \Gamma_{\xi}\,.
\label{eqs2}
\eea

The rest of corresponding relevant integrals are based on these two relations
given in Eq. (\ref{eqs1}) and Eq. (\ref{eqs2}).
So that after dropping those quartic divergences
and only keeping terms up to $O(p^4)$, we can get the following results about
the logarithm and trace of the terms $Tr\ln\Box_{\xi\,\xi}$ and
$Tr\ln\left (1-{\stackrel{\rightharpoonup}{X}^{\mu}} \Box^{-1}_{V\,V;\mu\nu} {\stackrel{\leftharpoonup}{X}^{\mu}} \Box^{-1}_{\xi\,\xi} \right )$
\bea
Tr\ln\Box_{\xi\,\xi}&=& \int_x \left [tr[\sigma_{0,\xi\xi}  \sigma_{2,\xi\xi}]
+ tr[ \sigma_{0,\xi\xi} \sigma_{4,\xi\xi}]\ssep
+ {2\over 3} ({1\over 4} {\Gamma^{a}_{\xi,\mn}} {\Gamma^{\mn,a}_{\xi}})+ {1\over 2} tr[\sigma_{2,\xi\xi} \sigma_{2,\xi\xi}]\right ]\,,\label{sclt}\\
Tr\ln\left (1-{\stackrel{\rightharpoonup}{X}^{\mu}} \Box^{-1}_{V\,V;\mu\nu} {\stackrel{\leftharpoonup}{X}^{\mu}} \Box^{-1}_{\xi\,\xi} \right )&=&{1\over {\bar \epsilon}} \int_x (p4t+p3t+p2t)\,.
\eea
 The
$\Gamma_{\mn}$ is the field strength
tensor corresponding to the affine connection
$\Gamma_{\mu}$. We have used the dimension regularization and the modified minimal
subtraction scheme to extract the divergent structures in this step.
The $p4t$ represents the
contributions of four propagators $tr(
{\stackrel{\rightharpoonup}{X}} \Box^{-1}_{W\,W} {\stackrel{\leftharpoonup}{X}} \Box'^{-1}_{\xi\,\xi}
{\stackrel{\rightharpoonup}{X}} \Box^{-1}_{W\,W} {\stackrel{\leftharpoonup}{X}} \Box'^{-1}_{\xi\,\xi}
)$, which reads
\bea
p4t&=&{g_{\mu\nu} g_{\mu'\nu'} \over 6} \left[ {g^{\alpha\beta\alpha'\beta'}\over 4}
 tr[2 {\stackrel{\rightharpoonup}{X}}^{\mu}_{\alpha\beta} {\stackrel{\leftharpoonup}{X}}^{\nu}_{\alpha'\beta'} {\stackrel{\rightharpoonup}{X}}^{\mu'}_{01} {\stackrel{\leftharpoonup}{X}}^{\nu'}_{01} \right. \nnb\\
&&\left. + 2 {\stackrel{\rightharpoonup}{X}}^{\mu}_{01} {\stackrel{\leftharpoonup}{X}}^{\nu}_{\alpha\beta} {\stackrel{\rightharpoonup}{X}}^{\mu'}_{\alpha'\beta'} {\stackrel{\leftharpoonup}{X}}^{\nu'}_{01}
+   {\stackrel{\rightharpoonup}{X}}^{\mu}_{\alpha\beta} {\stackrel{\leftharpoonup}{X}}^{\nu}_{01} {\stackrel{\rightharpoonup}{X}}^{\mu'}_{\alpha'\beta'} {\stackrel{\leftharpoonup}{X}}^{\nu'}_{01}
+   {\stackrel{\rightharpoonup}{X}}^{\mu}_{01} {\stackrel{\leftharpoonup}{X}}^{\nu}_{\alpha\beta} {\stackrel{\rightharpoonup}{X}}^{\mu'}_{01} {\stackrel{\leftharpoonup}{X}}^{\nu'}_{\alpha'\beta'}]\right.\nnb\\
&&\left. -  {g^{\alpha\beta\alpha'\beta'\alpha''\beta''}\over 16}
 tr[{\stackrel{\rightharpoonup}{X}}^{\mu}_{\alpha\beta} \sigma_{0,VV} {\stackrel{\leftharpoonup}{X}}^{\nu}_{\alpha'\beta'} {\stackrel{\rightharpoonup}{X}}^{\mu'}_{\alpha''\beta''} {\stackrel{\leftharpoonup}{X}}^{\nu'}_{01}
+   {\stackrel{\rightharpoonup}{X}}^{\mu}_{\alpha\beta} \sigma_{0,VV} {\stackrel{\leftharpoonup}{X}}^{\nu}_{\alpha'\beta'} {\stackrel{\rightharpoonup}{X}}^{\mu'}_{01} {\stackrel{\leftharpoonup}{X}}^{\nu'}_{\alpha''\beta''} \ssep
+   {\stackrel{\rightharpoonup}{X}}^{\mu}_{\alpha\beta}{\stackrel{\leftharpoonup}{X}}^{\nu}_{\alpha'\beta'}  \sigma_{0,\xi\xi} {\stackrel{\rightharpoonup}{X}}^{\mu'}_{\alpha''\beta''} {\stackrel{\leftharpoonup}{X}}^{\nu'}_{01}
+   {\stackrel{\rightharpoonup}{X}}^{\mu}_{\alpha\beta}{\stackrel{\leftharpoonup}{X}}^{\nu}_{\alpha'\beta'}  \sigma_{0,\xi\xi} {\stackrel{\rightharpoonup}{X}}^{\mu'}_{01} {\stackrel{\leftharpoonup}{X}}^{\nu'}_{\alpha''\beta''} \ssep
+   {\stackrel{\rightharpoonup}{X}}^{\mu}_{\alpha\beta} {\stackrel{\leftharpoonup}{X}}^{\nu}_{\alpha'\beta'} {\stackrel{\rightharpoonup}{X}}^{\mu'}_{\alpha''\beta''} \sigma_{0,VV} {\stackrel{\leftharpoonup}{X}}^{\nu'}_{01}
+   {\stackrel{\rightharpoonup}{X}}^{\mu}_{\alpha\beta} {\stackrel{\leftharpoonup}{X}}^{\nu}_{\alpha'\beta'} {\stackrel{\rightharpoonup}{X}}^{\mu'}_{01}              \sigma_{0,VV} {\stackrel{\leftharpoonup}{X}}^{\nu'}_{\alpha''\beta''} \ssep
+   {\stackrel{\rightharpoonup}{X}}^{\mu}_{\alpha\beta}{\stackrel{\leftharpoonup}{X}}^{\nu}_{\alpha'\beta'}  {\stackrel{\rightharpoonup}{X}}^{\mu'}_{\alpha''\beta''} {\stackrel{\leftharpoonup}{X}}^{\nu'}_{01}\sigma_{0,\xi\xi}
+   {\stackrel{\rightharpoonup}{X}}^{\mu}_{\alpha\beta}{\stackrel{\leftharpoonup}{X}}^{\nu}_{\alpha'\beta'}  {\stackrel{\rightharpoonup}{X}}^{\mu'}_{01} {\stackrel{\leftharpoonup}{X}}^{\nu'}_{\alpha''\beta''}\sigma_{0,\xi\xi} ]
\right ]\,.
\label{mixt0}
\eea
The $p3t$ represents the contributions of three propagators $tr(
{\stackrel{\rightharpoonup}{X}} \Box^{-1}_{W\,W} {\stackrel{\leftharpoonup}{X}} \Box'^{-1}_{\xi\,\xi} X_{\alpha\beta} d^{\alpha} d^{\beta} \Box'^{-1}_{\xi\,\xi})$, which reads
\bea
p3t&=&{1\over 24} g^{\alpha\beta\alpha'\beta'} g_{\mn}
tr[{\stackrel{\rightharpoonup}{X}}^{\mu}_{01}          \sigma_{0,VV} {\stackrel{\leftharpoonup}{X}}^{\nu}_{\alpha\beta} X_{\alpha'\beta'}
+  {\stackrel{\rightharpoonup}{X}}^{\mu}_{\alpha\beta} \sigma_{0,VV} {\stackrel{\leftharpoonup}{X}}^{\nu}_{01} X_{\alpha'\beta'} \wsep
+  {\stackrel{\rightharpoonup}{X}}^{\mu}_{01}          {\stackrel{\leftharpoonup}{X}}^{\nu}_{\alpha\beta} \sigma_{0,\xi\xi} X_{\alpha'\beta'}
+  {\stackrel{\rightharpoonup}{X}}^{\mu}_{\alpha\beta} {\stackrel{\leftharpoonup}{X}}^{\nu}_{01}          \sigma_{0,\xi\xi} X_{\alpha'\beta'} \wsep
+  {\stackrel{\rightharpoonup}{X}}^{\mu}_{01}          {\stackrel{\leftharpoonup}{X}}^{\nu}_{\alpha\beta} X_{\alpha'\beta'} \sigma_{0,\xi\xi}
+  {\stackrel{\rightharpoonup}{X}}^{\mu}_{\alpha\beta} {\stackrel{\leftharpoonup}{X}}^{\nu}_{01}          X_{\alpha'\beta'} \sigma_{0,\xi\xi}
] \wsep
- {1\over 4} g^{\alpha\beta} g_{\mn} tr[{\stackrel{\rightharpoonup}{X}}^{\mu}_{01} {\stackrel{\leftharpoonup}{X}}^{\nu}_{01} X_{\alpha\beta}]\,.
\eea
The $p2t$ represents the contributions of two propagators $tr(
{\stackrel{\rightharpoonup}{X}} \Box^{-1}_{W\,W} {\stackrel{\leftharpoonup}{X}} \Box'^{-1}_{\xi\,\xi})$, which can be further
divided into six groups:
\bea
p2t&=&t_{AA} + t_{AB} + t_{AC} + t_{BB} + t_{BC} + t_{CC}\,,\\
t_{AA}&=& { g^{\mn}\over 8} ({g^{\alpha\beta\alpha'\beta'\delta\gamma} \over 6} - {2 g^{\alpha\beta} g^{\alpha'\beta'\delta\gamma} \over 3}
-{g^{\alpha'\beta'} g^{\alpha\beta\delta\gamma} \over 3}+ g^{\alpha\beta} g^{\alpha'\beta'} g^{\delta\gamma})
tr[{\stackrel{\rightharpoonup}{X}}^{\mu}_{\alpha\beta} \sigma_{0,VV} D_{\delta}D_{\gamma}{\stackrel{\leftharpoonup}{X}}^{\nu}_{\alpha'\beta'}
\wsep+ { g^{\mn}\over 8} ({g^{\alpha\beta\alpha'\beta'\delta\gamma} \over 6} - { g^{\alpha\beta} g^{\alpha'\beta'\delta\gamma} \over 3}
-{2 g^{\alpha'\beta'} g^{\alpha\beta\delta\gamma} \over 3}+ g^{\alpha\beta} g^{\alpha'\beta'} g^{\delta\gamma})
tr[{\stackrel{\rightharpoonup}{X}}^{\mu}_{\alpha\beta} D_{\delta}D_{\gamma}{\stackrel{\leftharpoonup}{X}}^{\nu}_{\alpha'\beta'} \sigma_{0,\xi\xi}] \wsep
+ { g^{\alpha\beta\alpha'\beta'} \over 24} \left [
  g_{\mn} tr[{\stackrel{\rightharpoonup}{X}}^{\mu}_{\alpha\beta} \sigma_{0,VV} {\stackrel{\leftharpoonup}{X}}^{\nu}_{\alpha'\beta'} \sigma_{2,\xi\xi}
] + g_{\mu\mu'} g_{\nu\nu'} tr[
{\stackrel{\rightharpoonup}{X}}^{\mu}_{\alpha\beta}\sigma^{\mu'\nu'}_{2,VV}{\stackrel{\leftharpoonup}{X}}^{\nu}_{\alpha'\beta'} \sigma_{0,\xi\xi}] \right ]
\wsep + { g^{\alpha\beta\alpha'\beta'} \over 12} \left [
  g_{\mn} tr[{\stackrel{\rightharpoonup}{X}}^{\mu}_{\alpha\beta} \sigma_{0,\xi\xi} \sigma_{2,\xi\xi}  {\stackrel{\leftharpoonup}{X}}^{\nu}_{\alpha'\beta'}
] + g_{\mu\mu'} g_{\nu\nu'} tr[
{\stackrel{\rightharpoonup}{X}}^{\mu}_{\alpha\beta}  \sigma_{0,VV} \sigma^{\mu'\nu'}_{2,VV} {\stackrel{\leftharpoonup}{X}}^{\nu}_{\alpha'\beta'} ]
\right ]\,,\nnb\\
t_{AB}&=& { g_{\alpha'\alpha''} g_{\mu\nu} \over 4} \left [ (g^{\alpha\beta}g^{\alpha'\beta'} -{1\over 3} g^{\alpha\beta\alpha'\beta'} )
tr[{\stackrel{\rightharpoonup}{X}}^{\mu}_{\alpha\beta}\sigma_{0,VV}  D_{\beta'} {\stackrel{\leftharpoonup}{X}}^{\nu\alpha''}
- {\stackrel{\rightharpoonup}{X}}^{\mu\alpha''}       D_{\beta'} {\stackrel{\leftharpoonup}{X}}^{\nu}_{\alpha\beta} \sigma_{0,\xi\xi}]
\ssep +(g^{\alpha\beta}g^{\alpha'\beta'} -{2\over 3} g^{\alpha\beta\alpha'\beta'} )
tr[{\stackrel{\rightharpoonup}{X}}^{\mu}_{\alpha\beta} D_{\beta'} {\stackrel{\leftharpoonup}{X}}^{\nu\alpha''} \sigma_{0,\xi\xi}
- {\stackrel{\rightharpoonup}{X}}^{\mu\alpha''} \sigma_{0,VV} D_{\beta'} {\stackrel{\leftharpoonup}{X}}^{\nu}_{\alpha\beta}
] \right ] \,,\\
t_{AC}&=& - {g^{\alpha\beta} g_{\mn} \over 4}
tr[{\stackrel{\rightharpoonup}{X}}^{\mu}_{\alpha\beta}\sigma_{0,VV} {\stackrel{\leftharpoonup}{X}}^{\nu}_{01}
+  {\stackrel{\rightharpoonup}{X}}^{\mu}_{01}         \sigma_{0,VV} {\stackrel{\leftharpoonup}{X}}^{\nu}_{\alpha\beta}
+  {\stackrel{\rightharpoonup}{X}}^{\mu}_{\alpha\beta}\sigma_{0,VV} {\stackrel{\leftharpoonup}{X}}^{\nu}_{03Z}
+  {\stackrel{\rightharpoonup}{X}}^{\mu}_{03Z}        \sigma_{0,VV} {\stackrel{\leftharpoonup}{X}}^{\nu}_{\alpha\beta}\nnb\\
&&-\partial_{\alpha'} {\stackrel{\rightharpoonup}{X}}^{\mu}_{\alpha\beta} \sigma_{0,VV}{\stackrel{\leftharpoonup}{X}}^{\nu\alpha'}_{03Y}
-{\stackrel{\rightharpoonup}{X}}^{\mu\alpha'}_{03Y} \sigma_{0,VV} \partial_{\alpha'} {\stackrel{\leftharpoonup}{X}}^{\nu}_{\alpha\beta}\nnb\\
&&+{\stackrel{\rightharpoonup}{X}}^{\mu}_{\alpha\beta}{\stackrel{\leftharpoonup}{X}}^{\nu}_{01} \sigma_{0,\xi\xi}
+  {\stackrel{\rightharpoonup}{X}}^{\mu}_{01}         {\stackrel{\leftharpoonup}{X}}^{\nu}_{\alpha\beta} \sigma_{0,\xi\xi}
+  {\stackrel{\rightharpoonup}{X}}^{\mu}_{\alpha\beta}{\stackrel{\leftharpoonup}{X}}^{\nu}_{03Z} \sigma_{0,\xi\xi}
+  {\stackrel{\rightharpoonup}{X}}^{\mu}_{03Z}        {\stackrel{\leftharpoonup}{X}}^{\nu}_{\alpha\beta}\sigma_{0,\xi\xi}\nnb\\
&&-\partial_{\alpha'} {\stackrel{\rightharpoonup}{X}}^{\mu}_{\alpha\beta} {\stackrel{\leftharpoonup}{X}}^{\nu\alpha'}_{03Y}\sigma_{0,\xi\xi}
-{\stackrel{\rightharpoonup}{X}}^{\mu\alpha'}_{03Y} \partial_{\alpha'} {\stackrel{\leftharpoonup}{X}}^{\nu}_{\alpha\beta}\sigma_{0,\xi\xi}] \nnb\\
&&- {1\over 4} g^{\alpha\beta}
 tr[
  g_{\mu\nu} {\stackrel{\rightharpoonup}{X}}^{\mu}_{01} {\stackrel{\leftharpoonup}{X}}^{\nu}_{\alpha\beta} \sigma_{2,\xi\xi}
+ g_{\mu\mu'} g_{\nu\nu'} {\stackrel{\rightharpoonup}{X}}^{\mu}_{\alpha\beta} \sigma^{\mn'\nu'}_{2,VV} {\stackrel{\leftharpoonup}{X}}^{\nu}_{01}
]\,\nnb\\
&& + g_{\mu\nu} ({1\over 6} g^{\alpha\beta\alpha'\beta'} - {1\over 4} g^{\alpha\beta} g^{\alpha'\beta'})
tr [{\stackrel{\rightharpoonup}{X}}^{\mu}_{\alpha\beta} D_{\alpha'}D_{\beta'}{\stackrel{\leftharpoonup}{X}}^{\nu}_{01}
+{\stackrel{\rightharpoonup}{X}}^{\mu}_{01} D_{\alpha'}D_{\beta'}{\stackrel{\leftharpoonup}{X}}^{\nu}_{\alpha\beta}]\,,\\
t_{BB}&=& - {g_{\mn} g_{\alpha\beta} \over 4} tr[
 {\stackrel{\rightharpoonup}{X}}^{\mu\alpha} \sigma_{0,VV}{\stackrel{\leftharpoonup}{X}}^{\nu\beta}
+{\stackrel{\rightharpoonup}{X}}^{\mu\alpha} {\stackrel{\leftharpoonup}{X}}^{\nu\beta} \sigma_{0,\xi\xi}
]\,,\\
t_{BC}&=& { g^{\alpha\beta} g_{\alpha\alpha'} g_{\mn} \over 2}
tr[{\stackrel{\rightharpoonup}{X}}^{\mu\alpha'} D_{\beta} {\stackrel{\leftharpoonup}{X}}^{\nu}_{01}
-{\stackrel{\rightharpoonup}{X}}^{\mu}_{01} D_{\beta} {\stackrel{\leftharpoonup}{X}}^{\nu\alpha'}]\,,\\
t_{CC}&=&g_{\mn} tr[{\stackrel{\rightharpoonup}{X}}^{\mu}_{01} {\stackrel{\leftharpoonup}{X}}^{\nu}_{01}
+{\stackrel{\rightharpoonup}{X}}^{\mu}_{01} {\stackrel{\leftharpoonup}{X}}^{\nu}_{03Z}
+{\stackrel{\rightharpoonup}{X}}^{\mu}_{03Z} {\stackrel{\leftharpoonup}{X}}^{\nu}_{01}
-\partial_{\alpha'} {\stackrel{\rightharpoonup}{X}}^{\mu}_{01} {\stackrel{\leftharpoonup}{X}}^{\nu\alpha'}_{03Y}
-{\stackrel{\rightharpoonup}{X}}^{\mu\alpha'}_{03Y} \partial_{\alpha'} {\stackrel{\leftharpoonup}{X}}^{\nu}_{01}]\,.
\label{mixt1}
\eea
where the trace is to sum over the group indices,
and the high rank tensors $g^{\alpha\beta\gamma\delta}$ and $g^{\alpha\beta\gamma\delta\mu\nu}$
are symmetric on all indices and defined as
\bea
 g^{\alpha\beta\gamma\delta} &=& g^{\alpha\beta} g^{\gamma\delta} + g^{\alpha\gamma} g^{\beta\delta} + g^{\alpha\delta} g^{\beta\gamma}\,,\\
 g^{\alpha\beta\gamma\delta\mu\nu} &=& g^{\alpha\beta} g^{\gamma\delta\mu\nu} + g^{\alpha\gamma} g^{\beta\delta\mu\nu} + g^{\alpha\delta} g^{\gamma\beta\mu\nu}
+g^{\alpha \mu} g^{\beta\gamma\delta\nu} + g^{\alpha\nu} g^{\beta\gamma\delta\mu}\,.
\eea
The formula given in Eq. (\ref{mixt0}---\ref{mixt1}), can reduce to the ones given
the $SU(2)$ case \cite{our2}.

\subsection{The renormalization group equations}
Substitute the relevant matrices in the appendix A into the formula given
in Eqs. (\ref{mixt0}---\ref{mixt1}) and after some tedious algebraic manipulation,
we can extract out the complete RGEs of
the system, which will be provided in our next paper with
including the effective Higgs sector \cite{our3}.

To derive the RGEs, we have taken into account the contributions
of Fermions in the SM to the gauge couplings $g$ and $g'$. For the sake
of simplicity, we do not consider the fact that top quarks are
much heavier than other fermions (b quarks), so there is no
contribution of fermions to the ACs. Strictly speaking,
we should consider several possible cases for the EWCL of Higgs:
1) Higgs scalar is heavier than top quarks,
2) Higgs scalar is as heavy as top quarks,
and 3) Higgs scalar is lighter than top quarks. We will conduct such
a study in the future.

The complete RGEs, which are expressed in the ECs $C_i$,
are quite complicated and difficult to understand. We will provide in our
next paper. But here we would like to provide a simplifies version to
extract some helpful features about the RGEs.  
We can expand the effective couplings $C_i$ in $\al_i g^2 (\al_i g'^2)$
and neglect those higher powers of $\al_i g^2 (\al_i g'^2)$,
and get the following simplified version of RGEs
\bea
\label{svrgesa}
\dd{g} &\approx&{g^3 \over 2} \bbra{
- \frac{10}{3}   +
  \frac{1}{12 \rho}
+ \al_1 {g'}^2 +
  \frac{7 \al_2 {g'}^2}{6}  \wsep +
  \fmpng{\al_3}{  3 {g'}^2 \rho +
       g^2 \left( 1 + 31 \rho \right) }{2
     \rho}
- 2 \al_8 g^2 +
  \frac{9 \al_9 g^2}{2}
}\cma\\
\dd{g'} &\approx&{g'^3 \over 2} \bbra{
\frac{13}{2} -
  \frac{\rho}{3} + \frac{{\rho}^2}{12}
+ 2 \al_1 g^2 -
  \al_2 g^2 \left( -4 + \rho \right)
+ \frac{7 \al_3 g^2}{3}
}\cma\\
\dd{\al_1} &\approx&
  \frac{\rho}{6} - \frac{{\rho}^2}{12} +
2 \al_1 g^2 +
  \frac{\al_2 g^2 \left( 6 + \rho \right) }{2} -
  \frac{\al_3 g^2 \left( 22 + 3 \rho \right) }{6}
\wsep + \al_8 g^2 -
  \frac{\al_9 g^2 \left( -4 + \rho \right) }{2}
\cma\\
\dd{\al_2} &\approx&
- \frac{1}{24}
+
  \frac{1}{12 \rho} - \frac{\rho}{6} +
  \frac{{\rho}^2}{12}
- \frac{\al_1 g^2}{2}
\wsep +
  \mpng{\al_2 \left( -12 g^2
        \left( -1 + \rho \right)  +
       {g'}^2 \left( 46 + \rho \right)  \right) }{
24} \wsep+ \mpng{\al_3 \left( 24 {g'}^2
        \rho + g^2
        \left( -17 + 47 \rho + {\rho}^2 \right)
\right) }{24 \rho}
\wsep +
  \mpng{\al_4 \left( -2 {g'}^2 +
       g^2 \left( -4 + \rho \right)  \right) }{4}
\wsep +
{\bf   \al_5 \left( -{g'}^2 -
     \frac{g^2 \left( -4 + \rho \right) }{2} \right)}
 -
  \frac{\al_6 {g'}^2}{2} -
  \al_7 {g'}^2   +
  \frac{\al_9 g^2 \left( 18 + \rho \right) }{24}
\cma\\
\dd{\al_3} &\approx&
\frac{1}{24}
-
  \frac{1}{12 \rho}
+ \mpng{\al_1 {g'}^2
     \left( -2 + \rho \right) }{4}   +
  \mpng{\al_2 {g'}^2
     \left( -7 + 2 \rho \right) }{6}
\wsep + \mpng{\al_3
     \left( -6 {g'}^2
        \left( -1 - 8 \rho + {\rho}^2 \right)  +
       g^2 \left( -1 + 110 \rho + 14 {\rho}^2
\right)  \right) }{24 \rho}
\wsep - \mpng{\al_4
     \left( 5 {g'}^2 \rho +
       g^2 \left( 1 + 13 \rho \right)  \right) }{8
     \rho}
+ {\bf  \mpng{\al_5 \left( g^2 + 13 g^2 \rho +
       {g'}^2 \rho \right) }{4 \rho} }
\wsep - \frac{\al_6
     \left( 5 {g'}^2 \rho +
       g^2 \left( 1 + 13 \rho \right)  \right) }{8
     \rho}
+
  \mpng{\al_7 \left( {g'}^2 \rho +
       g^2 \left( 1 + 13 \rho \right)  \right) }{4
     \rho}
\wsep +
  \frac{\al_8 g^2 \rho}{4}
+
  \mpng{\al_9 \left( 12 {g'}^2 +
       g^2 \left( -3 + 2 \rho \right)  \right) }{6}
\cma\\
\dd{\al_4} &\approx&
\frac{1}{12}
-
  \frac{1}{12 {\rho}^2} - \frac{{\rho}^2}{12} +
  \mpng{\al_2 {g'}^2
     \left( 7 - 2 {\rho}^2 \right) }{24 \rho}
\wsep - \mpng{\al_3 g^2
     \left( -41 + 42 \rho + 2 {\rho}^2 \right) }
{24 \rho}
\wsep + \mpng{\al_4 \left( {g'}^2
        \left( -2 + 11 \rho \right)  +
       g^2 \left( 2 + 9 \rho - {\rho}^2 \right)
\right) }{2 \rho}
\wsep + {\bf \al_5
   \left( 3 {g'}^2 +
     g^2 \left( 3 + \rho \right)  \right) }
+
  \mpng{\al_6 \left( {g'}^2
        \left( -1 + \rho \right)  +
       g^2 \left( 1 + 2 \rho \right)  \right) }{\rho}
\wsep+ 2 \al_7 {g'}^2
+
  \mpng{\al_9 g^2
     \left( 7 + 80 \rho - 2 {\rho}^2 \right) }{24
     \rho}
\cma\\
\dd{\al_5} &\approx&
- \frac{1}{12}
-
  \frac{1}{24 {\rho}^2} + \frac{{\rho}^2}{12}
+
  \mpng{\al_2 {g'}^2
     \left( -2 + {\rho}^2 \right) }{12 \rho}
\wsep + \mpng{\al_3 g^2
     \left( -11 + 18 \rho + 2 {\rho}^2 \right) }
{24 \rho}
\wsep + \mpng{\al_4
     \left( -5 {g'}^2 \rho +
       g^2 \left( -1 - 2 \rho + {\rho}^2 \right)
\right) }{2 \rho}
+
 {\bf  \al_5 \left( - g^2
        \left( -1 + \rho \right)  -
     \frac{{g'}^2}{\rho} \right) }
\wsep -
  \mpng{\al_6 \left( g^2 + 2 {g'}^2 \right) }{2}    +
  \mpng{\al_7 \left( g^2 - {g'}^2 \right)
     \left( 1 + 2 \rho \right) }{\rho}
\wsep +
  \mpng{\al_9 g^2
     \left( -2 - 40 \rho + {\rho}^2 \right) }{12
     \rho}
\cma\\
\dd{\al_6} &\approx&
\frac{1}{12 {\rho}^2} - \frac{1}{12 \rho} -
  \frac{\rho}{12} + \frac{{\rho}^2}{12}
\wsep -
  \mpng{\al_2 \left( -24 g^2
        \left( -2 + \rho \right)  \rho +
       {g'}^2 \left( 7 - 20 \rho +
          6 {\rho}^2 + {\rho}^3 \right)  \right) }
{24 \rho}
\wsep +
  \mpng{\al_3 g^2
     \left( -62 + 53 \rho + 44 {\rho}^2 -
       2 {\rho}^3 \right) }{48 \rho}
\wsep +
  \mpng{\al_4 \left( {g'}^2
        \left( -16 + \rho \right)  \rho +
       g^2 \left( -2 - 3 \rho + 7 {\rho}^2
\right)  \right) }{4 \rho}
\wsep + {\bf \mpng{\al_5 \left( g^2
        \left( 9 - 2 \rho \right)  \rho +
       2 {g'}^2
        \left( -1 - 3 \rho + {\rho}^2 \right)
\right) }{2 \rho} }
\wsep +
  \mpng{\al_6 \left( {g'}^2
        \left( -4 + 2 \rho + {\rho}^2 \right)  -
       g^2 \left( -2 + 9 \rho + {\rho}^2 \right)
\right) }{4 \rho}
\wsep +
  \mpng{\al_7 \left( g^2
        \left( 2 + 11 \rho - 2 {\rho}^2 \right)  +
       2 {g'}^2
        \left( -2 - 2 \rho + {\rho}^2 \right)
\right) }{2 \rho}
\wsep -
  \mpng{\al_9 g^2
     \left( 7 + 92 \rho + 2 {\rho}^2 +
       {\rho}^3 \right) }{24 \rho} +
  \mpng{2 \al_a \left( -{g'}^2 +
       g^2 \left( 1 + 2 \rho \right)  \right) }{\rho}
\cma\\
\dd{\al_7} &\approx&
\frac{1}{24 {\rho}^2} + \frac{1}{12 \rho}
- \frac{\rho}{24} - \frac{{\rho}^2}{12}
\wsep +
  \mpng{\al_2 \left( -48 g^2
        \left( -2 + \rho \right)  \rho +
       {g'}^2 \left( 8 - 40 \rho +
          12 {\rho}^2 + 11 {\rho}^3 \right)
\right) }{48 \rho}
\wsep+
  \mpng{\al_3 g^2
     \left( 17 - 56 \rho - 8 {\rho}^2 +
       11 {\rho}^3 \right) }{48 \rho}
\wsep +
  \mpng{\al_4 \left( g^2
        \left( 11 - 6 \rho \right)  \rho +
       {g'}^2 \left( -2 + 11 \rho +
          2 {\rho}^2 \right)  \right) }{4 \rho}
\wsep+
{\bf   \mpng{\al_5 \left( g^2 \rho
        \left( 1 + 2 \rho \right)  +
       {g'}^2 \left( 1 + 3 \rho \right)  \right) }
{2 \rho} }
\wsep +
  \mpng{\al_6 \left( -2 g^2 \rho
        \left( 3 + \rho \right)  +
       {g'}^2 \left( -4 + 5 \rho +
          2 {\rho}^2 \right)  \right) }{4 \rho}
\wsep
+ \mpng{\al_7
     \left( 7 {g'}^2 \rho +
       g^2 \left( -1 - 17 \rho + 2 {\rho}^2
\right)  \right) }{2 \rho}
\wsep +
  \mpng{\al_9 g^2
     \left( 8 + 136 \rho + 28 {\rho}^2 +
       11 {\rho}^3 \right) }{48 \rho}
-
  \mpng{\al_a \left( {g'}^2 +
       g^2 \left( -1 + \rho \right)  \right) }{\rho}
\cma\\
\dd{\al_8} &\approx&
-
  \frac{1}{12 \rho} + \frac{{\rho}^2}{12}
 - \al_1 {g'}^2 -
  \frac{7 \al_2 {g'}^2}{6}
\wsep +
  \mpng{\al_3 \left( -3 {g'}^2 \rho +
       g^2 \left( -1 - \rho + 2 {\rho}^2 \right)
\right) }{2 \rho}
+ 6 \al_8 g^2
+
  \mpng{\al_9 g^2 \left( 3 + 2 \rho \right) }{2}
\cma\\
\dd{\al_9} &\approx&
  \frac{1}{12 \rho} - \frac{{\rho}^2}{12}
-
  \mpng{\al_1 {g'}^2
     \left( -2 + \rho \right) }{4}
+ \al_2 \left( g^2 + \frac{{g'}^2
        \left( 4 - 9 \rho \right) }{24} \right)
\wsep + \mpng{\al_3
     \left( 2 {g'}^2 {\rho}^2 +
       g^2 \left( 4 - 21 \rho - 9 {\rho}^2
\right)  \right) }{8 \rho}
\wsep +
  \mpng{\al_4 \left( 9 {g'}^2 \rho +
       g^2 \left( 1 + \rho - 2 {\rho}^2 \right)
\right) }{8 \rho}
\wsep +
{\bf  \mpng{\al_5 \left( 3 {g'}^2 \rho +
       g^2 \left( -1 - \rho + 2 {\rho}^2 \right)
\right) }{4 \rho} }
\wsep + \mpng{\al_6
     \left( 9 {g'}^2 \rho +
       g^2 \left( 1 + 13 \rho \right)  \right) }{8
     \rho}
\wsep -
  \mpng{\al_7 \left( -3 {g'}^2 \rho +
       g^2 \left( 1 + 13 \rho \right)  \right) }{4
     \rho}
-
  \mpng{\al_8 g^2 \left( 2 + \rho \right) }{4}
\wsep+
  \mpng{\al_9 \left( 16 {g'}^2 +
       g^2 \left( 36 - 7 \rho \right)  \right) }{8}
\cma\\
\dd{\al_a} &\approx&
\frac{-1}{16 {\rho}^2} + \frac{\rho}{8} -
  \frac{{\rho}^4}{16}
+
  \mpng{\al_2 {g'}^2
     \left( 1 - 3 {\rho}^3 \right) }{16 \rho}
\wsep -
  \mpng{\al_3 g^2
     \left( -5 + 7 \rho - 4 {\rho}^2 +
       {\rho}^3 \right) }{16 \rho}
\wsep -
  \mpng{\al_4 \left( g^2
        \left( -1 + \rho + {\rho}^2 \right)  +
       {g'}^2 \left( -4 + \rho +
          3 {\rho}^2 \right)  \right) }{4 \rho}
\wsep +
{\bf   \al_5 \left( -3 g^2 +
     \frac{{g'}^2}{\rho} -
     {g'}^2 \rho \right) }
\wsep +
  \mpng{\al_6 \left( {g'}^2
        \left( 10 - 7 \rho - 3 {\rho}^2 \right)  +
       g^2 \left( -4 + 10 \rho + 7 {\rho}^2
\right)  \right) }{4 \rho}
\wsep +
  \mpng{\al_7 \left( {g'}^2
        \left( 5 - 3 \rho - 2 {\rho}^2 \right)  +
       2 g^2 \left( -1 + 2 \rho + {\rho}^2
\right)  \right) }{2 \rho}
\wsep +
  \mpng{\al_9 g^2
     \left( 1 + 16 \rho - 8 {\rho}^2 -
       3 {\rho}^3 \right) }{16 \rho}
\wsep +
  \mpng{\al_a \left( 3 {g'}^2 +
       g^2 \left( -3 - 11 \rho + 2 {\rho}^2
\right)  \right) }{\rho}
\cma\\
\dd{v} &\approx&{v \over 2} \bbra{
\frac{-3 g^2}{4} - \frac{3 {g'}^2}{4} +
  \frac{g^2}{4 \rho} - g^2 \rho
}\cma\\
\dd{\rho}&\approx&
\frac{-9 g^2}{4} + \frac{3 g^2 \rho}{4} +
  \frac{3 {g'}^2 \rho}{4} +
  \frac{3 g^2 {\rho}^2}{2}
\,.
\label{svrgesb}
\eea
One feature about the RGEs given in Eqs. (\ref{svrgesa}---\ref{svrgesb})
is that quartic ACs ($\al_4$, $\al_5$, $\al_6$, $\al_7$,
and $\al_a$) will not contribute
directly to the $\beta$ functions of quadratic ACs ($\al_1$ and $\al_8$),
and the contribution is mediated by the triple ACs ($\al_2$,
$\al_3$, and $\al_9$), and vice versa (
such a statement only exist in the approximation we have
used). Similarly, quartic ACs do not contribute directly
to the $\beta$ function of gauge couplings $g$ and $g'$.

Another remarkable feature is that the contributions
of the quartic ACs dominate the $\beta$ functions of the
triple and quartic ACs and those of the
triple ACs dominate those of quadratic ACs,
if we estimate the contributions of ACs by
using the experimental
constraints given in Eq. (\ref{expcon}).

The third feature is that the ACs in the $\beta$ functions always
appear with gauge couplings ($g^2$ and $g'^2$) in the form
$\alpha_i g^2(g^{'2})$, and this fact
is related with the parameterization of the ACs in Eq. ( \ref{ewcle} ).
Actually, according to our modified power counting rule, the higher power of
these combinations should also belong to $O(p^0)$ and should appear
at the beta functions. Here just for the sake of simplicity and in order
to qualitatively understand the running behavior of the ACs, we have intendedly
omitted higher power terms. The complete RGEs will be provided in our coming
paper. However, the numerical analysis is based on the complete RGEs.

To get a negative $S$ parameter, large positive initial $\al_1$ at the
matching scale is one possible solution, as many new physics models propose.
Here, based on the RGEs of the $\al_1$, we regard it is possible
to generate a negative $S$ through the radiative
corrections. As indicated by the $\beta$ function
of $\al_1$, such a mechanism is possible either for
a large $\rho$ ( which make the leading
contributions smaller or even negative) or for large triple ACs,
especially the $\al_2$ and $\al_3$ due to their terms have a factor $3$
compared with $\al_8$ and $\al_9$ (only if they have different signs, and
$\al_2$ is negative ). But considering the
fact that $\rho$ is constrained by the $T$ parameter, and according
to the $\beta$ function of $\rho$, we conclude that a reasonable
way is large triple ACs. It is no hard to find solutions
to the negative $S$ parameter at $m_Z$ with
positive initial value at the matching scale
in the parameter space, and the region for such a solution
is quite large when considering the present experimental limits
on the ACs given in Eq. (\ref{expcon}). Such a new mechanism might be
a good news to some technicolor models, where
the theoretical predictions on $S$ contradict with the
present experimental measurement.

In the case when all ACs are of order ${1 \over (4 \pi)^2}$, we can neglect
those terms related with $\alpha_i$. In this limit, our result should reduce
to the result of the DM, which has the following form
\bea
\label{dmrgesa}
\dd{g}&=&- {13 \over 4} {g^3 \over 2}\,,\\
\dd{g'}&=&{25 \over 4} {g'^3 \over 2}\,\\
\dd{\al_1}& =& {1\over 12}\,,\\
\dd{\al_2} &=& - {1\over 24}\,,\\
\dd{\al_3} &=& - {1\over 24}\,,\\
\dd{\al_4} &=& - {1\over 12}\,,\\
\dd{\al_5} &=& - {1\over 24}\,,\\
\dd{v}    &=& - {3 \over 8} \, (2 g^2 + g'^2) \, v  \,,\\
\dd{\rho} &=& {3 \over 4} g'^2\,,
\label{dmrgesb}
\eea
while the rest of ACs vanish and do not develop,
according to the calculation of this method. These constants in the
$\beta$ functions of $\al_i$'s come from the contribution of Goldstone
bosons.

Below are several comments on the RGEs up to $O(p^4)$. 1)
The RGEs complete the EWCL as an
effective field theory method to effectively describe the
SM (without Higgs) below TeV, since the RGEs are one of the
basic ingredients of the effective field method.
2) Although the RGE method is essentially
equivalent to the calculations of infinite Feynman diagrams by keeping
to the leading logarithm results, it simplifies the
calculation of loop processes and makes the matching
procedure much easier. 3) Not like the DM, the RGE method includes the
contributions of all low energy DOFs, not only those
of Goldstone bosons, but also those of vector bosons and those of
mixing terms of these two kinds of bosons as well. As we have shown in the
$SU(2)$ case \cite{our2}, the last two kinds of contributions might make the
predictions of these method complete different. 4) The RGEs
also provide a new powerful tool for the comparative study on the possible
new physics near TeV region through the study of vector boson scatterings.
The effects of the heavy DOF
to the low energy physics are quite transparent in this method, only
the masses (which determines the matching scale) and the couplings with
the low energy DOF
play a part. Compared with the traditional procedure ( to formulate Feynman
rules, construct Feynman diagrams, calculate Feynman integral,
renormalize the full theory, and extract radiative corrections ),
the RGEs have integrated these necessary steps into
a powerful and ease-to-use chip.

About the simplified-version RGEs given in Eqs. (\ref{svrgesa}---\ref{svrgesb}),
we would like to point out that in the effective Lagrangian of the
Higgs model, the AC $\alpha_5$ can reach $O(1)$.
So the terms contain $\alpha_5$ can make the behavior of the
ACs quite different from the predictions made by
the DM\footnote{
Most of the past works have considered how to 
extract the non-decoupling contributions of heavy Higgs.
Therefore the tree-level contribution from the exchange of heavy Higgs should vanish
in the decoupling limit. Their results indeed have discarded the tree-level contribution
and are consistent with their assumptions. Since in the decoupling limit, the terms
proportional to $v^2/m_H^2$ indeed can be safely neglected. (The result
given by M. J. Herrero and E. R. Morales has correctly included the tree level
contribution.). But unfortunately, people use those results to consider the Higgs effects
even for the light Higgs case (say 200-600Gev), where  $v^2/m_H^2$ is not a
small number which can be safely neglected, and its radiative correction
is not small as expected from the NPC rule
}, as we will show in the next section.

\section{Numerical analysis: the application of
the RGE method to the Higgs effects}

To solve the differential equations of RGEs ( the numerical analysis in this section
is conducted by using the complete RGEs, while the simplified-version ones
given in Eqs. (\ref{svrgesa}---\ref{svrgesb}) can provide
a qualitative understanding to the behaviors of the ACs. Quantitatively,
the differences of results made by the complete and
the simplified are neglectable, at least for the
below case.), we need
some basic inputs, which are also called the boundary conditions.
The boundary conditions of the RGEs given in Eqs. (\ref{svrgesa}---\ref{svrgesb})
contain two classes:
1) the class of the low energy boundary conditions, where the parameters
$g$, $g'$, and $v_0$ are fixed from the experimental values.
We take the following experimental inputs
\bea
m_Z=91.18\,\, GeV,\,\,,\,\,m_W=80.33\,\, GeV,\,\,\sin\theta_W(m_Z) = 0.2312, \,\,\alpha_e(m_Z) = {1\over 128}\,,
\eea
with these inputs, according to the definition of $g$, $g'$, and $v_0$, we have
\bea
g(m_Z)=0.65,\,\,g'(m_Z)=0.36,\,\,v_0(m_Z)=246.708\,\,GeV\,.
\eea
2) the class of the matching-scale boundary conditions, where the
initial values of the ACs $\alpha_i$ are input by integrating
out the heavy DOF and matching the full theory with the
effective theory. For the heavy Higgs case, at the tree-level the matching
procedure yields
\bea
\alpha_5(m_0) = {v^2 \over 8 m_0^2}={1 \over 4 \lambda},\,\,\rho(m_0)=1\,,
\eea
while the rest of ACs vanish. For the RGE of DM, we will set all
ACs vanish expect that $\rho(m_0)=1$, in order to exaggeratingly 
demonstrate the differences of these two methods.

The relations between the ACs and the precision
test parameters, the quadratic vertices (S, T, and $U$) and the triple gauge vertices
are determined as \cite{appel, holdm}
\bea
S&=&-16 \pi \alpha_1\,,\\
T&=&{\rho -1 \over \alpha_{em}}\,,\\
U&=& - 16 \alpha_8\,,\\
g^Z_1-1&=& {1 \over c^2 -s^2} {\rho -1 \over 2} + {1 \over c^2 (c^2 -s^2)} e^2 \alpha_1 + {1 \over s^2 c^2} e^2 \alpha_3\,,\\
g^{\gamma}_1-1&=&0\,,\\
k_Z -1 &=& {1 \over c^2 -s^2} {\rho -1 \over 2} + {1 \over c^2 (c^2 -s^2)} e^2 \alpha_1
\wsep+ {1\over c^2} e^2 (\alpha_1 - \alpha_2) + {1 \over s^2 } e^2 (\alpha_3 -\alpha_8 + \alpha_9)\,,\\
k_{\gamma}-1&=& {1\over s^2} e^2 (-\alpha_1 + \alpha_2 + \alpha_3 - \alpha_8 + \alpha_9)\,,
\eea

Below we conduct a comparative study on the predictions of the
RGE method and the DM in the effective Lagrangian of the
Higgs model.

As we know, the Higgs scalar's effects include
both the decoupled mass square suppressed
part as shown in Eq. (\ref{init}),
and the nondecoupling logarithm part as shown
explicitly in the RGEs of the DM.
So we consider the following four cases to
trace the change of roles of these two competing
parts:
1) the light scalar case, with $m_0 = 150$ GeV,
where the decoupling mass-square-suppressed
part dominates;
2) the mediate heavy scalar case, with $m_0 = 300$ GeV,
where the decoupling mass-square-suppressed
part dominates;
3) the not too heavy scalar case, with $m_0 = 450$ GeV,
where both contributions are important;
4) the very massive scalar case, with $m_0 = 900$ GeV,
where the nondecoupling logarithm part dominates.

Fig. 1 is devoted to the first case, 
Fig. 2 to the second case, 
Fig. 3 to the third case, and
Fig. 4 to the third case. We also scan the region with the mass of
Higgs taken from $120$ GeV to $420$ GeV,
and the result is given in Fig. 5 In Fig. 6, we compare
the predictions of these two method to the ACs and 
to the precision test parameters, 1) the quadratic vertices parameters,
$S$ and $T$ (the $U$ parameter is quite small in both methods);
2) the triple gauge vertices, $g_1^Z-1$, $k_Z-1$, and $k_{\gamma}-1$.

Due to the large contributions of the leading terms and the
fact that quartic ACs do not contribute to their $\beta$ functions,
the gauge couplings $g$ and $g'$ are quite dull to the effects
of the ACs. So there is no viewable difference between these two
methods, and we omit the running of these two gauge couplings.

The differences of the AC $\al_1$ in these two methods
are small in these three cases we consider. The underlying reason
is due to the large leading contributions in its $beta$ function,
and due to the cancellation of the terms $\al_2$ and $\al_3$ (since
$\al_2$ and $\al_3$ have the same signs).

The differences of these two methods are dramatic in the ACs
$\al_2$, $\al_3$, and $\al_4$, when Higgs scalar is far from its
decoupling limit, similar to the
$SU(2)$ case \cite{our2}.
The differences of the AC $\al_5$ is the most dramatic, since the 
tree-level contribution is much larger than the 
loop corrections. As revealed in these figures, the heavier the Higgs,
the smaller the differences between these ACs.

For the ACs $\al_6$, $\al_7$, $\al_8$, $\al_9$, and $\al_a$,
due to the fact that the DM predicts a vanished value,
the differences are at the order of $10^{-4}$ in the first case (except $\al_8$
which is $10^{-6}$), at $10^{-5}$ in the second, third and fourth cases (except $\al_8$
which is $10^{-7}$). As the result of the accidental
symmetry---the custodial symmetry, the $\rho$ is always
near the unit in all these three cases.

When the Higgs scalar is far lighter than its decoupling limit,
the differences between these two method are much more dramatic,
as shown in Fig. 5. The tendency that the differences between these
two methods become smaller when the Higgs scalar is taken heavier has
been vividly outlined.

Such differences between these two methods
can not be meaningfully detected in LHC and LC for
the $\al_1$, $\al_6$, $\al_7$, $\al_8$, $\al_9$, and $\al_a$, but can be hopefully
detected for the  $\al_2$, $\al_3$, $\al_4$, and $\al_5$ when the Higgs
scalar is light. The detection power of these machines is estimated
as
\bea
\label{power}
(\al_1\,\,\al_8\,\,\beta) \sim O(0.001)-O(0.0001),\,\,\nnb\\
(\al_2\,\,\al_3\,\,\al_9)\sim O(0.01)-O(0.001),\,\,\nnb\\
(\al_4\,\,\al_5\,\,\al_6\,\,\al_7\,\,\al_a)\sim O(0.1)-O(0.01)\,.
\eea

From Fig. 6, we see that from the triple gauge vertices, 
if the Higgs scalar is relatively light (say 200 or 400 GeV or so),
it is definitely possible for the experiments to distinguish
the predictions of these two methods.

\section{Discussions and conclusions}

In this paper we have formulated the EWCL in its mass eigenstates,
and provided the relations between the complete basis
of the weak-interaction eigenstates and the one of the mass eigenstates.
We have modified the naive power rule in order to reliably extract the
large contributions of AOs and to control higher order contributions.
We also have studied the one-loop renormalization of the EWCL
and derived its RGEs to the same order. Theoretically,
the RGEs complete the EWCL as an effective theory to describe
the SM below a few TeV. We have studied the EWCl of the Higgs model and
have found that after taking into account the contributions
of Goldstone bosons, those of vector bosons, and those of mixing terms between
these two kinds of bosons as well,
the effects of Higgs can yield quite large triple ACs
when the Higgs scalar is light.

It is helpful to compare the result of the RGE method and the DM with
the underlying renormalizable theory, the SM, to see why the trilinear
couplings have such different behavior in these two methods 
in the lower energy region. The terms proportional to quartic couplings
in the beta function of trilinear couplings, is equivalent to the contribution of the
diagram (in unitary gauge) given in Fig. 7. While in the SM, this diagram
corresponds to the two diagrams (in unitary gauge) given in Fig. 8.
By contracting the Higgs line to a point, the Feynman diagrams in Fig. 8
reduce to the diagram in the Fig. 7.
In the SM, these two kinds of diagrams will contribute to the trilinear
couplings in both divergent and non-divergent terms.
The divergent part will contribute to both a small finite constant term
and a term with log dependence on the mass of Higgs, while
the non-divergent part is proportional to ${m_W^2 \over m_H^2}$.
The RGE method has correctly taken
into account both the log term and the non-divergent part, but has missed the
small finite constant term (by using the one-loop matching conditions and two loop RGEs,
this small finite term will be taken into account); 
while the DM has assumed that the $m_H^2$ is much
larger than $m_W^2$, so in this method the non-divergent part is
neglected in order to be consistent with its assumption, and only the log terms
and the small finite constant are taken into account. In the case when Higgs is not
too heavy, the non-divergent terms might be important, as the RGE method has revealed.
In the case when Higgs is heavy, {\it i.e.} approaching to the decoupling limit,
then only the log terms play the major part in the trilinear couplings, then
both these two method yield the almost same prediction, with a small finite
constant terms as difference. That is the reason why these two effective theory
methods give quite different predictions on the behavior of trilinear couplings,
when the Higgs is not too heavy. In the case when Higgs is relative light, 
the higher order operators (say, some $O(p^6)$ operators) might be also important,
and contributions from these operators will explain the difference between
the result of RGE and the exact computation in the underlying theory. According to
the direct computation, the relative ratio of the higher order operators contribution 
over the part computed by the RGE method can be approximately expressed 
as $\ln (1 + 2 m_W^2/m_H^2) - 1$, and considering the fact that such a ratio
is quite small, so we claim that the RGE method has efficiently summed over the most
important effects.

In the SM, the AC $\alpha_5$ is tightly related with Higgs mass.
While in the nonstandard Higgs model \cite{hht}, 
the ACs will have a lack dependence on the Higgs mass at the matching scale,
\bea
\alpha_5 ={k_s^2 \over 4 \lambda},\,\, \alpha_7 ={k_s k_s' \over 2 \lambda},\,\,
\alpha_a = {k_s' \over 4 \lambda},\,\,
\alpha_i (i\neq5,\,\,7\,\,a)=0 \,.
\eea
With the RGEs, we can careful explore the effects of quartic couplings to
low energy dynamics, in this model.

The RGEs will greatly simplify the procedure for us to study the effects of
the new physics beyond the SM, due to the fact that
the tree-level matching conditions are enough for the calculation
while the one-loop contributions have been efficiently summed
up by the RGEs. As the applications
of the RGEs, it is easier to comparatively study the possible
effects to the vector bosonic sector of the SM which might come
from various possible
new physics candidates, SUSY models, TC models, or ED models.

To establish the modified power counting rule and
to derive the RGEs, we have assumed that all ACs are of $O(1)$.
By assuming ACs of $O(1)$,
the EL might be limited in the realistic application, due to the fact
that the amplitude of the longitudinal components of vector bosons (Goldstones)
scattering processes at
higher energy regions might violet the unitarity condition once the
momentum of vector boson goes a little higher than the mass
of vector bosons. However, considering the
fact that the parameter space of the effective theories
should be composed by both the ultraviolet cutoff $\Lambda_{UV}$ and
the ACs at that scale, the condition for the
violation of unitarity just imposes a helpful correlation on
the matching scale and the magnitude of ACs. If the magnitude of ACs is smaller,
then the $\Lambda_{UV}$ can be larger, vice versa.
So our assumption and the RGE method have relatively more flexibility
to match with an unknown underlying theory from $m_W$ to $4\, \pi \,v$
than the specific assumption by assuming ACs are tiny and the cutoff is
at $4\, \pi \,v$. As the matter of fact, for the case when the ACs is large,
before the unitarity condition is
actually violated, new particles or new resonances might
have been found. Therefore, new effective theories should be formulated to
include new particles, and new RGEs should be derived.

In order to simplify the calculations,
we have not included the contributions of the Fermion, especially the RGEs of
the case with the top quarks much heavier than the bottom quarks and the
EW symmetry is further broken from the fermionic part, where the anomaly term and
terms violating $C$, $P$, and $CP$ might be important after taking into
account radiative corrections.

The real world might prefer a light Higgs, as predicted in the SUSY models.
So it seems necessary to take the Higgs scalar as one of the basic
blocks of the SM, and consider the corresponding EWCL with not only the
effective bosonic sector but also the effective Higgs sector \cite{ewhiggs}.
We will consider such a case in our future works \cite{our3}.

The one-loop RGEs still can not reach to the precision which can be 
approached by the future's colliders, we will provide two-loop RGEs
of EWCL in our future works \cite{our3}.

\acknowledgements
One of the author, Q. S. Yan, would like to thank
Professor C. D. L\"u in the theory division of IHEP of CAS
for helpful discussion, and Dr. H. J. He for his
helpful informations on the related references.
And special thanks to Prof. Y. P. Kuang and Prof. Q. Wang
from physics department of Tsinghua university,
for their kind help to ascertain some important points
and to improve the representation of this paper.
To manipulate the algebraic
calculation and extract the $\beta$ functions of RGEs,
we have used the FeynCalc \cite{fcalc}.
The work of Q. S. Yan is supported by
the Chinese Postdoctoral Science Foundation
and the CAS K. C. Wong Postdoctoral Research Award Foundation.
The work of D. S. Du is supported by the
National Natural Science Foundation of China.

\section{Appendix: the related matrices}
In this appendix, we provide the related matrices of the quadratic terms of
the ${\cal L}_{1-loop}$.
The basic definitions include,
1) the physics fields $W^+$, $W^-$, $Z$, and $A$ from the interaction eigenstates:
\bea
W^+&=&{1\over \sqrt{2}} \left( W^1 - i W^2 \right)\,,\,\,W^-={1\over \sqrt{2}} \left( W^1 + i W^2 \right)\,,\nnb\\
Z  &=& {1\over 2} \left( \sin\theta_W B - \cos\theta_W W^3 \right)\,,\,\,A  = {1\over 2} \left( \cos\theta_W B + \sin\theta_W W^3 \right)\,,
\eea
2) the Lie algebra generator matrices:
\bea
T^+&=&{1\over \sqrt{2}} \left( T^1 + i T^2 \right)\,,\,\,T^-={1\over \sqrt{2}} \left( T^1 - i T^2 \right)\,,\nnb\\
\left[T^+, T^3\right]&=& - T^+\,,\,\,\left[T^-, T^3\right]= T^-\,,\nnb\\
\left[T^+, T^-\right]&=& T^3\,,\,\,tr[T^+ T^-] = tr[T^3 T^3]={1\over 2 }\,.
\eea

There are some definitions to simplify the expressions:
\bea
H_+^{\mu\nu} &=& d^{\mu} W_+^{\nu} + d^{\nu} W_+^{\mu}\,,\,\,H_-^{\mu\nu} = d^{\mu} W_-^{\nu} + d^{\nu} W_-^{\mu}\,,\nnb\\
H_Z^{\mu\nu} &=& \partial^{\mu} Z^{\nu} + \partial^{\nu} Z^{\mu}\,,\,\,SF_+^{\mu\nu}=W_+^{\mu} Z^{\nu} + W_+^{\nu} Z^{\mu} \,,\nnb\\
SF_-^{\mu\nu}&=&Z^{\mu} W_-^{\nu} + Z^{\nu} W_-^{\mu}\,,\,\,SF_Z^{\mu\nu}=W_+^{\mu} W_-^{\nu} + W_+^{\nu} W_-^{\mu}\,,\nnb\\
SW_+^{\mu\nu}&=&W_+^{\mu} W_+^{\nu}\,,\,\,SW_-^{\mu\nu}=W_-^{\mu} W_-^{\nu}\,,\nnb\\
SZ_Z^{\mu\nu}&=&Z^{\mu} Z^{\nu}\,,\,\,SFP= 2 W_+ \cdot Z\,,\nnb\\
SFM&=& 2 W_- \cdot Z\,,\,\,SFZ= 2 W_+ \cdot W_-\,,\nnb\\
OW^+&=& W_+ \cdot W_+\,,\,\,OW^-= W_- \cdot W_- \,,\,\,
OZZ=Z\cdot Z\,.
\eea

Below are those related matrices of the quadratic terms.
The field strength of $\Gamma_V$ is given as
\begin{displaymath}
\Gamma^{\mn}_{VV} =\left (\begin{array}{cccc}
\Gamma^{\mn}_{AA}  &\Gamma^{\mn}_{AZ}  &\Gamma^{\mn}_{AW^+}  &\Gamma^{\mn}_{AW^-}\\
\Gamma^{\mn}_{ZA}  &\Gamma^{\mn}_{ZZ}  &\Gamma^{\mn}_{ZW^+}  &\Gamma^{\mn}_{ZW^-}\\
\Gamma^{\mn}_{W^-A}&\Gamma^{\mn}_{W^-Z}&\Gamma^{\mn}_{W^-W^+}&\Gamma^{\mn}_{W^-W^-}\\
\Gamma^{\mn}_{W^+A}&\Gamma^{\mn}_{W^+Z}&\Gamma^{\mn}_{W^+W^+}&\Gamma^{\mn}_{W^+W^-}\\
\end{array}\right)
\end{displaymath}
and its components read
\bea
\Gamma^{\mn}_{AA}&=&\Gamma^{\mn}_{AZ}=\Gamma^{\mn}_{ZA}=\Gamma^{\mn}_{ZZ}=0\,,\nnb\\
\Gamma^{\mn}_{AW^+}&=&i {(2 e + C_5) \over 4 \sqrt{C_1}} W_-^{\mu  \nu}
+{(2 e + C_5) C_7 \over 8 \sqrt{C_1}} F_-^{\mu  \nu}
\,,\nnb\\
\Gamma^{\mn}_{ZW^+}&=&-
i {2 e C_2 + C_2 C_5 + C_1 (-C_6 + C_7) \over 4 \sqrt{g_Z} C_1} W_-^{\mu \nu}
\wsep -C_7 {2 e C_2 + C_2 C_5 + C_1 (-C_6 + C_7)
\over 8 \sqrt{g_Z} C_1} F_-^{\mu \nu}\,,\nnb\\
\Gamma^{\mn}_{W^-W^+}&=&-i e A^{\mu  \nu} +
 {i C_7 \over 2} Z^{\mu  \nu} \wsep-
 {g_Z C_1 (2 e + C_5)^2 + \left (2 e C_2 + C_2 C_5 + C_1 (-C_6 + C_7) \right )^2
   \over 16 g_Z C_1^2} F_Z^{\mu  \nu}
\,,\nnb\\
\Gamma^{\mn}_{AW^-}&=&\Gamma^{\dagger \mn}_{AW^+}\,,\,\,
\Gamma^{\mn}_{ZW^-}=\Gamma^{\dagger \mn}_{ZW^+}\,,
\Gamma^{\mn}_{W^-A}=-\Gamma^{\mn}_{AW^-}\,,\nnb\\
\Gamma^{\mn}_{W^+A}&=&-\Gamma^{\mn}_{AW^+}\,,
\Gamma^{\mn}_{W^+W^-}=\Gamma^{\dagger \mn}_{W^-W^+}\,,\,\,
\Gamma^{\mn}_{W^-W^-}=\Gamma^{\mn}_{W^+W^+}=0\,.
\eea

The field strength of $\Gamma_\xi$ is given as
\begin{displaymath}
\Gamma^{\mn}_{\xi\xi} =\left (\begin{array}{ccc}
\Gamma^{\mn}_{\xi_Z \xi_Z}&\Gamma^{\mn}_{\xi_Z\xi^{+}}&\Gamma^{\mn}_{\xi_Z\xi^-}\\
\Gamma^{\mn}_{\xi^- \xi_Z}&\Gamma^{\mn}_{\xi^-\xi^{+}}&\Gamma^{\mn}_{\xi^-\xi^-}\\
\Gamma^{\mn}_{\xi^+ \xi_Z}&\Gamma^{\mn}_{\xi^+\xi^{+}}&\Gamma^{\mn}_{\xi^+\xi^-}\\
\end{array}\right)
\end{displaymath}
and its components read
\bea
\Gamma^{\mn}_{\xi_Z \xi_Z}&=&\Gamma^{\mn}_{\xi^+\xi^{+}}=\Gamma^{\mn}_{\xi^-\xi^{-}}=0\,,\nnb\\
\Gamma^{\mn}_{\xi_Z\xi^{+}}&=& i { g \over 2 \sqrt{\rho}} W_-^{\mu  \nu}+{-2 e^2 G + g^2 G \rho \over
  4 g \sqrt{\rho}} F_-^{\mu  \nu}
\,,\nnb\\
\Gamma^{\mn}_{\xi^-\xi^{+}}&=&-i e A^{\mu  \nu} +
 i \sbra{-{ e^2 G \over g^2} + {1 \over 2} G \rho} Z^{\mu  \nu}
-
{ g^2 \over 4 \rho} F_Z^{\mu  \nu} \,,\nnb\\
\Gamma^{\mn}_{\xi^+\xi^{-}}&=&\Gamma^{\dagger \mn}_{\xi^-\xi^{+}}\,,\,\,
\Gamma^{\mn}_{\xi_Z\xi^{-}} = \Gamma^{\dagger \mn}_{\xi_Z\xi^{+}}\,,\nnb\\
\Gamma^{\mn}_{\xi^{-}\xi_Z}&=&-\Gamma^{\mn}_{\xi_Z\xi^{-}}\,,\,\,
\Gamma^{\mn}_{\xi^{+}\xi_Z}=-\Gamma^{\mn}_{\xi_Z\xi^{+}}\,.
\eea

The $\sigma^{\mn}_{VV}=S\sigma^{\mn}_{VV} + A\sigma^{\mn}_{VV}$,
and the symmetric matrix $S\sigma^{\mn}_{VV}$ is determined as
\begin{displaymath}
S\sigma^{\mn}_{VV} =\left (\begin{array}{cccc}
S\sigma^{\mn}_{AA}&S\sigma^{\mn}_{AZ}&S\sigma^{\mn}_{AW^+}&S\sigma^{\mn}_{AW^-}\\
S\sigma^{\mn}_{ZA}&S\sigma^{\mn}_{ZZ}&S\sigma^{\mn}_{ZW^+}&S\sigma^{\mn}_{ZW^-}\\
S\sigma^{\mn}_{W^-A}&S\sigma^{\mn}_{W^-Z}&S\sigma^{\mn}_{W^-W^+}&S\sigma^{\mn}_{W^-W^-}\\
S\sigma^{\mn}_{W^+A}&S\sigma^{\mn}_{W^+Z}&S\sigma^{\mn}_{W^+W^+}&S\sigma^{\mn}_{W^+W^-}\\
\end{array}\right)
\end{displaymath}
and its components are listed as
\bea
S\sigma^{\mn}_{AA} &=& -\wpng{-4 e^2 + C_5^2}{4 C_1} SF_Z^{\mu  \nu} \wsep+
 \wpng{-12 e^2 + 4 e C_5 + C_5^2}{
  16 C_1} SFZ g^{\mu  \nu} \cma\nnb\\
S\sigma^{\mn}_{AZ} &=& \wpng{-4 e^2 C_2 + C_5 (C_2 C_5 - C_1 C_6) - 2 e C_1 C_7
   }{4 \sqrt{g_Z} C_1^{3 \over 2}} SF_Z^{\mu  \nu} \wsep +
 \mpng{12 e^2 C_2 - C_5 \left (C_2 C_5 + C_1 (-C_6 + C_7) \right ) \wsep +
    e \left (-4 C_2 C_5 + 2 C_1 (C_6 + 3 C_7) \right )}{16 \sqrt{g_Z} C_1^{3 \over 2}}
SFZ g^{\mu\nu} \cma\nnb\\
S\sigma^{\mn}_{AW^+}&=& \wpng{(2 e - C_5) C_7 }{8 \sqrt{C_1}} SF_-^{\mu  \nu} \wsep+
 \wpng{(-2 e + C_5) C_7 }{16 \sqrt{C_1}} SFM g^{\mu  \nu} \wsep+
i \wpng{2 e - C_5}{4 \sqrt{C_1}}  H_-^{\mu  \nu}
\cma\nnb\\
S\sigma^{\mn}_{ZZ}&=& {4 C_a \over g_Z} OZZ g^{\mu \nu} \wsep +
 \wpng{4 e^2 C_2^2 - C_2^2 C_5^2 + 2 C_1 C_2 C_5 C_6  + 4 e C_1 C_2 C_7 -
    C_1^2 (C_6^2 - 4 C_9)}{4 g_Z C_1^2} SF_Z^{\mu  \nu}
\wsep + \mpng{-12 e^2 C_2^2 + C_2^2 C_5^2 +
    2 C_1 C_2 C_5 (-C_6 + C_7) \wsep + 4 e C_2 \left ( C_2 C_5 - C_1 (C_6 + 3 C_7)\right ) +
    C_1^2 (C_6^2 - 2 C_6 C_7 + C_7^2 + 16 C_8)}{16 g_Z C_1^2}
SFZ g^{\mu  \nu} \wsep + 8 {C_a \over g_Z } SZ^{\mu  \nu}
\cma\nnb\\
S\sigma^{\mn}_{ZW^+}&=& \wpng{-2 e C_2 C_7 + C_2 C_5 C_7 + C_1 (-C_6 C_7 + 8 C_8 + 4 C_9 )
   }{8 \sqrt{g_Z} C_1} SF_-^{\mu  \nu} \wsep +
 \wpng{2 e C_2 C_7 - C_2 C_5 C_7 + C_1 (C_6 C_7 - C_7^2 + 8 C_9)
   }{16 \sqrt{g_Z} C_1 } SFM g^{\mu  \nu} \wsep +
 i \wpng{2 g_Z f_{zw} C_1 + C_2 C_5 - C_1 C_6}{4 \sqrt{g_Z} C_1}
 H_-^{\mu\nu}
\cma\\
S\sigma^{\mn}_{W^{-}W^{+}}&=&({C_7^2 \over 4} + C_8) OZZ g^{\mu \nu} +
 (-{g_Z f_{zw}^2\over 2} - {e^2 \over 2 C_1} + C_b + 2 C_c) SF_Z^{\mu
   \nu} \wsep + \wpng{\left (2 e C_2 + C_2 C_5 + C_1 (-C_6 + C_7) \right )^2 \ssep +
    g_Z C_1 (4 e^2 + 4 e C_5 + C_5^2 + 32 C_1 C_b)
   }{32 g_Z C_1^2} SFZ g^{\mu  \nu} \wsep +
 (-{C_7^2\over 4} + C_9) SZ^{\mu \nu}
\cma\\
S\sigma^{\mn}_{W^{-}W^{-}}&=&-\mpng{\left (2 e C_2 + C_2 C_5 + C_1 (-C_6 + C_7)\right )^2 \wsep +
     g_Z C_1 (4 e^2 + 4 e C_5 + C_5^2 - 32 C_1 C_c)
    }{16 g_Z C_1^2} OW^+ g^{\mu \nu} \wsep +
 (g_Z f_{zw}^2 + {e^2 \over C_1} + 2 C_b) SW_+^{\mu  \nu}
\cma\\
S\sigma^{\mn}_{AW^-}&=& S\sigma^{\dagger \mn}_{AW^+}\cma\,\,
S\sigma^{\mn}_{ZA}= S\sigma^{\mn}_{AZ}\cma\,\,
S\sigma^{\mn}_{ZW^{-}}=S\sigma^{\dagger \mn}_{ZW^+}\cma\\
S\sigma^{\mn}_{W^{-}A}&=&S\sigma^{\mn}_{AW^-}\cma\,\,
S\sigma^{\mn}_{W^{-}Z}=S\sigma^{\mn}_{ZW^{-}}\cma\,\,
S\sigma^{\mn}_{W^{+}A}=S\sigma^{\mn}_{AW^{+}}\cma\\
S\sigma^{\mn}_{W^{+}Z}&=&S\sigma^{\mn}_{ZW^{+}}\cma\,\,
S\sigma^{\mn}_{W^{+}W^{-}}=S\sigma^{\mn}_{W^{-}W^{+}}\,\,
S\sigma^{\mn}_{W^{+}W^{+}}=S\sigma^{\dagger \mn}_{W^{-}W^{-}}\,.
\eea
Here $\dagger$ means to change the field $W^{\pm} \rightarrow W^{\mp}$
and $i \rightarrow -i$.

The antisymmetric matrix $A\sigma^{\mn}_{VV}$
is given  as
\begin{displaymath}
A\sigma^{\mn}_{VV} =\left (\begin{array}{cccc}
A\sigma^{\mn}_{AA}  &A\sigma^{\mn}_{AZ}  &A\sigma^{\mn}_{AW^+}   &A\sigma^{\mn}_{AW^-}\\
A\sigma^{\mn}_{ZA}  &A\sigma^{\mn}_{ZZ}  &A\sigma^{\mn}_{ZW^+}   &A\sigma^{\mn}_{ZW^-}\\
A\sigma^{\mn}_{W^-A}&A\sigma^{\mn}_{W^-Z}&A\sigma^{\mn}_{W^-W^+} &A\sigma^{\mn}_{W^-W^-}\\
A\sigma^{\mn}_{W^+A}&A\sigma^{\mn}_{W^+Z}&A\sigma^{\mn}_{W^+W^+} &A\sigma^{\mn}_{W^+W^-}\\
\end{array}\right)
\end{displaymath}
and its components are listed as
\bea
A\sigma^{\mn}_{AA} &=&0\cma\,\,
A\sigma^{\mn}_{AZ} =0\cma\nnb\\
A\sigma^{\mn}_{AW^+}&=&\wpng{(6 e + C_5) C_7}{8 \sqrt{C_1}} F_-^{\mu  \nu}+
 i \wpng{6 e + C_5}{4 \sqrt{C_1}} W_-^{\mu  \nu}
 \cma\nnb\\
A\sigma^{\mn}_{ZZ}&=&0 \cma\nnb\\
A\sigma^{\mn}_{ZW^+}&=& \wpng{-6 e C_2 C_7 - C_2 C_5 C_7 + C_1 (C_6 C_7 + 8 C_8 - 4 C_9)
   }{8 \sqrt{g_Z} C_1} F_-^{\mu  \nu} \wsep +
 i \wpng{2 g_Z f_{zw} C_1 - 4 e C_2 - C_2 C_5 + C_1 C_6 - 2 C_1 C_7
   }{4 \sqrt{g_Z} C_1} W_-^{\mu  \nu}
\nnb\cma\\
A\sigma^{\mn}_{W^{-}W^{+}}&=&-i {2 e + C_5 \over 2} A^{\mu  \nu} +
 \left (-{g_Z f_{zw}^2\over 2} \ssep- {e^2 \over 2 C_1} + C_b - 2 C_c \right ) F_Z^{\mu
   \nu} - i {C_6 - C_7 \over 2}  Z^{\mu\nu}
\nnb\cma\\
A\sigma^{\mn}_{W^{-}W^{-}}&=&0\cma\,\,
A\sigma^{\mn}_{AW^-}= A\sigma^{\dagger \mn}_{AW^+}\cma\,\,
A\sigma^{\mn}_{ZA}=   - A\sigma^{\mn}_{AZ}\cma\nnb\\
A\sigma^{\mn}_{ZW^{-}}&=&  A\sigma^{\dagger \mn}_{ZW^+}\cma\,\,
A\sigma^{\mn}_{W^{-}A}=- A\sigma^{\mn}_{AW^-}\cma\,\,
A\sigma^{\mn}_{W^{-}Z}=-A\sigma^{\mn}_{ZW^{-}}\cma\nnb\\
A\sigma^{\mn}_{W^{+}A}&=&-A\sigma^{\mn}_{AW^{+}}\cma\,\,
A\sigma^{\mn}_{W^{+}Z}=-A\sigma^{\mn}_{ZW^{+}}\cma\,\,
A\sigma^{\mn}_{W^{+}W^{-}}=-A\sigma^{\mn}_{W^{-}W^{+}}\,,\nnb\\
A\sigma^{\mn}_{W^{+}W^{+}}= 0\,.
\eea

The symmetric $\sigma_{2,\xi\xi}$ is given  as
\begin{displaymath}
\sigma_{2,\xi\xi} =\left (\begin{array}{ccc}
\sigma_{2,\xi_Z \xi_Z}&\sigma_{2,\xi_Z\xi^{+}}&\sigma_{2,\xi_Z\xi^-}\\
\sigma_{2,\xi^- \xi_Z}&\sigma_{2,\xi^-\xi^{+}}&\sigma_{2,\xi^-\xi^-}\\
\sigma_{2,\xi^+ \xi_Z}&\sigma_{2,\xi^+\xi^{+}}&\sigma_{2,\xi^+\xi^-}\\
\end{array}\right)
\end{displaymath}
and its matrix components are listed as
\bea
\sigma_{2,\xi_Z \xi_Z}&=&{g^2 \over 4 \rho} SFZ\cma\,\,
\sigma_{2,\xi_Z \xi^+}={g G \sqrt{\rho} \over 8} SFM\cma\\
\sigma_{2,\xi^- \xi^+}&=&{G^2 \rho^2 \over 4} OZZ + {g^2 \over 8 \rho} SFZ\cma\,\,
\sigma_{2,\xi^- \xi^-}=- {g^2 \over 4 \rho} OW^+ \cma\\
\sigma_{2,\xi_Z \xi^-}&=&\sigma_{2,\xi_Z,\xi^+}^{\dagger}\cma\,\,
\sigma_{2,\xi^- \xi_Z}=\sigma_{2,\xi_Z \xi^-}\cma\,\,
\sigma_{2,\xi^+ \xi_Z}=\sigma_{2,\xi_Z \xi^+}\cma\nnb\\
\sigma_{2,\xi^+ \xi^+}&=&\sigma_{2,\xi^+ \xi^+}^{\dagger}\cma\,\,
\sigma_{2,\xi^+ \xi^-}=\sigma_{2,\xi^- \xi^+}\cma
\eea

One feature about the $\sigma^{\mn}_{VV}$ and $\sigma_{2,\xi\xi}$
is remarkable. There is no terms like $\partial \cdot Z$, $d \cdot W^{\pm}$
in the components, the basic reason is due to the gauge fixing terms
we have chosen in Eq. (\ref{gftew}).

Considering the fact that only the diagonal components
of the matrix $\sigma_4$ contribute meaningfully to the
renormalization up to $O(p^4)$, we only list
those we concern
\bea
\rho \,v^2\, \sigma_{4,\xi_Z \xi_Z}&=& H_1 \sbra{{4 g^2 \over G^2}  C_4 - {2 \over G} C_7 } \wsep +
 {\cal L}_3 \sbra{{-5 g^2 - 4 G^2 \over G^4}  C_4 + 2 {2 g^2 + G^2}{g^2 G^3}  C_7 +
   {2 \over g^2 G^2} C_8 - {1 \over g^2 G^2} C_9} \wsep +
 {\cal L}_5 \sbra{-4 {g^2 - G^2 \over g^2 G^2} C_1 - 8 {e \over g^2 G} C_2+ {4 \over G^2} C_3 -
   4 {e \over g^4} C_5+ {4 \over g^2 G} C_6- {8 \over g^4} C_b } \wsep +
 {\cal L}_6 \sbra{-6 {g^4 \over G^6} C_4+ 6 {g^2 \over G^5} C_7+ {4 \over G^4} C_8 - {2\over G^4} C_9 } \wsep +
 {\cal L}_7 \sbra{4 {g^2 - G^2 \over g^2 G^2} C_1 + 8 {e \over g^2 G} C_2- 4 {C_3 \over G^2} C_3 +
   2 {3 g^4 + 2 g^2 G^2 \over G^6} C_4+ 4 {e \over g^4}  C_5 \wsep- {4 \over g^2 G} C_6  -
   2 {4 g^2 + G^2 \over G^5} C_7 - 4 {g^2 + G^2 \over g^2 G^4} C_8-
   2 {-3 g^2 + 2 G^2 \over g^2 G^4} C_9 + {8 \over g^4} C_b} \wsep +
 {\cal L}_8 \sbra{-4 {1 \over G^2} C_4 + 2 {1 \over g^2 G} C_7} \wsep +
 {\cal L}_9 \sbra{{5 g^2 + 4 G^2\over G^4} C_4- 2 {2 g^2 + G^2\over g^2 G^3} C_7-
  {2 \over g^2 G^2} C_8+ {1 \over g^2 G^2} C_9} \wsep +
 {\cal L}_a \sbra{-2 {g^2 - G^2 \over g^2 G^2} C_1- 4 {e \over g^2 G} C_2+ {2 \over G^2} C_3-
   4 {g^2\over G^4} C_4- 2 {e \over g^4} C_5+ {2 \over g^2 G} C_6 \wsep +
   2 {g^2 + G^2 \over G^5} C_7+ {4 \over g^2 G^2} C_8+
   4 {-g^2 + G^2 \over g^2 G^4} C_9 - {4 \over g^4} C_b}
\cma\\
v^2 \sigma_{4,\xi^- \xi^+}&=& H_1 \sbra{4 {g^2 - G^2\over G^2} C_1+ 8 {e \over G} C_2-
   4 {g^2 \over G^2} C_3+ 2 {g^2 + 2 G^2 \rho \over G^2 \rho} C_4- {1 \over G \rho} C_7} \wsep +
 H_2 \mbra{-4 {(-g^2 + G^2) (-2 + \rho)\over G^2} C_1+
   4 {e (2 g^2 - G^2) (-2 + \rho) \over g^2 G} C_2 \wsep +
   4 {(-g^2 + G^2) (-2 + \rho) \over G^2} C_3+ 2 {e (-2 + \rho)\over g^2} C_5-
   2 {(g^2 - G^2) (-2 + \rho)\over g^2 G} C_6} \wsep +
 {\cal L}_1 \mbra{2 {2 G^2 + 2 g^2 \rho - G^2 \rho \over g^2 G^2} C_1 \wsep -
   2 \wpng{2 g^2 G^2 + 2 {e^2 G^4\over g^2} + g^4 \rho + e ^2 G^2 \rho -
      3 {e^2 G^4 \rho\over g^2} }{ e  G^5} C_2 \wsep - 4 {\rho \over G^2} C_3+
   \wpng{-2 {e^2 G^2 \over g^2} - g^2 \rho + {e^2 G^2 \rho \over g^2} }{e g^2 G^2} C_5-
   2 {(-1 + \rho) \over g^2 G} C_6} \wsep +
 {\cal L}_2 \mbra{{6 g^2 - 6 G^2 - 5 g^2 \rho + 3 G^2 \rho \over g^2 G^2} C_1 +
   2 \wpng{6 {e^2 G^2\over g^2} + g^2 \rho - 4 {e^2 G^2 \rho \over g^2}}{e G^3} C_2 \wsep +
   {-6 + 5 \rho \over G^2} C_3 + \mpng{-2 g^4 - 2 e ^2 G^2 + 2 g^2 G^2 +
      6 {e^2 G^4 \over g^2} + g^4 \rho \wsep + e ^2 G^2 \rho - 3 {e^2 G^4 \rho \over g^2}}{
    e  g^2 G^4} C_5 + 2 {-3 + 2 \rho  \over g^2 G} C_6 - 2 {-2 + \rho \over g^4} C_b +
   4 {-2 + \rho\over g^4}  C_c} \wsep+
{\cal L}_3 \mbra{4 {(g^2 - G^2)^2 \over g^2 G^4} C_1+ 8 e  {g^2 - G^2\over g^2 G^3} C_2 +
   4 {-g^2 + G^2\over G^4} C_3- 2 {g^2 - 2 g^2 \rho + 3 G^2 \rho \over
    G^4 \rho} C_4 \wsep + 2 e  {g^2 - G^2\over g^4 G^2} C_5+
   2 {-g^2 + G^2 \over g^2 G^3} C_6 + {g^2 - 2 g^2 \rho + 3 G^2 \rho \over
    g^2 G^3 \rho} C_7 } \wsep
+{\cal L}_4 \mbra{6 {(g^2 - G^2) \rho \over g^2 G^2} C_1+
   12 e  {\rho \over g^2 G} C_2- 6 {\rho \over G^2} C_3+ 6 {e \rho }{\over g^4} C_5 \wsep -
   6 {\rho \over g^2 G} C_6+ 4 {\rho \over g^4} C_b- 8 {\rho \over g^4} C_c}
 +{\cal L}_5 \mbra{-2 {(g^2 - G^2) (1 + 3 \rho^2)\over g^2 G^2 \rho} C_1\wsep -
  4 \wpng{ {e G \over g} + 3 {e G \rho^2 \over g} }{ g G^2 \rho} C_2 +
   2 {1 + 3 \rho^2 \over G^2 \rho} C_3- 2 \wpng{{e G \over g} + 3 {e G \rho^2 \over g }}{
    g^3 G \rho} C_5 \wsep + 2 {1 + 3 \rho^2 \over g^2 G \rho} C_6-
   4 {1 + \rho^2 \over g^4 \rho} C_b + 8 {\rho \over g^4} C_c} \wsep +
 {\cal L}_6 \mbra{-2 \wpng{(g^2 - G^2) (2 g^4 - 4 g^2 G^2 + 2 G^4 + 3 G^4 \rho) }{g^2 G^6} C_1 \wsep -
   4 e  \wpng{2 g^4 - 4 g^2 G^2 + 2 G^4 + 3 G^4 \rho }{g^2 G^5} C_2 \wsep +
   2 \wpng{2 g^4 - 4 g^2 G^2 + 2 G^4 + 3 G^4 \rho}{ G^6} C_3-
   2 \wpng{g^4 - 4 g^4 \rho + 4 g^2 G^2 \rho}{G^6 \rho} C_4 \wsep -
   2 e  \wpng{2 g^4 - 4 g^2 G^2 + 2 G^4 + 3 G^4 \rho}{g^4 G^4} C_5\wsep +
   2 \wpng {2 g^4 - 4 g^2 G^2 + 2 G^4 + 3 G^4 \rho}{g^2 G^5} C_6\wsep+
   \wpng{g^2 - 8 g^2 \rho + 8 G^2 \rho }{G^5 \rho} C_7 -
   8 \wpng{-g^2 + G^2}{g^2 G^4} C_9 \wsep - 4 \wpng{-2 g^4 + 4 g^2 G^2 - 2 G^4 + G^4 \rho}{
     g^4 G^4} C_b \wsep + 8 \wpng{2 g^4 - 4 g^2 G^2 + 2 G^4 + G^4 \rho}{g^4 G^4} C_c} \wsep +
 {\cal L}_7 \mbra{2 (g^2 - G^2) \wpng{G^4 + 10 g^4 \rho - 16 g^2 G^2 \rho + 6 G^4 \rho +
      4 g^2 G^2 \rho^2 + G^4 \rho^2}{g^2 G^6 \rho} C_1 \wsep +
   4 e  \wpng{G^4 + 10 g^4 \rho - 16 g^2 G^2 \rho + 6 G^4 \rho + 4 g^2 G^2 \rho^2 +
      G^4 \rho^2}{g^2 G^5 \rho} C_2 \wsep -
   2 \wpng{G^4 + 10 g^4 \rho - 16 g^2 G^2 \rho + 6 G^4 \rho + 4 g^2 G^2 \rho^2 +
      G^4 \rho^2}{G^6 \rho} C_3 \wsep - 2 g^2 \wpng{-g^2 - G^2 + 4 g^2 \rho - 4 G^2 \rho}{
     G^6 \rho} C_4 \wsep + 2 e  \wpng{G^4 + 8 g^4 \rho - 14 g^2 G^2 \rho + 6 G^4 \rho +
      3 g^2 G^2 \rho^2 + G^4 \rho^2}{g^4 G^4 \rho} C_5\wsep -
   2 \wpng{G^4 + 8 g^4 \rho - 14 g^2 G^2 \rho + 6 G^4 \rho + 3 g^2 G^2 \rho^2 +
      G^4 \rho^2}{g^2 G^5 \rho}  C_6\wsep +
   \wpng{-2 g^2 - G^2 + 12 g^2 \rho - 12 G^2 \rho + 2 G^2 \rho^2}{G^5 \rho} C_7 \wsep+
   2 \wpng{-G^2 + 8 g^2 \rho - 8 G^2 \rho + 2 G^2 \rho^2}{g^2 G^4 \rho}  C_8 \wsep +
   2 \wpng{g^2 - G^2}{g^2 G^4 \rho} C_9 +
   4 \wpng{G^4 + 6 g^4 \rho - 12 g^2 G^2 \rho + 6 G^4 \rho \ssep+ 2 g^2 G^2 \rho^2 -
      G^4 \rho^2}{g^4 G^4 \rho} C_b - 8 {\rho \over g^4} C_c} \wsep +
 {\cal L}_8 \mbra{-4 {(g^2 - G^2) (3 + \rho)\over g^2 G^2} C_1 -
   8 e  {3 + \rho\over g^2 G} C_2 + 4 {3 + \rho\over G^2} C_3 -
   2 {g^2 + 6 G^2 \rho\over g^2 G^2 \rho} C_4 \wsep - 2 {e \rho \over g^4} C_5 +
   2 {\rho\over g^2 G} C_6+ {1 \over g^2 G \rho} C_7} \wsep +
 {\cal L}_9 \mbra{-(g^2 - G^2) \wpng{4 g^2 - 8 G^2 - 5 G^2 \rho}{g^2 G^4} C_1 \wsep -
   2 e  \wpng{4 g^2 - 8 G^2 - 5 G^2 \rho}{g^2 G^3} C_2 \wsep +
   \wpng{4 g^2 - 8 G^2 - 5 G^2 \rho}{ G^4} C_3+
   2 \wpng{g^2 - 2 g^2 \rho + 3 G^2 \rho}{G^4 \rho} C_4 \wsep -
   2 e  \wpng{g^2 - 2 G^2 - 2 G^2 \rho}{g^4 G^2} C_5+
   2 \wpng{g^2 - 2 G^2 - 2 G^2 \rho}{g^2 G^3}  C_6 \wsep +
   \wpng{-g^2 + 2 g^2 \rho - 3 G^2 \rho}{g^2 G^3 \rho)} C_7
  + 2 {\rho \over g^4} C_b-
   4 {\rho \over g^4} C_c} \wsep +
 {\cal L}_a \mbra{-  \fwpng{(g^2 - G^2)}{G^4 + 16 g^4 \rho - 24 g^2 G^2 \rho + 8 G^4 \rho +
       8 g^2 G^2 \rho^2 - 4 G^4 \rho^2}{g^2 G^6 \rho} C_1 \wsep -
   2 e  \wpng{G^4 + 16 g^4 \rho - 24 g^2 G^2 \rho + 8 G^4 \rho + 8 g^2 G^2 \rho^2 -
      4 G^4 \rho^2}{g^2 G^5 \rho} C_2 \wsep +
   \wpng{G^4 + 16 g^4 \rho - 24 g^2 G^2 \rho + 8 G^4 \rho + 8 g^2 G^2 \rho^2 -
      4 G^4 \rho^2}{G^6 \rho}  C_3- 2 {g^2 \over G^4 \rho} C_4 \wsep -
   e  \wpng{G^4 + 12 g^4 \rho - 20 g^2 G^2 \rho + 8 G^4 \rho + 6 g^2 G^2 \rho^2 -
      4 G^4 \rho^2}{g^4 G^4 \rho} C_5\wsep+
   \wpng{G^4 + 12 g^4 \rho - 20 g^2 G^2 \rho + 8 G^4 \rho + 6 g^2 G^2 \rho^2 -
      4 G^4 \rho^2}{g^2 G^5 \rho} C_6 \wsep +
   \wpng{g^2 + G^2 - 4 g^2 \rho + 4 G^2 \rho - 2 G^2 \rho^2 }{G^5 \rho}  C_7 \wsep +
   2 \wpng{G^4 + 12 g^4 \rho - 32 g^2 G^2 \rho + 20 G^4 \rho + 4 g^2 G^2 \rho^2 -
      6 G^4 \rho^2}{g^2 G^6 \rho} C_8 \wsep +
   \fwpng{2 (g^2 - G^2)} {-G^2 + 12 g^2 \rho - 16 G^2 \rho + 4 G^2 \rho^2}{
    g^2 G^6 \rho}  C_9 \wsep + {16 \over G^6} \left( 4 g^2 - 4 G^2 + G^2 \rho \right) C_a \wsep -
   2 \wpng{G^4 + 16 g^4 \rho - 32 g^2 G^2 \rho + 16 G^4 \rho + 4 g^2 G^2 \rho^2 -
      4 G^4 \rho^2}{g^4 G^4 \rho} C_b \wsep- 16 {(g^2 - G^2)^2 }{g^4 G^4} C_c}
\cma\\
\sigma_{4,\xi^+ \xi^-}&=& \sigma_{4,\xi^- \xi^+}\,.
\eea

The $sS$ matrix is a symmetric matrix about is Lorentz indices,
and is given  as
\begin{displaymath}
v^2 \, sS_{\xi\xi}^{\ab} =\left (\begin{array}{ccc}
sS_{\xi_Z \xi_Z}&sS_{\xi_Z\xi^{+}}&sS_{\xi_Z\xi^-}\\
sS_{\xi^- \xi_Z}&sS_{\xi^-\xi^{+}}&sS_{\xi^-\xi^-}\\
sS_{\xi^+ \xi_Z}&sS_{\xi^+\xi^{+}}&sS_{\xi^+\xi^-}\\
\end{array}\right)
\end{displaymath}
and its components read
\bea
sS_{\xi_Z \xi_Z}&=&16 {C_a \over G^2} OZZ g^{\alpha \beta} \wsep +
 4 \wpng{- e^2 G C_4 + g^2 (G C_4 - C_7) + G C_9}{G^3}
SF_Z^{\alpha \beta}+ 32 {C_a \over
  G^2} SZ^{\alpha  \beta} \cma\\
sS_{\xi_Z \xi^{+}}&=&- 2 \wpng{e^2 G C_4 + g^2 (-G C_4 + C_7) + G (2 C_8 + C_9
  }{g G^2} SF_-^{\alpha  \beta}\cma\\
sS_{\xi^-\xi^+}&=&2 \wpng{e^2 G (C_1 - C_3) - e (2 g^2 C_2 + G C_5) \ssep + g^2 (G C_3 + C_6) +
    2 G (C_b + 2 C_c)}{g^2 G} SF_Z^{\alpha  \beta} \wsep +
 4 \left ( C_4 - {e^2 \over g^2}  C_4 - {1\over G}C_7 + {1\over g^2} C_9 \right ) SZ^{\alpha
   \beta}
\cma\\
sS_{\xi^-\xi^-}&=&-4 \wpng{e^2 G (C_1 - C_3) - e (2 g^2 C_2 + G C_5) \ssep + g^2 (G C_3 + C_6) - 2 G C_b
  }{g^2 G} SW_+^{\alpha  \beta}\cma\\
sS_{\xi_Z \xi^-}&=&sS_{\xi_Z\xi^{+}}^{\dagger}\cma\,\,
sS_{\xi^- \xi_Z}=sS_{\xi_Z\xi^-}\cma\,\,
sS_{\xi^+\xi_Z}=sS_{\xi_Z\xi^+}\cma\nnb\\
sS_{\xi^+\xi^+}&=&sS_{\xi^-\xi^-}^{\dagger}\cma\,\,
sS_{\xi^+\xi^-}=sS_{\xi^-\xi^+}^{\ab}\,.
\eea
The $sA$ matrix is an antisymmetric matrix about its Lorentz
indices, and is given  as
\begin{displaymath}
v^2 \, sA_{\xi\xi}^{\ab} =\left (\begin{array}{ccc}
sA_{\xi_Z \xi_Z}&sA_{\xi_Z\xi^{+}}&sA_{\xi_Z\xi^-}\\
sA_{\xi^- \xi_Z}&sA_{\xi^-\xi^{+}}&sA_{\xi^-\xi^-}\\
sA_{\xi^+ \xi_Z}&sA_{\xi^+\xi^{+}}&sA_{\xi^+\xi^-}\\
\end{array}\right)
\end{displaymath}
and its components read
\bea
sA_{\xi_Z \xi_Z}&=&0\cma\\
sA_{\xi_Z \xi^{+}}&=&-
 2 i \frac{2 g^2 C_4 - G C_7}{g G^2}  W_-^{\alpha  \beta}\wsep+
2 \wpng{-e^2 G C_4 + g^2 (G C_4 - 2 C_7) + G (-2 C_8 + C_9)
   }{g G^2} F_-^{\alpha  \beta}
\cma\\
sA_{\xi^-\xi^{+}}&=&-2 i \wpng{-2 e G C_1 + 2 g^2 C_2 + G C_5}{g^2 G}  A^{\alpha\beta} \wsep -
 2 i \frac{-2 e G C_2 + 2 g^2 C_3 + G C_6}{g^2 G} Z^{\alpha \beta}
\wsep -
 2 \wpng{e^2 G (C_1 - C_3) - 2 e (g^2 C_2 + G C_5) \ssep+ g^2 (G C_3 + 2 C_6) -
    2 G (C_b - 2 C_c) }{g^2 G} F_Z^{\alpha  \beta} \cma\\
sA_{\xi^-\xi^-}&=&0\cma\,\,
sA_{\xi_Z\xi^-}=sA_{\xi_Z \xi^+}^{\dagger}\cma\,\,
sA_{\xi^-\xi_Z}=-sA_{\xi_Z\xi^-}\cma\nnb\\
sA_{\xi^+\xi_Z}&=&-sA_{\xi_Z \xi^+}\cma\,\,
sA_{\xi^+\xi^+}=0\cma\,\,
sA_{\xi^+\xi^-}=-sA_{\xi^-\xi^{+}}\cma
\eea

The ${\widetilde S^{\mu}_{\ab}}$ matrix is antisymmetric
on $\ab$, and is given  as
\begin{displaymath}
{\widetilde S^{\mu}_{\ab}} =\left (\begin{array}{ccc}
0& i cs_{A\xi_W}X^{-,\mu}_{\ab} & - i cs_{A\xi_W} X^{+,\mu}_{\ab}\\
0& i cs_{Z\xi_W}X^{-,\mu}_{\ab} & - i cs_{Z\xi_W} X^{+,\mu}_{\ab}\\
-i cs_{W\xi_Z} X^{+,\mu}_{\ab}&-i cs_{W \xi_W} X^{Z,\mu}_{\ab}&0\\
i cs_{W\xi_Z} X^{-,\mu}_{\ab}&0&i cs_{W \xi_W} X^{Z,\mu}_{\ab}\\
\end{array}\right)
\end{displaymath}
where the relevant definitions are
\bea
X^{-,\mu}_{\ab}&=&W^{-}_{\al} g^{\mu}_{\be} + W^{-}_{\be} g^{\mu}_{\al} - 2 g_{\al\be} W^{-,\mu}\cma\\
X^{+,\mu}_{\ab}&=&W^{+}_{\al} g^{\mu}_{\be} + W^{+}_{\be} g^{\mu}_{\al} - 2 g_{\al\be} W^{+,\mu}\cma\\
X^{Z,\mu}_{\ab}&=&Z_{\al} g^{\mu}_{\be} + Z^{\be} g^{\mu}_{\al} - 2 g_{\al\be} Z^{\mu}\cma\\
v cs_{A\xi_W}&=&-{-2 e  G C_1 + 2 g^2 C_2 + G C_5 \over 2 g G }\cma\\
v cs_{Z\xi_W}&=&-{-2 e  G C_2 + 2 g^2 C_3 + G C_6 \over 2 g G}\cma\\
v cs_{W\xi_Z}&=&-{2 g^2 C_4 - G C_7\over 2 G^2}\cma\\
v cs_{W\xi_W}&=&{2 g^2 C_4 - G C_7 \over 2 g G}\,.
\eea

The ${\widetilde A^{\mu}_{\ab}}$ matrix is
antisymmetric on $\ab$, and is given  as
\begin{displaymath}
 {\widetilde A^{\mu}_{\ab}} =\left (\begin{array}{ccc}
0& i cs_{A\xi_W} A^{-,\mu}_{\ab} & - i cs_{A\xi_W} A^{+,\mu}_{\ab}\\
0& i cs_{Z\xi_W} A^{-,\mu}_{\ab} & - i cs_{Z\xi_W} A^{+,\mu}_{\ab}\\
-i cs_{W\xi_Z} A^{+,\mu}_{\ab}&-i cs_{W \xi_W} A^{Z,\mu}_{\ab}&0\\
i cs_{W\xi_Z}  A^{-,\mu}_{\ab}&0&i cs_{W \xi_W} A^{Z,\mu}_{\ab}
\end{array}\right)
\end{displaymath}
\bea
A^{-,\mu}_{\ab}&=&W^{-}_{\be} g^{\mu}_{\al} - W^{-}_{\al} g^{\mu}_{\be}\cma\\
A^{+,\mu}_{\ab}&=&W^{+}_{\be} g^{\mu}_{\al} - W^{+}_{\al} g^{\mu}_{\be}\cma\\
A^{Z,\mu}_{\ab}&=&Z_{\be} g^{\mu}_{\al}     - Z^{\al} g^{\mu}_{\be}\cma
\eea

The ${\widetilde X^{\mu \al}_1}$ is a $4 \times 3$ matrix
\begin{displaymath}
v \, {\widetilde X^{\mu \al}_1}=\left (\begin{array}{ccc}
X_{1,A\xi_Z}^{\mu \al}  &X_{1,A\xi^+}^{\mu \al}  &X_{1,A\xi^-}^{\mu \al}\\
X_{1,Z\xi_Z}^{\mu \al}  &X_{1,Z\xi^+}^{\mu \al}  &X_{1,Z\xi^-}^{\mu \al}\\
X_{1,W^-\xi_Z}^{\mu \al}&X_{1,W^-\xi^+}^{\mu \al}&X_{1,W^-\xi^-}^{\mu \al}\\
X_{1,W^+\xi_Z}^{\mu \al}&X_{1,W^+\xi^+}^{\mu \al}&X_{1,W^+\xi^-}^{\mu \al}\\
\end{array}\right)\,,
\end{displaymath}
and its components read
\bea
X_{1,A\xi_Z}^{\mu \al}&=&e {2 g^2 C_4 - G C_7) \over G^2}  SF_Z^{\alpha  \mu}+
 e {-2 g^2 C_4 + G C_7 \over G^2} SFZ g^{\alpha  \mu}
\cma\\
X_{1,A\xi^+}^{\mu \al}&=&e \sbra{-{g C_4\over G} + {C_7 \over 2 g}} F_-^{\alpha  \mu} +
 e \sbra{{g C_4 \over G} - {C_7\over 2 g}} SF_-^{\alpha  \mu} \wsep +
 e \sbra{-{g C_4 \over G} + {C_7\over 2 g}} SFM g^{\alpha  \mu}
\cma\\
X_{1,Z\xi_Z}^{\mu \al}&=&{8 C_a \over G} OZZ g^{\alpha  \mu}+
 {-g^2 C_7  + 2 G C_9 \over G^2}  SF_Z^{\alpha  \mu}\wsep+
 {(g^2 C_7 + 2 G C_8 \over G^2} SFZ g^{\alpha  \mu}+
 {16 C_a \over G } SZ^{\alpha  \mu}
\cma\\
X_{1,Z\xi^+}^{\mu \al}&=&\sbra{{3 g C_7 \over 2 G} + {2 C_8 - C_9 \over g}} F_-^{\alpha  \mu} -
 {g^2 C_7 + 4 G C_8 + 2 G C_9 \over
  2 g G}  SF_-^{\alpha  \mu} \wsep + \sbra{{g C_7\over 2 G} - {C_9\over g}} SFM g^{\alpha
   \mu} + i \sbra{{2 g C_4\over G} - {C_7\over g}}
  W_-^{\alpha  \mu}
\cma\\
X_{1,W^-\xi_Z}^{\mu\al}&=&-{3 g^2 C_7 + 4 G C_8 - 2 G C_9 \over 2 G^2} F_+^{\alpha  \mu}
  + {g^2 C_7 + 4 G C_8 + 2 G C_9 \over 2 G^2}  SF_+^{\alpha \mu} \wsep -
 {g^2 C_7 - 2 G C_9 \over 2 G62} SFP g^{\alpha  \mu}+
 i {-2 g^2 C_4 + G C_7 \over G^2} W_+^{\alpha  \mu}
\cma\\
X_{1,W^-\xi^+}^{\mu\al}&=&-i {-2 e G C_1 + 2 g^2 C_2 + G C_5 \over g G} A^{\alpha
    \mu} \wsep + {3 e G C_5 - 3 g^2 C_6 + 4 G (C_b - 2 C_c) \over 2 g G}
   F_Z^{\alpha  \mu} +
 -\sbra{{g C_7\over G} + {2 C_8\over g}} OZZ g^{\alpha  \mu} \wsep +
 {e G C_5 - g^2 C_6 - 4 G (C_b + 2 C_c) \over 2 g G} SF_Z^{\alpha
    \mu}  -
 {e G C_5 - g^2 C_6 + 4 G C_b\over
  2 g G} SFZ g^{\alpha  \mu} \wsep + \sbra{{g C_7\over G} - {2 C_9\over g}} SZ^{\alpha  \mu} \wsep -
 i {-2 e G C_2 + 2 g^2 C_3 + G C_6 \over g G} Z^{\alpha
    \mu}
\cma\\
X_{1,W^-\xi^-}^{\mu\al}&=&\wpng{e C_5 - {g^2 C_6\over G} - 4 C_c}{ g} OW^+ g^{\alpha \mu} \wsep +
 \wpng{-e C_5 + {g^2 C_6 \over G} - 4 C_b}{g} SW_+^{\alpha  \mu}
\cma\\
X_{1,A\xi^-}^{\mu \al}&=&X_{1,A\xi^+}^{\mu \al,\dagger}\cma\,\,
X_{1,Z\xi^-}^{\mu \al}=X_{1,Z\xi^+}^{\mu\al,\dagger}\cma\nnb\\
X_{1,W^+\xi_Z}^{\mu\al}&=&X_{1,W^-\xi_Z}^{\mu\al,\dagger}\cma\,\,
X_{1,W^+\xi^+}^{\mu\al}=X_{1,W^-\xi^+}^{\mu\al,\dagger}\cma\,\,
X_{1,W^+\xi^-}^{\mu\al}=X_{1,W^-\xi^-}^{\mu\al,\dagger}\,.
\eea

The ${\widetilde X^{\mu \al}_2}$ is a $4 \times 3$ matrix
\begin{displaymath}
v \, {\widetilde X^{\mu \al}_2}=\left (\begin{array}{ccc}
X_{2,A\xi_Z}^{\mu \al}  &X_{2,A\xi^+}^{\mu \al}  &X_{2,A\xi^-}^{\mu \al}\\
X_{2,Z\xi_Z}^{\mu \al}  &X_{2,Z\xi^+}^{\mu \al}  &X_{2,Z\xi^-}^{\mu \al}\\
X_{2,W^-\xi_Z}^{\mu \al}&X_{2,W^-\xi^+}^{\mu \al}&X_{2,W^-\xi^-}^{\mu \al}\\
X_{2,W^+\xi_Z}^{\mu \al}&X_{2,W^+\xi^+}^{\mu \al}&X_{2,W^+\xi^-}^{\mu \al}\\
\end{array}\right)\,,
\end{displaymath}
and its components read
\bea
X_{2,A\xi_Z}^{\mu \al}&=&0
\cma\\
X_{2,A\xi^+}^{\mu \al}&=&\mpng{-4 e^2 G C_2 - G^2 C_5 + g^2 (2 G C_2 + C_5) \wsep+
    e \left (-4 g^2 C_1 + G (2 G C_1 + C_7)\right)} {g G}
F_-^{\alpha \mu} \wsep + i 2 \mpng{g^2 C_2 + e G (-C_1 + C_4)}{g G}
   W_-^{\alpha  \mu}
\cma\\
X_{2,Z\xi_Z}^{\mu \al}&=&0
\cma\\
X_{2,Z\xi^+}^{\mu \al}&=&\mpng{2 e (-2 g^2 + G^2) C_2 - 4 e^2 G C_3 \wsep - G^2 C_6 + g^2 (2 G C_3 + C_6 - C_7)}{g G}
   F_-^{\alpha  \mu} \wsep -
 2 i {e G C_2 + g^2 (-C_3 + C_4) \over g G} W_-^{\alpha  \mu}
\cma\\
X_{2,W^-\xi_Z}^{\mu\al}&=&\sbra{-{2 g^2 C_4\over G} + C_7} F_+^{\alpha  \mu}
\cma\\
X_{2,W^-\xi^+}^{\mu\al}&=&2 i {g^2 C_2 + e G (-C_1 + C_4) \over g G} A^{\alpha  \mu}
  + \sbra{ 2 g C_4 - {e C_5 \over g} + {g C_6 \over G}} F_Z^{\alpha
   \mu} \wsep - 2 i \mpng{e G C_2 + g^2 (-C_3 + C_4)}{g G}
   Z^{\alpha  \mu}
\cma\\
X_{2,W^-\xi^-}^{\mu\al}&=&0
\cma\,\,
X_{2,A\xi^-}^{\mu \al}=X_{2,A\xi^+}^{\mu \al,\dagger}\cma\,\,
X_{2,Z\xi^-}^{\mu \al}=X_{2,Z\xi^+}^{\mu\al,\dagger}\cma\nnb\\
X_{2,W^+\xi_Z}^{\mu\al}&=&X_{2,W^-\xi_Z}^{\mu\al,\dagger}\cma\,\,
X_{2,W^+\xi^+}^{\mu\al}=X_{2,W^-\xi^+}^{\mu\al,\dagger}\cma\,\,
X_{2,W^+\xi^-}^{\mu\al}=X_{2,W^-\xi^-}^{\mu\al,\dagger}\,.
\eea

The matrix ${\widetilde X}^{\mu}_{01}$ is given  as
\begin{displaymath}
{\widetilde X}^{\mu}_{01}=\left (\begin{array}{ccc}
Z_{01,A\xi_Z}^{\mu }  &Z_{01,A\xi^+}^{\mu }  &Z_{01,A\xi^-}^{\mu }\\
Z_{01,Z\xi_Z}^{\mu }  &Z_{01,Z\xi^+}^{\mu }  &Z_{01,Z\xi^-}^{\mu }\\
Z_{01,W^-\xi_Z}^{\mu }&Z_{01,W^-\xi^+}^{\mu }&Z_{01,W^-\xi^-}^{\mu }\\
Z_{01,W^+\xi_Z}^{\mu }&Z_{01,W^+\xi^+}^{\mu }&Z_{01,W^+\xi^-}^{\mu }\\
\end{array}\right)\,,
\end{displaymath}
and its components read
\bea
Z_{01,A\xi_Z}^{\mu}&=&0
\cma\,\,
Z_{01,A\xi^+}^{\mu }=-{i\over 4} g (2 e + C_5) v W_-^{ \mu}
\cma\nnb\\
Z_{01,Z\xi_Z}^{\mu }&=&0
\cma\,\,
Z_{01,Z\xi^+}^{\mu }=-{i\over 4 g} (2 e^2 G + g^2 C_6) v W_-^{ \mu}
\cma\nnb\\
Z_{01,W^-\xi_Z}^{\mu }&=&{i\over 2} f_{zw} G \rho v W_+^{ \mu}
\cma\\
Z_{01,W^-\xi^+}^{\mu }&=&-{i\over 4 g} (2 e^2 G + g^2 C_7) v Z^{ \mu}
\cma\\
Z_{01,W^-\xi^-}^{\mu }&=&0
\cma\,\,
Z_{01,A\xi^-}^{\mu}=Z_{01,A\xi^+}^{\mu,\dagger}\cma\,\,
Z_{01,Z\xi^-}^{\mu }=Z_{01,Z\xi^+}^{\mu,\dagger}\cma\\
Z_{01,W^+\xi_Z}^{\mu }&=&Z_{01,W^-\xi_Z}^{\mu,\dagger }\cma\,\,
Z_{01,W^+\xi^+}^{\mu }=Z_{01,W^-\xi^-}^{\mu,\dagger }\cma\,\,
Z_{01,W^+\xi^-}^{\mu }=Z_{01,W^-\xi^+}^{\mu,\dagger }\cma
\eea

The matrix ${\widetilde X}^{\mu}_{03}$ is given  as
\begin{displaymath}
v {\widetilde X}^{\mu}_{01}=\left (\begin{array}{ccc}
Z_{03,A\xi_Z}^{\mu }  &Z_{03,A\xi^+}^{\mu }  &Z_{03,A\xi^-}^{\mu }\\
Z_{03,Z\xi_Z}^{\mu }  &Z_{03,Z\xi^+}^{\mu }  &Z_{03,Z\xi^-}^{\mu }\\
Z_{03,W^-\xi_Z}^{\mu }&Z_{03,W^-\xi^+}^{\mu }&Z_{03,W^-\xi^-}^{\mu }\\
Z_{03,W^+\xi_Z}^{\mu }&Z_{03,W^+\xi^+}^{\mu }&Z_{03,W^+\xi^-}^{\mu }\\
\end{array}\right)\,,
\end{displaymath}
and its components read
\bea
Z_{03,A\xi_Z}^{\mu}&=&0
\cma\\
Z_{03,A\xi^+}^{\mu }&=&i \sbra{ {e g C_7)\over G} + {2 e C_8\over g}} OZZ \, \,W_-^{ \mu} \wsep+
 i \sbra{- {2  e^3 \over g} C_1 + 4 {-g^5 + g^3 G^2\over G^3}  C_2- 2 {e g^3 \over G^2} C_3+
   {3 \over 2} {-g^3 + g G^2 \over G^2 } C_5 \wsep - {3 \over 2} {e g \over G} C_6
+ 2 {e \over g} C_b}
  SFZ \, \, W_-^{ \mu}
\cma\\
Z_{03,Z\xi_Z}^{\mu }&=&0
\cma\\
Z_{03,Z\xi^+}^{\mu }&=&i \sbra{{2 e^2 G C_8 \over g^3} + {2 e^2 G C_9\over g^3}} OZZ\,\,
  W_-^{ \mu} \wsep +
i \sbra{- g {2 g^4 - 3 g^2 G^2 + G^4 \over G^3} C_1- 2 e g {2 g^2 - G^2 \over G^2} C_2
  - {-2 g^5 + g^3 G^2 \over G^3} C_3 \wsep - {e\over 2} {3 g^2 - 2 G^2 \over g G} C_5
 + {1 \over 2} {3 g^3 - 2 g G^2 \over G^2} C_6
+ 2 {e^2 G \over g^3} C_b} SFZ \,\,
  W_-^{ \mu}
\cma\\
Z_{03,W^-\xi_Z}^{\mu }&=&0
\cma\\
Z_{03,W^-\xi^+}^{\mu }&=&i\sbra{{2 e^2 G C_8\over g^3} + {2 e^2 G C_9\over g^3}} OZZ \,\,
  Z^{ \mu} \wsep + i
 \sbra{{e\over 2}{-2 g^2 + G^2 \over g G} C_5+ {1\over 2} {2 g^3 - g G^2 \over G^2}  C_6+
   {g \over 2} C_7 + {2 e^2 G \over g^3} C_b} SFZ \,\, Z^{ \mu}
\cma\\
Z_{03,W^-\xi^-}^{\mu }&=&0\cma\,\,
Z_{03,A\xi^-}^{\mu}=Z_{03,A\xi^+}^{\mu,\dagger}\cma\,\,
Z_{03,Z\xi^-}^{\mu }=Z_{03,Z\xi^+}^{\mu,\dagger}\cma\\
Z_{03,W^+\xi_Z}^{\mu }&=&Z_{03,W^-\xi_Z}^{\mu,\dagger }\cma\,\,
Z_{03,W^+\xi^+}^{\mu }=Z_{03,W^-\xi^-}^{\mu,\dagger }\cma\,\,
Z_{03,W^+\xi^-}^{\mu }=Z_{03,W^-\xi^+}^{\mu,\dagger }\cma
\eea

In these tilded quantities, there are also no terms like
$\partial \cdot Z$ and $d\cdot W^{\pm}$. The unbroken
$U_{em}$ symmetry is explicit in these matrices.

\vskip 1 cm
{\bf \Large Figures and Captions}:
\begin{figure}
     \begin{minipage}[t]{5.5cm}
     \epsfig{file=al1015.epsi,width=5.5cm}
     \mbox{ }\hfill\hspace{1cm}(a)\hfill\mbox{ }
     \end{minipage}
     \hspace{2cm}
     \begin{minipage}[t]{5.5cm}
     \epsfig{file=al2015.epsi,width=5.5cm}
     \mbox{ }\hfill\hspace{1cm}(b)\hfill\mbox{ }
     \end{minipage}
\vskip 0.05truein
     \begin{minipage}[t]{5.5cm}
     \epsfig{file=al3015.epsi,width=5.5cm}
     \mbox{ }\hfill\hspace{1cm}(c)\hfill\mbox{ }
     \end{minipage}
     \hspace{2cm}
     \begin{minipage}[t]{5.5cm}
     \epsfig{file=al4015.epsi,width=5.5cm}
     \mbox{ }\hfill\hspace{1cm}(d)\hfill\mbox{ }
     \end{minipage}
\vskip 0.05truein
     \begin{minipage}[t]{5.5cm}
     \epsfig{file=al5015.epsi,width=5.5cm}
     \mbox{ }\hfill\hspace{1cm}(e)\hfill\mbox{ }
     \end{minipage}
     \hspace{2cm}
     \begin{minipage}[t]{5.5cm}
     \epsfig{file=al6015.epsi,width=5.5cm}
     \mbox{ }\hfill\hspace{1cm}(f)\hfill\mbox{ }
     \end{minipage}
\vskip 0.05truein
     \begin{minipage}[t]{5.5cm}
     \epsfig{file=al7015.epsi,width=5.5cm}
     \mbox{ }\hfill\hspace{1cm}(g)\hfill\mbox{ }
     \end{minipage}
     \hspace{2cm}
     \begin{minipage}[t]{5.5cm}
     \epsfig{file=al8015.epsi,width=5.5cm}
     \mbox{ }\hfill\hspace{1cm}(h)\hfill\mbox{ }
     \end{minipage}
\vskip 0.05truein
     \begin{minipage}[t]{5.5cm}
     \epsfig{file=al9015.epsi,width=5.5cm}
     \mbox{ }\hfill\hspace{1cm}(i)\hfill\mbox{ }
     \end{minipage}
     \hspace{2cm}
     \begin{minipage}[t]{5.5cm}
     \epsfig{file=ala015.epsi,width=5.5cm}
     \mbox{ }\hfill\hspace{1cm}(j)\hfill\mbox{ }
     \end{minipage}
     \caption{\it
The $x$ axe is $t=\ln{\mu\over m_Z}$, and $m_0=150$ GeV.
The solid lines are for the RGE method,
while the dashed lines for the DM.
The fig1.a---fig1.j are for the ACs
$\al_1$---$\al_a$, respectively.}
\label{fig1}
\end{figure}

\begin{figure}
     \begin{minipage}[t]{5.5cm}
     \epsfig{file=al1030.epsi,width=5.5cm}
     \mbox{ }\hfill\hspace{1cm}(a)\hfill\mbox{ }
     \end{minipage}
     \hspace{2cm}
     \begin{minipage}[t]{5.5cm}
     \epsfig{file=al2030.epsi,width=5.5cm}
     \mbox{ }\hfill\hspace{1cm}(b)\hfill\mbox{ }
     \end{minipage}
\vskip 0.05truein
     \begin{minipage}[t]{5.5cm}
     \epsfig{file=al3030.epsi,width=5.5cm}
     \mbox{ }\hfill\hspace{1cm}(c)\hfill\mbox{ }
     \end{minipage}
     \hspace{2cm}
     \begin{minipage}[t]{5.5cm}
     \epsfig{file=al4030.epsi,width=5.5cm}
     \mbox{ }\hfill\hspace{1cm}(d)\hfill\mbox{ }
     \end{minipage}
\vskip 0.05truein
     \begin{minipage}[t]{5.5cm}
     \epsfig{file=al5030.epsi,width=5.5cm}
     \mbox{ }\hfill\hspace{1cm}(e)\hfill\mbox{ }
     \end{minipage}
     \hspace{2cm}
     \begin{minipage}[t]{5.5cm}
     \epsfig{file=al6030.epsi,width=5.5cm}
     \mbox{ }\hfill\hspace{1cm}(f)\hfill\mbox{ }
     \end{minipage}
\vskip 0.05truein
     \begin{minipage}[t]{5.5cm}
     \epsfig{file=al7030.epsi,width=5.5cm}
     \mbox{ }\hfill\hspace{1cm}(g)\hfill\mbox{ }
     \end{minipage}
     \hspace{2cm}
     \begin{minipage}[t]{5.5cm}
     \epsfig{file=al8030.epsi,width=5.5cm}
     \mbox{ }\hfill\hspace{1cm}(h)\hfill\mbox{ }
     \end{minipage}
\vskip 0.05truein
     \begin{minipage}[t]{5.5cm}
     \epsfig{file=al9030.epsi,width=5.5cm}
     \mbox{ }\hfill\hspace{1cm}(i)\hfill\mbox{ }
     \end{minipage}
     \hspace{2cm}
     \begin{minipage}[t]{5.5cm}
     \epsfig{file=ala030.epsi,width=5.5cm}
     \mbox{ }\hfill\hspace{1cm}(j)\hfill\mbox{ }
     \end{minipage}
     \caption{\it
The $x$ axe is $t=\ln{\mu\over m_Z}$, and $m_0=300$ GeV.
The solid lines are for the RGE method,
while the dashed lines for the DM.
The fig2.a---fig2.j are for the ACs
$\al_1$---$\al_a$, respectively.}
\label{fig2}
\end{figure}

\begin{figure}
     \begin{minipage}[t]{5.5cm}
     \epsfig{file=al1045.epsi,width=5.5cm}
     \mbox{ }\hfill\hspace{1cm}(a)\hfill\mbox{ }
     \end{minipage}
     \hspace{2cm}
     \begin{minipage}[t]{5.5cm}
     \epsfig{file=al2045.epsi,width=5.5cm}
     \mbox{ }\hfill\hspace{1cm}(b)\hfill\mbox{ }
     \end{minipage}
\vskip 0.05truein
     \begin{minipage}[t]{5.5cm}
     \epsfig{file=al3045.epsi,width=5.5cm}
     \mbox{ }\hfill\hspace{1cm}(c)\hfill\mbox{ }
     \end{minipage}
     \hspace{2cm}
     \begin{minipage}[t]{5.5cm}
     \epsfig{file=al4045.epsi,width=5.5cm}
     \mbox{ }\hfill\hspace{1cm}(d)\hfill\mbox{ }
     \end{minipage}
\vskip 0.05truein
     \begin{minipage}[t]{5.5cm}
     \epsfig{file=al5045.epsi,width=5.5cm}
     \mbox{ }\hfill\hspace{1cm}(e)\hfill\mbox{ }
     \end{minipage}
     \hspace{2cm}
     \begin{minipage}[t]{5.5cm}
     \epsfig{file=al6045.epsi,width=5.5cm}
     \mbox{ }\hfill\hspace{1cm}(f)\hfill\mbox{ }
     \end{minipage}
\vskip 0.05truein
     \begin{minipage}[t]{5.5cm}
     \epsfig{file=al7045.epsi,width=5.5cm}
     \mbox{ }\hfill\hspace{1cm}(g)\hfill\mbox{ }
     \end{minipage}
     \hspace{2cm}
     \begin{minipage}[t]{5.5cm}
     \epsfig{file=al8045.epsi,width=5.5cm}
     \mbox{ }\hfill\hspace{1cm}(h)\hfill\mbox{ }
     \end{minipage}
\vskip 0.05truein
     \begin{minipage}[t]{5.5cm}
     \epsfig{file=al9045.epsi,width=5.5cm}
     \mbox{ }\hfill\hspace{1cm}(i)\hfill\mbox{ }
     \end{minipage}
     \hspace{2cm}
     \begin{minipage}[t]{5.5cm}
     \epsfig{file=ala045.epsi,width=5.5cm}
     \mbox{ }\hfill\hspace{1cm}(j)\hfill\mbox{ }
     \end{minipage}
     \caption{\it
The $x$ axe is $t=\ln{\mu\over m_Z}$, and $m_0=450$ GeV.
The solid lines are for the RGE method,
while the dashed lines for the DM.
The fig3.a---fig3.j are for the ACs
$\al_1$---$\al_a$, respectively.
}
\label{fig3}
\end{figure}

\begin{figure}
     \begin{minipage}[t]{5.5cm}
     \epsfig{file=al1090.epsi,width=5.5cm}
     \mbox{ }\hfill\hspace{1cm}(a)\hfill\mbox{ }
     \end{minipage}
     \hspace{2cm}
     \begin{minipage}[t]{5.5cm}
     \epsfig{file=al2090.epsi,width=5.5cm}
     \mbox{ }\hfill\hspace{1cm}(b)\hfill\mbox{ }
     \end{minipage}
\vskip 0.05truein
     \begin{minipage}[t]{5.5cm}
     \epsfig{file=al3090.epsi,width=5.5cm}
     \mbox{ }\hfill\hspace{1cm}(c)\hfill\mbox{ }
     \end{minipage}
     \hspace{2cm}
     \begin{minipage}[t]{5.5cm}
     \epsfig{file=al4090.epsi,width=5.5cm}
     \mbox{ }\hfill\hspace{1cm}(d)\hfill\mbox{ }
     \end{minipage}
\vskip 0.05truein
     \begin{minipage}[t]{5.5cm}
     \epsfig{file=al5090.epsi,width=5.5cm}
     \mbox{ }\hfill\hspace{1cm}(e)\hfill\mbox{ }
     \end{minipage}
     \hspace{2cm}
     \begin{minipage}[t]{5.5cm}
     \epsfig{file=al6090.epsi,width=5.5cm}
     \mbox{ }\hfill\hspace{1cm}(f)\hfill\mbox{ }
     \end{minipage}
\vskip 0.05truein
     \begin{minipage}[t]{5.5cm}
     \epsfig{file=al7090.epsi,width=5.5cm}
     \mbox{ }\hfill\hspace{1cm}(g)\hfill\mbox{ }
     \end{minipage}
     \hspace{2cm}
     \begin{minipage}[t]{5.5cm}
     \epsfig{file=al8090.epsi,width=5.5cm}
     \mbox{ }\hfill\hspace{1cm}(h)\hfill\mbox{ }
     \end{minipage}
\vskip 0.05truein
     \begin{minipage}[t]{5.5cm}
     \epsfig{file=al9090.epsi,width=5.5cm}
     \mbox{ }\hfill\hspace{1cm}(i)\hfill\mbox{ }
     \end{minipage}
     \hspace{2cm}
     \begin{minipage}[t]{5.5cm}
     \epsfig{file=ala090.epsi,width=5.5cm}
     \mbox{ }\hfill\hspace{1cm}(j)\hfill\mbox{ }
     \end{minipage}
     \caption{\it
The $x$ axe is $t=\ln{\mu\over m_Z}$, and $m_0=900$ GeV.
The solid lines are for the RGE method,
while the dashed lines for the DM.
The fig4.a---fig4.j are for the ACs $\al_1$---$\al_a$, respectively.
}
\label{fig4}
\end{figure}

\begin{figure}
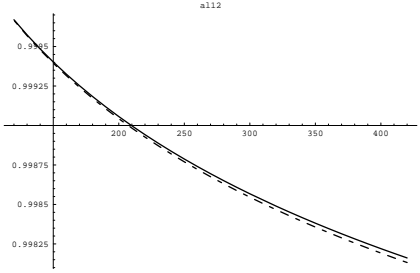

     \begin{minipage}[t]{5.5cm}
     \epsfig{file=al1cmp.epsi,width=5.5cm}
     \mbox{ }\hfill\hspace{1cm}(a)\hfill\mbox{ }
     \end{minipage}
     \hspace{2cm}
     \begin{minipage}[t]{5.5cm}
     \epsfig{file=al2cmp.epsi,width=5.5cm}
     \mbox{ }\hfill\hspace{1cm}(b)\hfill\mbox{ }
     \end{minipage}
\vskip 0.05truein
     \begin{minipage}[t]{5.5cm}
     \epsfig{file=al3cmp.epsi,width=5.5cm}
     \mbox{ }\hfill\hspace{1cm}(c)\hfill\mbox{ }
     \end{minipage}
     \hspace{2cm}
     \begin{minipage}[t]{5.5cm}
     \epsfig{file=al4cmp.epsi,width=5.5cm}
     \mbox{ }\hfill\hspace{1cm}(d)\hfill\mbox{ }
     \end{minipage}
\vskip 0.05truein
     \begin{minipage}[t]{5.5cm}
     \epsfig{file=al5cmp.epsi,width=5.5cm}
     \mbox{ }\hfill\hspace{1cm}(e)\hfill\mbox{ }
     \end{minipage}
     \hspace{2cm}
     \begin{minipage}[t]{5.5cm}
     \epsfig{file=al6cmp.epsi,width=5.5cm}
     \mbox{ }\hfill\hspace{1cm}(f)\hfill\mbox{ }
     \end{minipage}
\vskip 0.05truein
     \begin{minipage}[t]{5.5cm}
     \epsfig{file=al7cmp.epsi,width=5.5cm}
     \mbox{ }\hfill\hspace{1cm}(g)\hfill\mbox{ }
     \end{minipage}
     \hspace{2cm}
     \begin{minipage}[t]{5.5cm}
     \epsfig{file=al8cmp.epsi,width=5.5cm}
     \mbox{ }\hfill\hspace{1cm}(h)\hfill\mbox{ }
     \end{minipage}
\vskip 0.05truein
     \begin{minipage}[t]{5.5cm}
     \epsfig{file=al9cmp.epsi,width=5.5cm}
     \mbox{ }\hfill\hspace{1cm}(i)\hfill\mbox{ }
     \end{minipage}
     \hspace{2cm}
     \begin{minipage}[t]{5.5cm}
     \epsfig{file=alacmp.epsi,width=5.5cm}
     \mbox{ }\hfill\hspace{1cm}(j)\hfill\mbox{ }
     \end{minipage}
\vskip 0.05truein
     \begin{minipage}[t]{5.5cm}
     \epsfig{file=rhocmp.epsi,width=5.5cm}
     \mbox{ }\hfill\hspace{1cm}(j)\hfill\mbox{ }
     \end{minipage}
     \caption{\it
The $\al_i(m_Z)$ with the scanning of the Higgs mass
from $120$ GeV to $420$ GeV.
The solid lines are for the RGE method,
while the dashed lines for the DM.
The fig5.a---fig5.i are for the ACs $\al_1$---$\al_a$, respectively.
The fig5.j is for $\rho$.}
\label{fig5}
\end{figure}

\begin{figure}
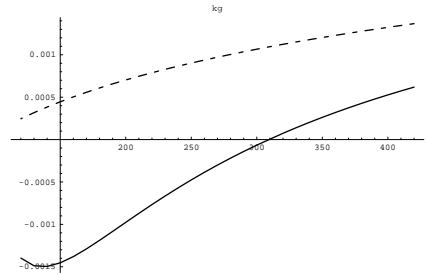

     \begin{minipage}[t]{5.5cm}
     \epsfig{file=stscan.epsi,width=5.5cm}
     \mbox{ }\hfill\hspace{1cm}(a)\hfill\mbox{ }
     \end{minipage}
     \hspace{2cm}
     \begin{minipage}[t]{5.5cm}
     \epsfig{file=gzscan.epsi,width=5.5cm}
     \mbox{ }\hfill\hspace{1cm}(b)\hfill\mbox{ }
     \end{minipage}
\vskip 0.05truein
     \begin{minipage}[t]{5.5cm}
     \epsfig{file=kzscan.epsi,width=5.5cm}
     \mbox{ }\hfill\hspace{1cm}(c)\hfill\mbox{ }
     \end{minipage}
     \hspace{2cm}
     \begin{minipage}[t]{5.5cm}
     \epsfig{file=kgscan.epsi,width=5.5cm}
     \mbox{ }\hfill\hspace{1cm}(d)\hfill\mbox{ }
     \end{minipage}
     \caption{\it
The precision test parameters at $m_Z$
with the scanning of the Higgs mass
from $120$ GeV to $420$ GeV.
The solid lines are for the RGE method,
while the dashed lines for the DM.
The fig6.a is for the $S$-$T$ plate, where the $S$ and $T$
are represented by the $x$ and $y$ axes, respectively.
The fig6.b, fig6.c and fig6.d are
for the triple gauge vertices $g_1^Z-1$, $k_Z-1$,
and $k_{\gamma}-1$, respectively.
While for these three figures,
the $x$ axe is the varying of the mass of the Higgs.
}
\label{fig6}
\end{figure}
\newpage
\begin{figure}
\begin{minipage}[t]{11 cm}
     \epsfig{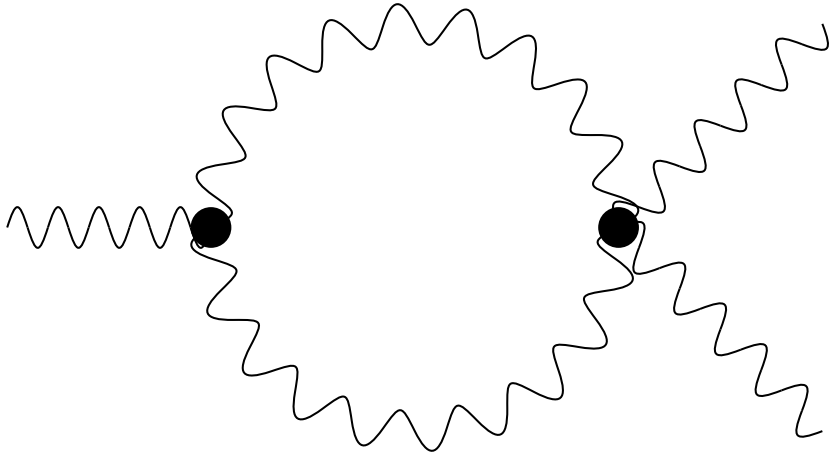}
     \mbox{ }\hfill\hspace{5.5cm}\hfill\mbox{ }
\end{minipage}
\caption{\it The related Feynman diagram (in unitary gauge) in the effective theory which contribute
to the trilinear couplings.}
\label{fig7}
\end{figure}
\newpage
\begin{figure}
\begin{minipage}[t]{11 cm}
     \epsfig{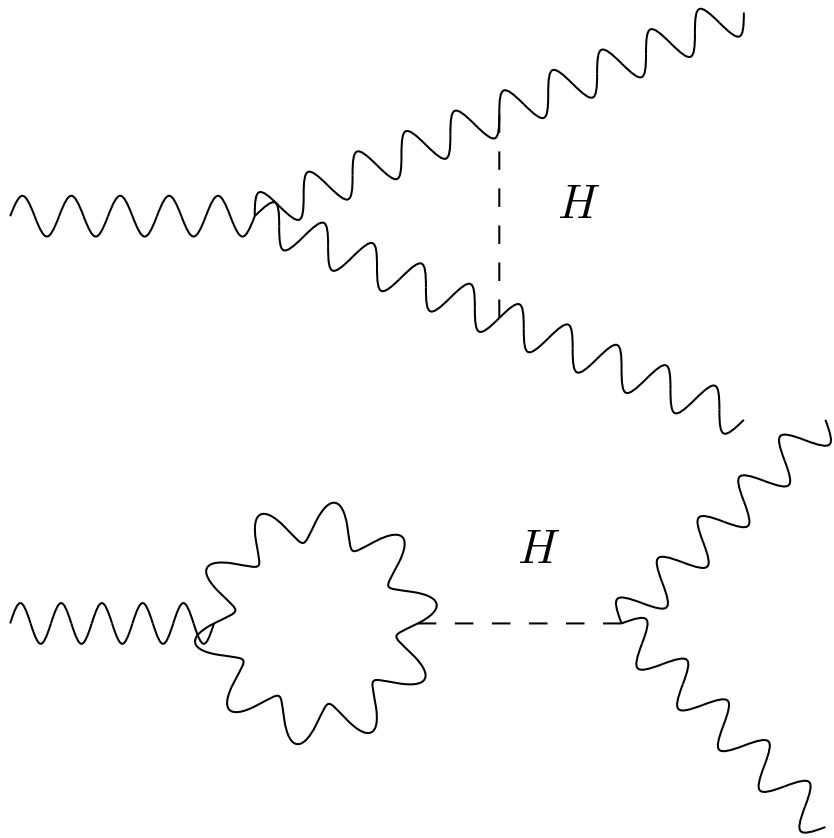}
     \mbox{ }\hfill\hspace{5.5cm}\hfill\mbox{ }
\end{minipage}
\caption{\it The related Feynman diagram (in unitary gauge) in the renormalizable theory which contribute
to the trilinear couplings.}
\label{fig8}
\end{figure}

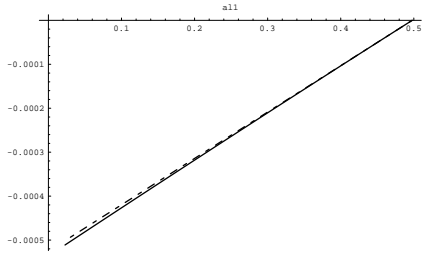
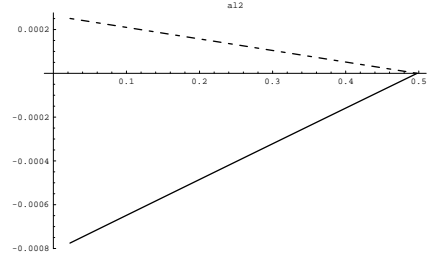
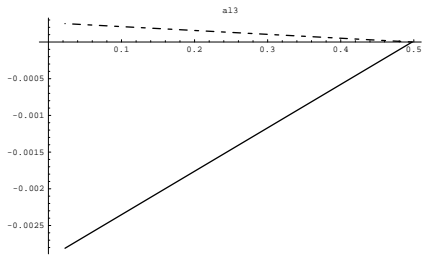
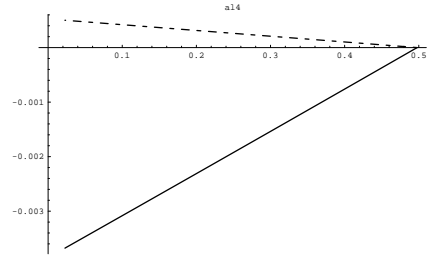
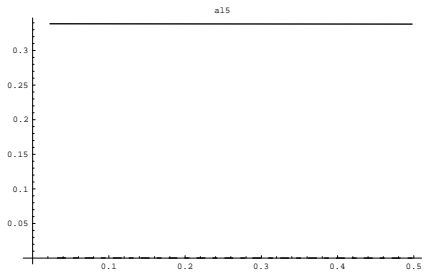
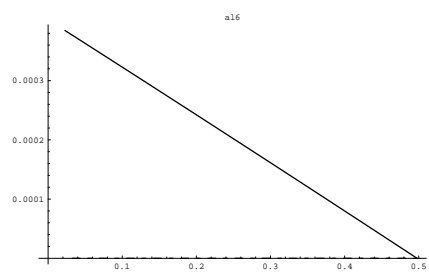
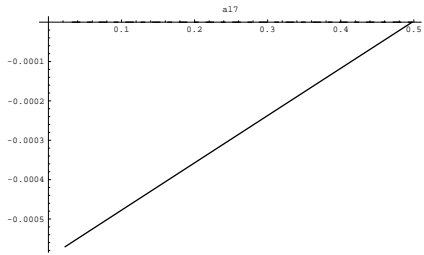
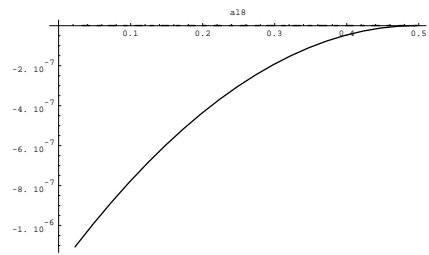
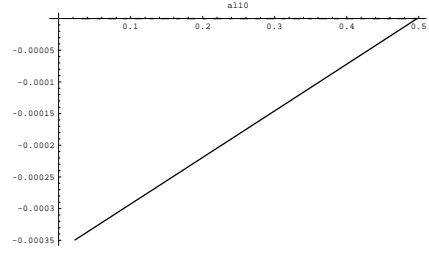
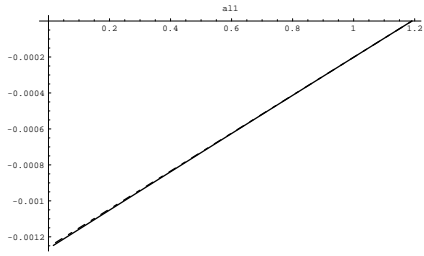
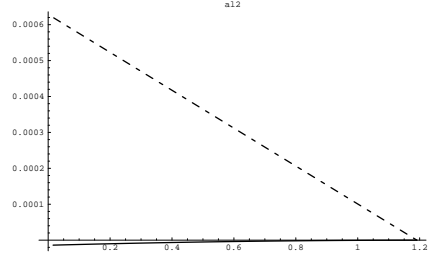
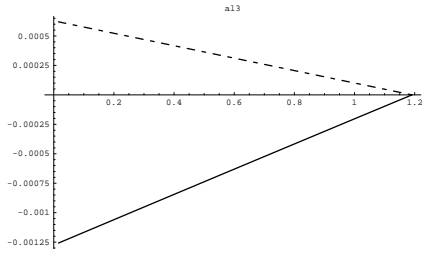
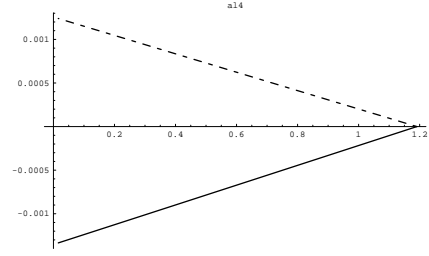
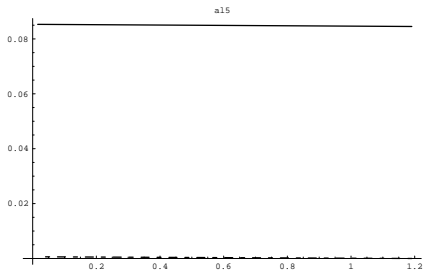
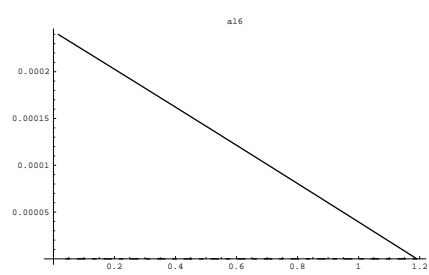
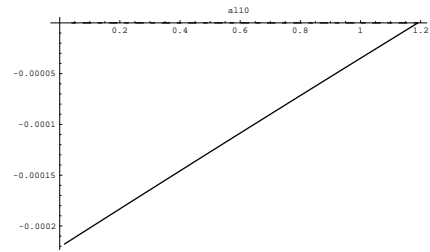
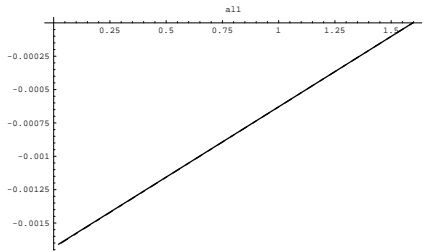
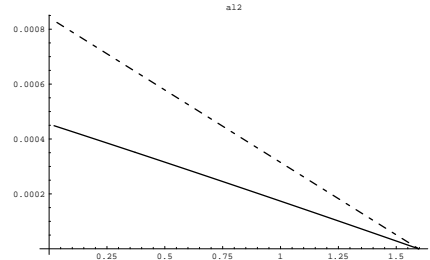
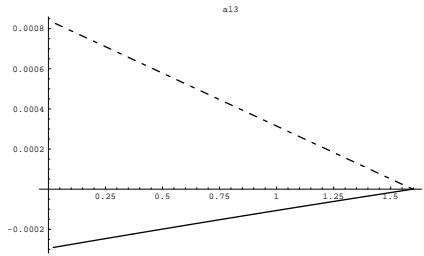
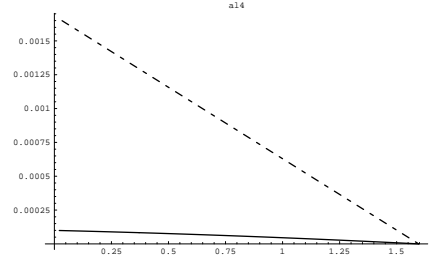
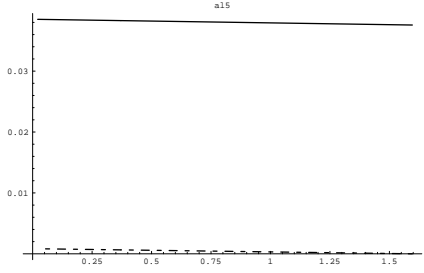
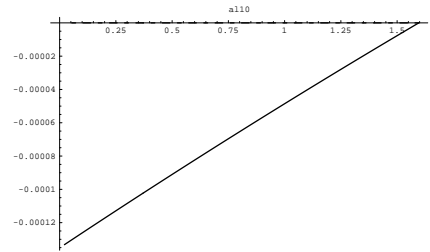
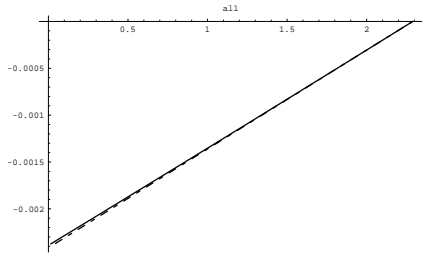
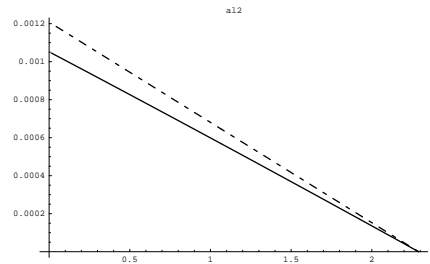
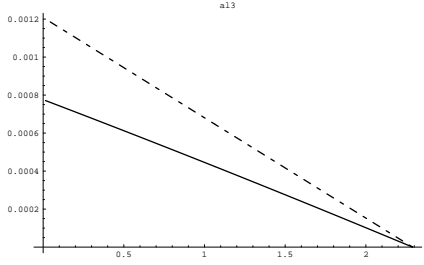
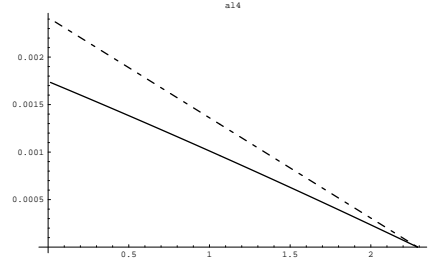
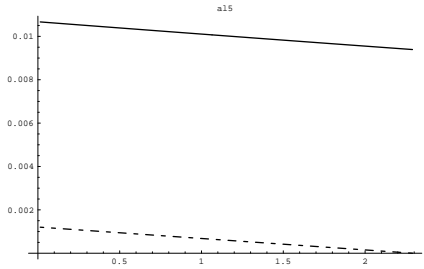
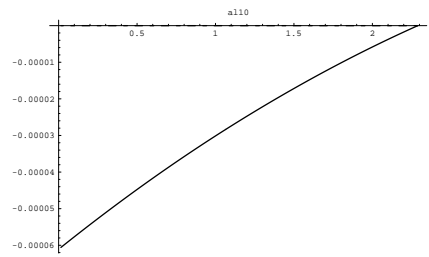
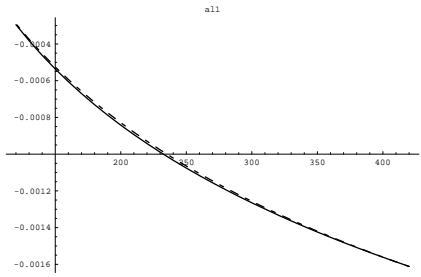
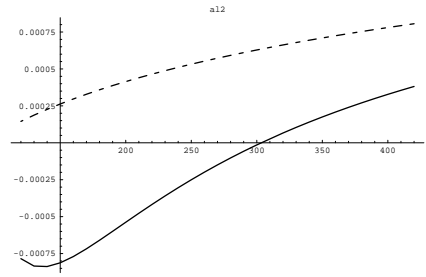
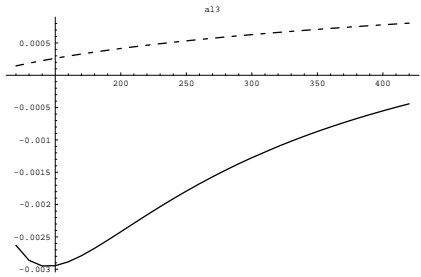
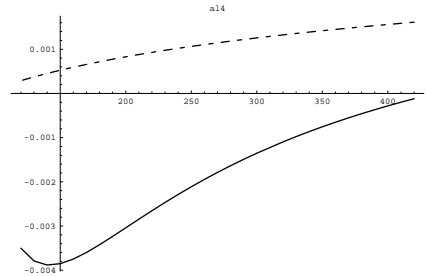
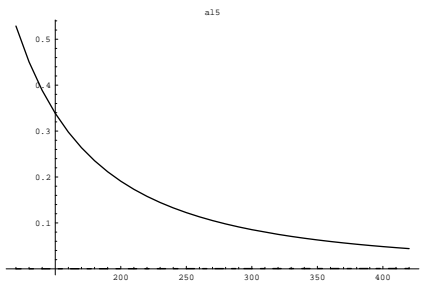
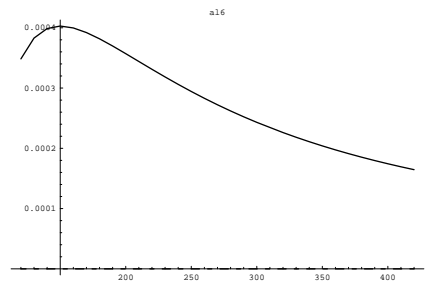
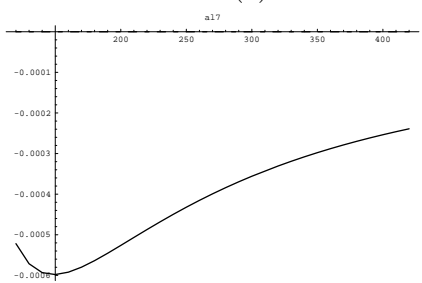
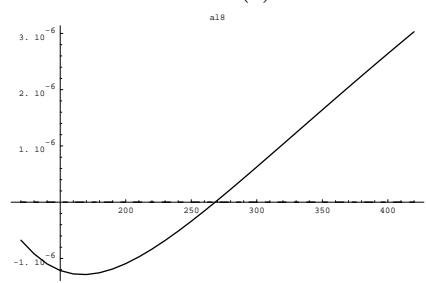
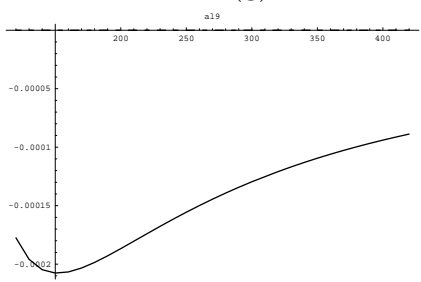
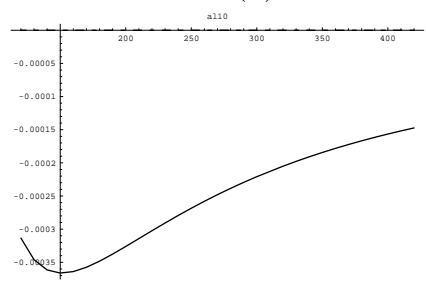
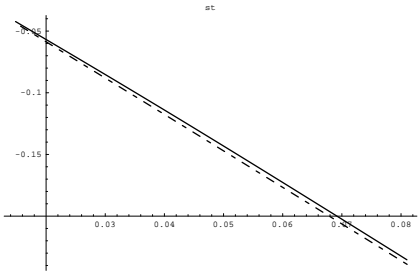
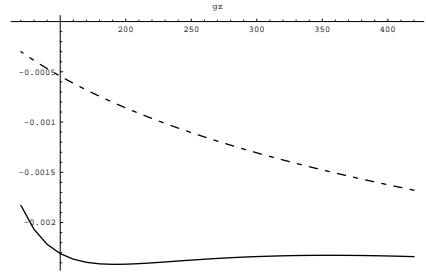
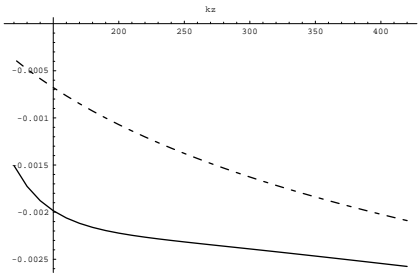
\end{document}